\documentclass[sn-mathphys,Numbered,10pt]{sn-jnl}

\usepackage[english]{babel}

\usepackage{graphicx}%
\usepackage{multirow}%
\usepackage{amsmath,amssymb,amsfonts}%
\usepackage{amscd}
\usepackage{soul}
\usepackage{amsthm}%
\usepackage{mathrsfs}%
\usepackage[title]{appendix}%
\usepackage{xcolor}%
\usepackage{textcomp}%
\usepackage{manyfoot}%
\usepackage{booktabs}%
\usepackage{algorithm}%
\usepackage{algorithmicx}%
\usepackage{algpseudocode}%
\usepackage{listings}%

\usepackage{xypic}
\usepackage{enumitem}

\newtheoremstyle{theoremstyle}
  {1em} 
  {1em} 
  {\itshape} 
  {} 
  {\bfseries} 
  {.} 
  {.5em} 
  {} 

\newtheoremstyle{definitionstyle}
  {1em} 
  {1em} 
  {} 
  {} 
  {\bfseries} 
  {.} 
  {.5em} 
  {} 

\theoremstyle{theoremstyle}
\newtheorem{theorem}{Theorem}[section]
\newtheorem{proposition}[theorem]{Proposition}
\newtheorem{lemma}[theorem]{Lemma}
\newtheorem{corollary}[theorem]{Corollary}

\theoremstyle{definitionstyle}
\newtheorem{definition}[theorem]{Definition}

\newcommand{\R}{\mathbb{R}}

\newcommand{\cA}{\mathcal{A}}

\newcommand{\overbar}[1]{\mkern 1.5mu\overline{\mkern-1.5mu#1\mkern-1.5mu}\mkern 1.5mu}

\makeatletter
\newcommand\RedeclareMathOperator{%
  \@ifstar{\def\rmo@s{m}\rmo@redeclare}{\def\rmo@s{o}\rmo@redeclare}%
}
\newcommand\rmo@redeclare[2]{%
  \begingroup \escapechar\m@ne\xdef\@gtempa{{\string#1}}\endgroup
  \expandafter\@ifundefined\@gtempa
     {\@latex@error{\noexpand#1undefined}\@ehc}%
     \relax
  \expandafter\rmo@declmathop\rmo@s{#1}{#2}}
\newcommand\rmo@declmathop[3]{%
  \DeclareRobustCommand{#2}{\qopname\newmcodes@#1{#3}}%
}
\@onlypreamble\RedeclareMathOperator
\makeatother

\RedeclareMathOperator{\Re}{Re}
\RedeclareMathOperator{\Im}{Im}
\RedeclareMathOperator{\c}{c}

\DeclareMathOperator{\End}{End}

\newcommand{\vs}[0]{\vspace{2mm}}

\newcommand{\mattwo}[4]{
\left(
\begin{array}{cc}
#1 & #2 \\
#3 & #4 \\
\end{array}
\right)
}

\newcommand{\smallmattwo}[4]{
\left(
\begin{smallmatrix}
#1 & #2 \\
#3 & #4 \\
\end{smallmatrix}
\right)
}

\newcommand{\wh}[1]{\widehat{#1}}
\newcommand{\til}[1]{\widetilde{#1}}

\begin{document}

\title[A quantization of moduli spaces of 3-dimensional gravity]{A quantization of moduli spaces of 3-dimensional gravity}

\author[1]{\fnm{Hyun Kyu} \sur{Kim}}\email{hkim@kias.re.kr}

\author[2]{\fnm{Carlos} \sur{Scarinci}}\email{cscarinci@dgist.ac.kr}

\affil[1]{\orgdiv{School of Mathematics}, \orgname{Korea Institute for Advanced Study (KIAS)},\\ \orgaddress{\street{85 Hoegi-ro Dongdaemun-gu}, \city{Seoul} \postcode{02455}, \country{Republic of Korea}}}

\affil[2]{\orgname{Daegu Gyeongbuk Institute of Science and Technology (DGIST)},\\ \orgaddress{\street{333 Techno Jungang-daero, Dalseong-gun}, \city{Daegu} \postcode{42988}, \country{Republic of Korea}}}


\abstract{
	We construct a quantization of the moduli space $\mathcal{GH}_\Lambda(S\times\mathbb{R})$ of maximal globally hyperbolic Lorentzian metrics on $S\times \mathbb{R}$ with constant sectional curvature $\Lambda$, for a punctured surface $S$. Although this moduli space is known to be symplectomorphic to the cotangent bundle of the Teichm\"uller space of $S$ independently of the value of $\Lambda$, we define geometrically natural classes of observables leading to $\Lambda$-dependent quantizations. Using special coordinate systems, we first view $\mathcal{GH}_\Lambda(S\times\mathbb{R})$ as the set of points of a cluster $\mathscr{X}$-variety valued in the ring of generalized complex numbers $\mathbb{R}_\Lambda = \mathbb{R}[\ell]/(\ell^2+\Lambda)$. We then develop an $\mathbb{R}_\Lambda$-version of the quantum theory for cluster $\mathscr{X}$-varieties by establishing $\mathbb{R}_\Lambda$-versions of the quantum dilogarithm function. As a consequence, we obtain three families of projective unitary representations of the mapping class group of $S$. For $\Lambda <0$ these representations recover those of Fock and Goncharov, while for $\Lambda\geq 0$ the representations are new.
}


\maketitle

\tableofcontents

\setlength{\parskip}{10pt plus 2pt minus 1pt}

\section{Introduction}

\subsection{3d gravity and Teichm\"uller theory}

Three dimensional General Relativity, also known as 3d gravity, is a simple yet non-trivial toy model describing the gravitational force in physics \cite{DJ84,DJtH84}. In its essence, it consists of the study of Lorentzian solutions of the Einstein equation on 3-dimensional manifolds, their classification and the properties of their moduli space.
On one hand, the theory is locally trivial: Einstein 3-manifolds have constant sectional curvature and are, therefore, locally isometric to a model homogeneous geometry, depending only on the value of a cosmological constant $\Lambda\in\mathbb{R}$. On the other hand, global and asymptotic properties of the solutions make the theory rich enough to accommodate important physical phenomena such as point-particles, black-holes and holography \cite{tHo88,BTZ92,BH86,Wit07}.
In addition, the problem of quantization of the gravitational force on 3-dimensional manifolds reduces from the difficult realm of quantum field theory to the better tamed realm of quantum mechanics, thus opening up the possibility of establishing a well defined and mathematically rigorous theory of quantum gravity.

Under mild assumptions on causality, and with an appropriate choice of boundary conditions, the moduli space of 3d gravity is shown to be a finite dimensional symplectic manifold, closely related to the Teichm\"uller space of a Riemann surface \cite{Wit88,Mon89,Mes07,Sca99}. This has motivated the development of a quantum theory of Teichm\"uller spaces, whose first major results were established by Kashaev \cite{Kas98} and independently by Chekhov and Fock \cite{FC99,Foc97}. These works build on global parametrizations of the Teichm\"uller space of a punctured surface in terms of Penner's lambda lengths and Thurston's shear coordinates \cite{Pen87,Thu98}, a simple combinatorial description of the Weil-Petersson symplectic structure \cite{Wol83,FC99,Foc97} and of the mapping class group action in terms of these coordinates, and special properties of the so-called {\it quantum dilogarithm function}, studied by Faddeev and Kashaev \cite{FK94}. These were later generalized by Fock and Goncharov to a more general theory of quantum cluster varieties and their representations \cite{FG09a}.

Quantum Teichm\"uller theory, however, is not a theory of 3d quantum gravity. In fact, the moduli space of 3-dimensional Einstein metrics is not identical to the Teichm\"uller space but rather to a certain geometric bundle over that space. This follows from reformulations of 3d gravity as a Chern-Simons gauge theory of connections on principal bundles as developed by Ach\'ucarro-Townsend and Witten \cite{AT86,Wit88} or, equivalently, as a constrained Hamiltonian system for 2-dimensional Riemannian metrics developed by Moncrief \cite{Mon89}.
From a more geometric perspective, it is a result of the classification of constant curvature Lorentzian 3-manifolds initiated by Mess \cite{Mes07,Sca99,Bar05,BB09a}, obtained by employing tools from low dimensional topology and geometry first developed by Thurston in his study of 3-dimensional hyperbolic geometry \cite{Thu97,EM87}.
Specifically, when the ambient topological 3-manifold is $M = S\times \mathbb{R}$, for an oriented surface $S$, the moduli space $\mathcal{GH}_\Lambda(S\times \mathbb{R})$ of maximal globally hyperbolic metrics with constant curvature $\Lambda\in\mathbb{R}$ can be parametrized by the bundle $\mathcal{ML}(S)$ of measured geodesic laminations over the Teichm\"uller space $\mathcal{T}(S)$ via Lorentzian counterparts of Thurston's bending construction or, equivalently, by other closely related bundles depending on the value of $\Lambda$ via earthquakes and grafting:
\begin{align*}
	\mathcal{GH}_\Lambda(S\times\mathbb{R}) = \mathcal{ML}(S) =
	\begin{cases}
		\mathcal{T}(S)\times \overbar{\mathcal{T}}(S), & \Lambda <0,
		\cr
		T\mathcal{T}(S),                               & \Lambda = 0,
		\cr
		\mathcal{CP}(S),                               & \Lambda > 0.
	\end{cases}
\end{align*}
Here $\overbar{\mathcal{T}}(S) = \mathcal{T}(S^\mathrm{op})$ denotes the Teichm\"uller space of $S$ with reversed orientation and $\mathcal{CP}(S)$ denotes the moduli space of complex projective structures on $S$. Under these last parametrizations, the gravitational symplectic structure on $\mathcal{GH}_\Lambda(S\times\mathbb{R})$ is given respectively by: the difference of Weil-Petersson symplectic forms $\omega_{WP}$ on each copy of $\mathcal{T}(S)$ for $\Lambda < 0$; the canonical cotangent bundle symplectic form on $T^*\mathcal{T}(S)$ for $\Lambda =0$, under the identification between $T^*\mathcal{T}(S)$ and $T\mathcal{T}(S)$ induced by $\omega_{WP}$; and the imaginary part of the Goldman symplectic form $\omega_G$ on $\mathcal{CP}(S)$ for $\Lambda > 0$. Note, in particular, that such parametrizations also induce a natural action of the mapping class group ${\rm MCG}(S) = \mathrm{Diff}^+(S) / \mathrm{Diff}(S)_0$ of $S$ on the moduli space $\mathcal{GH}_\Lambda(S\times\mathbb{R})$ of 3d gravity.

The bundle of measured geodesic laminations $\mathcal{ML}(S)$ can be seen as a universal parameter space for 3d gravity, independent of the value of the cosmological constant $\Lambda$ \cite{BB09a}. What is more, for distinct values $\Lambda,\Lambda'\in\mathbb{R}$, one has a canonical homeomorphism $\mathcal{GH}_\Lambda(S\times\mathbb{R})\to \mathcal{GH}_{\Lambda'}(S\times\mathbb{R})$, factoring through $\mathcal{ML}(S)$ and preserving the gravitational symplectic structures \cite{KS09,SS18}. In other words, in terms of the grafting parametrization, the gravitational symplectic structure can be seen as a $\Lambda$-independent quantity on $\mathcal{ML}(S)$. This may seem to imply that a quantum theory of gravity in 3 dimensions can be formulated independently of the cosmological constant as a quantization of $\mathcal{ML}(S) = T^*\mathcal T(S)$. This is indeed correct, however it cannot be the whole story since a quantum theory of gravity should also encompass information about the underlying classical geometry which does depend on the value of $\Lambda$. We contend that this apparent inconsistency is naturally resolved in the context of deformation quantization where, besides the choice of a symplectic/Poisson manifold, one must also select an appropriate class of observables to be quantized. In the present paper, we will choose observables arising as suitable versions of the trace-of-monodromy functions, also known as Wilson loops.

In \cite{MS16} Meusburger and the second author introduced coordinate systems on the moduli space $\mathcal{GH}_\Lambda(S\times \mathbb{R})$, for punctured surfaces $S$, via certain analytic continuations of the shear coordinates on the Teichm\"uller space $\mathcal{T}(S)$. One starts with the choice of an ideal triangulation $T$ of the surface $S$, that is, a maximal collection of isotopy classes of non-intersecting simple paths running between the punctures of $S$, called ideal arcs, dividing the surface into ideal triangles. For each $i\in T$, one obtains a pair of coordinate functions $x_i,y_i$ on $\mathcal{GH}_\Lambda(S\times \mathbb{R})$, interpreted as shearing and bending parameters along the ideal arc $i$ on a Cauchy surface in $S\times\R$.
The functions $x_i$ and $y_i$ are real valued, but are more naturally viewed as the real and the imaginary parts of a single coordinate function $z_i:= x_i + \ell y_i$, taking values in the ring of {\it generalized complex numbers}
\begin{align*}
	\mathbb{R}_\Lambda = \mathbb{R}[\ell]/(\ell^2+\Lambda).
\end{align*}
This ring is isomorphic to $\mathbb{R}^2$ equipped with the multiplication rule $(x,y)\cdot (u,v) = (xu-\Lambda yv, xv+yu)$. It naturally arises in the study of 3-dimensional constant curvature Lorentzian geometries since the group of isometries of the corresponding model spacetimes are isomorphic to $\mathrm{PSL}_2(\mathbb{R}_\Lambda)$. Moreover, in terms of generalized complex numbers, the relation between earthquakes and grafting can be interpreted via an $\mathbb{R}_\Lambda$-analytic continuation of the measure on geodesic laminations, providing a unified description of the Lorentzian bending construction in terms of generalized complex earthquakes \cite{McM98,Meu06,Meu07}.

The symplectic structure on $\mathcal{GH}_\Lambda(S\times \mathbb{R})$, or more precisely the corresponding Poisson brackets, are shown to behave nicely in terms of these coordinates. They can be simply described as $\{Z_i,Z_j\} = \ell \varepsilon_{ij}  Z_i Z_j$, where $Z_i := \exp(x_i+ \ell y_i)$ are the exponentiated versions of the $\mathbb{R}_\Lambda$-valued coordinates and $\varepsilon_{ij}$ are integers encoding the combinatorics of the ideal triangulation $T$ via
\begin{align*}
	\varepsilon_{ij} & = a_{ij} - a_{ji},                                                      \\
	a_{ij}           & = \mbox{ the number of corners of ideal triangles }
	\cr
	                 & \quad\;\;\;  \mbox{delimited by $i$ on the right and $j$ on the left };
\end{align*}
we call $\varepsilon = \varepsilon_T = \{\varepsilon_{ij}\}_{i,j\in T}$ the {\it exchange matrix} of $T$.

\vs

Since there is no canonical choice of ideal triangulation $T$ for a given punctured surface $S$, one must keep track of the dependence of all constructions on the choice of $T$. It is well known that any change of ideal triangulations $T\leadsto T'$ is generated by simple moves called {\it mutations} (or {\it flips}), which change the triangulation one arc at a time. Then, when $T'$ is obtained from $T$ by applying the mutation at an arc $k$, the exponentiated $\mathbb{R}_\Lambda$-valued coordinates transform according to the following rational formulas:
\begin{align}
	\label{eq:intro_Z_transformation}
	Z_i' = \left\{
	\begin{array}{ll}
		Z_k^{-1}                                                      & \mbox{if $i=k$},     \\
		Z_i(1+Z_k^{-{\rm sgn}(\varepsilon_{ik})})^{-\varepsilon_{ik}} & \mbox{if $i\neq k$},
	\end{array}
	\right.
\end{align}
where ${\rm sgn}(a)=1$ if $a>0$ and ${\rm sgn}(a)=-1$ if $a<0$. The exchange matrix transforms as
\begin{align}
	\label{eq:intro_varepsilon_transformation}
	\varepsilon'_{ij} = \left\{
	\begin{array}{ll}
		-\varepsilon_{ij}                                                                                        & \mbox{if $k\in \{i,j\}$},     \\
		\varepsilon_{ij} + \frac{1}{2}(\varepsilon_{ik}|\varepsilon_{kj}| + |\varepsilon_{ik}| \varepsilon_{kj}) & \mbox{if $k \notin \{i,j\}$}.
	\end{array}
	\right.
\end{align}
One can immediately recognize the above coordinate transformation formula in eq.\eqref{eq:intro_Z_transformation} as the cluster $\mathscr{X}$-mutation formula for the cluster $\mathscr{X}$-varieties of Fock and Goncharov \cite{FG06,FG09b}. On the other hand, the coordinate variables $Z_i$ and their Poisson brackets are intrinsically different from those in the usual cluster variety setting, taking values in $\R_\Lambda$ instead of $\R$. This difference prevents a direct application of the results of Fock and Goncharov \cite{FG09a} on quantization of cluster varieties to the context of 3d gravity, and calls for an appropriate modification of their methods.

\subsection{The quantization problem}

Let us now precisely formulate the problem of quantization of the moduli space $\mathcal{GH}_\Lambda(S\times \mathbb{R})$.
In general, for a Poisson manifold $M$, a {\it quantization} consists of a complex Hilbert space $\mathscr{H}$, a ring of classical observables  $\mathcal{A} \subset C^\infty(M)$ to be quantized, and a quantization map
$$
	{\bf Q}^\hbar~:~\mathcal{A} \to \{\mbox{self-adjoint operators on $\mathscr{H}$}\}
$$
depending real analytically on a quantum parameter $\hbar\in\R$. These are required to satisfy
\begin{itemize}
	\item[\rm 1)] ${\bf Q}^\hbar$ is $\mathbb{R}$-linear,

	\item[\rm 2)] ${\bf Q}^\hbar(1) = {\rm Id}$,

	\item[\rm 3)]
	      $
		      [{\bf Q}^\hbar(f), {\bf Q}^\hbar(g)] = {\rm i} \hbar {\bf Q}^\hbar(\{f,g\}) + o(\hbar) \quad \mbox{as $\hbar \to 0$}.
	      $
\end{itemize}
If, moreover, there is a symmetry group acting on $M$, preserving the Poisson structure and the classical ring $\mathcal{A}$, one would also like the resulting quantization to be equivariant under such an action;  see \S\ref{sec:a_deformation_quantization_problem} for more details.

We often regard $\mathcal{A}\subset C^\infty(M;\mathbb{C})$ as a subring of the ring of complex-valued functions on $M$, equipped with the natural $*$-structure, and decompose the quantization map as
$$
	\xymatrix{
	\mathcal{A} \ar[r]^-{\wh{Q}^\hbar} & \mathcal{A}^\hbar \ar[r]^-{\pi^\hbar} & \big\{\mbox{densely-defined operators on $\mathscr{H}$}\big\}.
	}
$$
Here $\mathcal{A}^\hbar$ is an associative non-commutative $*$-algebra over $\mathbb{C}$ that deforms the classical algebra $\mathcal{A}$ and the map $\wh{Q}^\hbar$ called a deformation quantization map, while $\pi^\hbar$ is a $*$-representation of the quantum algebra $\mathcal{A}^\hbar$ on the Hilbert space $\mathscr{H}$.
When $M$ admits a coordinate system for which the Poisson bracket is simple, one is usually able to quantize the coordinate functions, and perhaps also the polynomial ring generated by these functions. However, such a quantization will in general depend on the choice of the coordinate system and one must establish some compatibility statement for the quantizations resulting from different choices. Choosing an appropriate classical algebra to be quantized, that is big enough to contain interesting functions on the manifold but well-behaved enough so that its quantization is independent of the choice of coordinate systems, then becomes a crucial part of the quantization problem.

\vs

In the case of Teichm\"uller spaces of puncture surfaces, or more generally cluster $\mathscr{X}$-varieties, the ring of classical observables implicitly chosen in \cite{FG09a} is the ring of universally Laurent polynomials in the real-valued cluster $\mathscr{X}$-variables. Generalizing this construction, here we propose the classical ring $\mathcal{A}$ to be the ring of  $\mathbb{R}_\Lambda$-valued functions on the moduli space $\mathcal{GH}_\Lambda(S\times \mathbb{R})$ that are universally Laurent in the coordinate functions $Z_i$.
More precisely, for each triangulation $T$, let $\mathcal{A}_{T}$ denote the Laurent polynomial ring with generators $Z_i$, $i\in T$. Then, for each change of triangulations $T\leadsto T'$, consider the associated classical coordinate transformation map
$$
	\mu_{T,T' } : {\rm Frac}(\mathcal{A}_{T' }) \to {\rm Frac}(\mathcal{A}_{T})
$$
between the corresponding fields of fractions. This can be defined by the composition of coordinate transformations of the form in eq.\eqref{eq:intro_Z_transformation}. Note that they satisfy the consistency equations
$$
	\mu_{T,T'} \circ \mu_{T',T''} = \mu_{T,T''},
$$
for all triples $T,T',T''$. We can then define the classical ring $\cA$ as the ring of universally Laurent functions
$$
	\mathbb{L}_T = \bigcap_{T'} \mu_{T,T'} (\mathcal{A}_{T'}) ~\subset~ \mathcal{A}_T ~ \subset ~ {\rm Frac}(\mathcal{A}_T).
$$
For a pair of distinct triangulations $T,T'$, the rings $\mathbb{L}_T$ and $\mathbb{L}_{T'}$ are naturally isomorphic under the map $\mu_{T,T'}$.

\vs

We can now formulate the quantization problem of the moduli space $\mathcal{GH}_\Lambda(S\times \mathbb{R})$ as the following steps:
\begin{enumerate}
	\item[\rm (Q1)] For each ideal triangulation $T$, construct an associative non-commutative $*$-algebra $\mathcal{A}^\hbar_T$ that deforms the classical coordinate algebra $\mathcal{A}_T$;

	\item[\rm (Q2)] For each change of triangulations $T\leadsto T'$, construct a quantum coordinate change isomorphism $\mu^\hbar_{T,T'} : {\rm Frac}(\mathcal{A}^\hbar_{T'}) \to {\rm Frac}(\mathcal{A}^\hbar_T)$ between the corresponding skew-fields of fractions, that recovers the classical map $\mu_{T,T'}$ as $\hbar\to 0$, and such that the consistency equations
	      $$
		      \mu^\hbar_{T,T'} \circ \mu^\hbar_{T',T''} = \mu^\hbar_{T,T''}
	      $$
	      hold for all triples $T,T',T''$;

	\item[\rm (Q3)] For each $T$ define the quantum universally Laurent algebra
	      $$
		      \mathbb{L}^\hbar_T = \bigcap_{T'} \mu^\hbar_{T,T'}(\mathcal{A}^\hbar_{T'}) ~\subset~ \mathcal{A}^\hbar_T ~\subset~ {\rm Frac}(\mathcal{A}^\hbar_T)
	      $$
	      and construct a deformation quantization map
	      $$
		      \wh{Q}^\hbar_T ~:~ \mathbb{L}_T \to \mathbb{L}^\hbar_T
	      $$
	      that is compatible with the isomorphisms $\mu_{T,T'}$ and $\mu^\hbar_{T,T'}$, in the sense that
	      $$
		      \wh{Q}^\hbar_T \circ \mu_{T,T'} = \mu^\hbar_{T,T'} \circ \wh{Q}^\hbar_{T'};
	      $$

	\item[\rm (Q4)] For each $T$ construct a $*$-representation of $\mathbb{L}^\hbar_T$ on a dense (Schwartz) subspace $\mathscr{S}_T$ of a Hilbert space $\mathscr{H}_T$
	      $$
		      \pi_T = \pi^\hbar_T : \mathbb{L}^\hbar_T \to \End(\mathscr{S}_T);
	      $$

	\item[\rm (Q5)] For each change $T\leadsto T'$ construct a corresponding unitary intertwining operator
	      $$
		      {\bf K}_{T,T'} = {\bf K}^\hbar_{T,T'} ~:~ \mathscr{H}_{T'} \to \mathscr{H}_T
	      $$
	      that sends $\mathscr{S}_{T'}$ to $\mathscr{S}_T$, satisfies the intertwining equations
	      $$
		      {\bf K}_{T,T'} \circ \pi_{T'}(u) = \pi_T(\mu^\hbar_{T,T'}(u)) \circ {\bf K}_{T,T'}, \qquad \forall u \in \mathbb{L}^\hbar_{T'},
	      $$
	      and satisfies the consistency equations up to multiplicative constants:
	      $$
		      {\bf K}_{T,T'} \circ {\bf K}_{T',T''} = c_{T,T',T''}{\bf K}_{T,T''}.
	      $$
\end{enumerate}
Another stipulation is the equivariance under the mapping class group ${\rm MCG}(S)$. For this, we also require $\mathcal{A}^\hbar_T$, $\mu^\hbar_{T,T'}$ and ${\bf K}_{T,T'}$ to be invariant under the action of ${\rm MCG}(S)$; so they must only depend on the underlying exchange matrices for the relevant ideal triangulations $T,T'$.

The following is the main result of the present paper.
\begin{theorem}[main theorem]
	There exists a solution to the above quantization problem (Q1)--(Q5) for the moduli spaces $\mathcal{GH}_\Lambda(S\times \mathbb{R})$ of 3d gravity.
\end{theorem}

\subsection{Sketch of the construction}

As hinted above, one of the previous works that motivated the present paper is the quantization of the Teichm\"uller space $\mathcal{T}(S)$ of a punctured surface $S$, which is an example of a cluster $\mathscr{X}$-variety. In fact, $\mathcal{T}(S)$ is covered by positive-real coordinate systems enumerated by ideal triangulations $T$, whose coordinate functions $X_i$ (with values in $\mathbb{R}^+ = \mathbb{R}_{>0}$) transform under the change of triangulations as eq.\eqref{eq:intro_Z_transformation}.
We recall that the quantization problem (Q1)--(Q5) for the Teichm\"uller\ space is solved in \cite{FC99,Foc97,FG09a,Kim21b,CKKO20,AK17}, but note that much of these results can be extended to general cluster $\mathscr{X}$-varieties \cite{FG09a}. Namely, instead of an ideal triangulation of a surface, one could begin with an arbitrary skew-symmetric $N\times N$ integer matrix $\varepsilon=(\varepsilon_{ij})_{i,j=1,\ldots,N}$ and define an associated {\it seed} $\Gamma = (\varepsilon,\{X_i\}_{i=1}^N)$, where $X_i$ are now formal commuting variables. These can be thought of as coordinate functions on a split algebraic torus $(\mathbb{G}_m)^N$, whose regular functions form the algebra of Laurent polynomials in $X_i$'s. One can then define new seeds by applying {\it seed mutations} $\mu_k$, labeled by $k\in\{1,2\ldots,N\}$. This produces a new seed $\Gamma' = \mu_k(\Gamma) = (\varepsilon',\{X_i'\}_{i=1}^N)$, with $\varepsilon'$ given by the formula in eq.\eqref{eq:intro_varepsilon_transformation}, called the quiver mutation, and with $X_i'$'s given by a formula as in eq.\eqref{eq:intro_Z_transformation}, called the cluster $\mathscr{X}$-mutation. Starting from an initial seed $\Gamma = (\varepsilon, \{X_i\}_{i=1}^N)$, one produces more seeds by repeatedly applying the mutations. The split algebraic tori corresponding to these seeds are glued with one another along the mutation maps to define the cluster $\mathscr{X}$-variety $\mathscr{X}_{|\varepsilon|}$. Here $|\varepsilon|$ denotes the equivalence class of the initial exchange matrix $\varepsilon$ under quiver mutations.

In \cite{FG09a} Fock and Goncharov quantize the set $\mathscr{X}_{|\varepsilon|}(\mathbb{R}^+)$ of positive-real points of the cluster $\mathscr{X}$-variety. More precisely, they provide a solution to the problems (Q1), (Q2), (Q4) and (Q5) above, while a deformation quantization map (Q3) can be constructed using the `theta' basis of the ring of universally Laurent elements \cite{GHKK18} and the quantized theta basis \cite{DM21}. This will be explained in more details in \S\ref{sec:a_deformation_quantization_problem}. We note that these works on theta bases are not in general completely constructive. However, when the exchange matrix $\varepsilon$ comes from an ideal triangulation of a punctured surface $S$, so that the set of positive real points $\mathscr{X}_{|\varepsilon|}(\mathbb{R}^+)$ recovers the Teichm\"uller space, an explicit construction of a basis of the ring of universally Laurent elements is described by Fock and Goncharov \cite{FG06} in the classical (commutative) context, and by Allegretti and the first author \cite{AK17} in the quantum (non-commutative) context.
Meanwhile, the results of \cite{MS16} suggest an identification of the moduli space $\mathcal{GH}_\Lambda(S\times \mathbb{R})$ as the set $\mathscr{X}_{|\varepsilon|}(\mathbb{R}_\Lambda^+)$ of $\mathbb{R}_\Lambda^+$-points for such $\varepsilon$'s,  where $\mathbb{R}_\Lambda^+ = \exp(\mathbb{R}_\Lambda)$. In fact, $\mathcal{GH}_\Lambda(S\times \mathbb{R})$ can be seen as a symplectic leaf of $\mathscr{X}_{|\varepsilon|}(\mathbb{R}_\Lambda^+)$ consisting of points satisfying the constraint equations
\begin{align}
	\label{eq:intro_constraint}
	\prod_{i \in T} Z_i^{\theta_i} =1
\end{align}
associated to elements $(\theta_i)_{i\in T}$ in the kernel of the exchange matrix, i.e. $\sum_{j\in T} \varepsilon_{ij} \theta_j=0$, $\forall i\in T$; see \S\ref{subsec:generalized_shear_coordinates} and \S\ref{subsec:geometric_structures_on_the_set_of_R_Lambda_positive_points}. This allows us to formulate our quantization problem in the general language of  cluster $\mathscr{X}$-varieties. Namely, for {\it any} initial exchange matrix $\varepsilon$, we aim to quantize the set of $\mathbb{R}_\Lambda^+$-points of the cluster $\mathscr{X}$-variety $\mathscr{X}_{|\varepsilon|}$.

\vs

Note that parts (Q1)--(Q2) constitute the construction of a quantum cluster $\mathscr{X}$-variety, be it for $\mathbb{R}^+$ or $\mathbb{R}_\Lambda^+$. The classical and quantum algebras for our quantization problem for $\mathscr{X}_{|\varepsilon|}(\mathbb{R}^+_\Lambda)$ are isomorphic to the tensor products of two copies of those for $\mathscr{X}_{|\varepsilon|}(\mathbb{R}^+)$ when $\Lambda=-1,1$. Hence the solution to (Q1)--(Q3) for $\mathbb{R}^+$ can be used to obtain a solution to (Q1)--(Q3) for $\mathbb{R}^+_\Lambda$ directly. It can be said that we set up our quantization problem exactly in such a way that the algebraic steps can be solved using previous results for the $\mathbb{R}^+$-points of a cluster variety. The $\Lambda=0$ case is more subtle, but can be treated in a similar way; see \S\ref{sec:a_deformation_quantization_problem}.

\vs

Parts (Q4)--(Q5), constituting the construction of a projective $*$-representation of the above obtained quantum cluster $\mathscr{X}$-variety, can be viewed as a major new contribution of the present paper. In dealing with these constructions, one of the main hurdles arises when considering representations of algebras over $\mathbb{R}_\Lambda = \mathbb{R}[\ell]/(\ell^2+\Lambda)$; namely, how do we represent elements $x+\ell y$ on a quantum Hilbert space? To solve this problem, for each ideal triangulation $T$ of $S$, or more generally for a cluster $\mathscr{X}$-seed $\Gamma$, we first recall the Hilbert space $\mathring{\mathscr{H}}_T= L^2(\mathbb{R}^T, \wedge_{i\in T} dt_i)$ used in the Fock-Goncharov quantization and then consider the representation on a doubled version
$$
	\mathscr{H}_T := \mathbb{C}^2 \otimes \mathring{\mathscr{H}}_T \cong \mathring{\mathscr{H}}_T \oplus \mathring{\mathscr{H}}_T.
$$
Here we use the standard representation of $\mathbb{R}_\Lambda$ on the tensor factor $\mathbb{C}^2$ by
$$
	x+ \ell y \mapsto \mattwo{x}{-\Lambda y}{y}{x} \in {\rm End}(\mathbb{C}^2).
$$
In particular, we set $\wh{\ell} = \smallmattwo{0}{-\Lambda}{1}{0} \otimes {\rm Id}$, which is the operator on $\mathscr{H}_T$ representing the imaginary element $\ell\in\mathbb{R}_\Lambda$.

For each triangulation $T$, the quantum algebra $\mathcal{A}^\hbar_{T}$ is defined as the non-commutative algebra generated by the elements $\wh{Z}_i$, $i\in T$, and their inverses, modulo the relations $\wh{Z}_i \wh{Z}_j = e^{2\pi {\rm i} \ell \hbar \varepsilon_{ij}} \wh{Z}_j \wh{Z}_i$. We then construct a representation $\pi_{T}$ of $\mathcal{A}^\hbar_{T}$ on the doubled Hilbert space $\mathscr{H}_T$ by
\begin{align*}
	\pi_{T} (\wh{Z}_i) = \exp({\bf z}_i),
	 &  &
	{\bf z}_i = \mattwo{ {\bf x}_i }{ -\Lambda \hbar {\bf y}_i }{\hbar {\bf y}_i }{{\bf x}_i}
		=  {\rm Id}_{\mathbb{C}^2}\otimes {\bf x}_i + \wh{\ell} \hbar \, ({\rm Id}_{\mathbb{C}^2} \otimes {\bf y}_i),
\end{align*}
where ${\bf x}_i,{\bf y}_i$ are self-adjoint operators on the (single) Hilbert space $\mathring{\mathscr{H}}_T$ that satisfy the Heisenberg relations $[{\bf x}_i, {\bf y}_j] = \pi {\rm i} \varepsilon_{ij}$. More specifically, we take
$$
	{\bf x}_i = - \pi {\rm i} \frac{\partial}{\partial t_i}, \qquad {\bf y}_i = \sum_{j\in T} \varepsilon_{ij} t_j, \qquad \forall i \in T,
$$
which can also be essentially found in Fock-Goncharov's work \cite{FG09a}.  One can easily verify that the desired relations $\pi_{T}(\wh{Z}_i) \pi_{T}(\wh{Z}_j) = e^{2\pi {\rm i} \wh{\ell} \hbar \varepsilon_{ij}} \pi_{T}(\wh{Z}_j) \pi_{T}(\wh{Z}_i)$ hold, so we indeed obtain a representation $\pi_{T}$ of the quantum coordinate ring $\mathcal{A}^\hbar_T$ for $T$, and of the universally Laurent subalgebra $\mathbb{L}^\hbar_{T}$. This provides a solution to (Q4); see \S\ref{subsec:representations_on_doubled_Hilbert_spaces} for more details, in particular for a discussion about the Schwartz space $\mathscr{S}_T$.

\vs

Finally, the last but most important part of our program is (Q5). We focus on the case when $T\leadsto T'$ is a flip at an arc $k$, i.e. for a pair of seeds $\Gamma,\Gamma'$ related by a single mutation $\mu_k = \mu_{T,T'} = \mu_{\Gamma,\Gamma'}$. Following Fock-Goncharov's construction of the mutation intertwiner for the quantization of $\mathscr{X}_{|\varepsilon|}(\mathbb{R}^+)$, we present an answer for the unitary mutation intertwiner ${\bf K}_{T,T'} : \mathscr{H}_{T'} \to \mathscr{H}_T$ as the composition
$$
	\xymatrix{
	\mathscr{H}_{T'} \ar[r]^-{{\bf K}'_{T,T'}} & \mathscr{H}_T \ar[r]^-{{\bf K}^\sharp_{T,T'}} & \mathscr{H}_T.
	}
$$
Here the `monomial transformation part' ${\bf K}'_{T,T'} : \mathscr{H}_{T'} \to \mathscr{H}_T$ is induced by a simple linear transformation $\mathbb{R}^T \to \mathbb{R}^{T'}$ (see \S\ref{subsec:FG_mutation_intertwiner} and \S\ref{subsec:mutation_intertwiner}), and the `automorphism part' ${\bf K}^\sharp_{T,T'}$ is given in terms of a version of functional calculus for $\mathbb{R}_\Lambda$-valued functions (see \S\ref{subsec:representations_on_doubled_Hilbert_spaces}), as
\begin{align}
	\label{eq:intro_our_K_sharp}
	{\bf K}^\sharp_{T,T'} = \Phi^{\ell\hbar}({\bf x}_k + \wh{\ell} \hbar {\bf y}_k) \, \Phi^{-\ell\hbar}({\bf x}_k - \wh{\ell} \hbar {\bf y}_k).
\end{align}
The functions $\Phi^{\pm \ell\hbar}$ stand for $\mathbb{R}_\Lambda$-versions of the quantum dilogarithm of Faddeev and Kashaev \cite{FK94}, which we develop as follows.

The crude version of the quantum dilogarithm is the {\it compact quantum dilogarithm}
$$
	\psi^{\bf q}(z) = \prod_{i=1}^\infty (1+{\bf q}^{2i-1} z)^{-1}
$$
which makes sense as an honest meromorphic function on $\mathbb{C}$ if the parameter ${\bf q}$ is a complex number satisfying $|{\bf q}|<1$. This version can be used to heuristically describe the quantum mutation $\mu^\hbar_{T,T'}$. However, in Chekhov-Fock-Goncharov's mutation intertwiner for $\mathscr{X}_{|\varepsilon|}(\mathbb{R}^+)$ \cite{FC99,FG09a}, one has the condition $|{\bf q}|=1$, so they use the {\it non-compact quantum dilogarithm}
\begin{equation}\label{eq:intro_non-compact_quantum_dilog}
	\Phi^\hbar(z) = \exp\left( - \frac{1}{4} \int_\Omega \frac{e^{-{\rm i} p z}}{\sinh(\pi p) \sinh(\pi \hbar p)} \frac{dp}{p} \right).
\end{equation}
This is defined for real parameters $\hbar \in \mathbb{R}^+$ as an analytic function on the strip $|{\rm Im}(z)| < \pi(1+\hbar)$ in the complex plane, and then analytically continued to a meromorphic function on $\mathbb{C}$. Here $\Omega$ is the contour in $\mathbb{C}$ along the real line that avoids the origin along a small half circle above the origin. This contour integral formula was known already 100 years ago \cite{Bar01}, and was revived in \cite{FK94} for applications to mathematical physics. It enjoys many remarkable properties, and was used in Fock-Goncharov's solution \cite{FG09a} to the automorphism part operator ${\bf K}^\sharp_{T,T'}$ as $\Phi^\hbar({\bf x}_k+\hbar {\bf y}_k) (\Phi^\hbar({\bf x}_k - \hbar {\bf y}_k))^{-1}$, which in particular is a unitary operator satisfying the intertwining equations and the consistency equations.

For our purposes, we need some extension of the quantum dilogarithm function to account for the dependence on the ring of generalized complex numbers $\mathbb{R}_\Lambda$. Noting that the only difference between the Poisson brackets among the $\mathbb{R}^+$-valued coordinates of Fock and Goncharov and those among our $\mathbb{R}_\Lambda^+$-valued coordinates is given by a single factor of $\ell\in\R_\Lambda$, our first proposal for extending the quantum dilogarithm function is given by replacing the real parameter $\hbar$ by the purely imaginary parameter $\ell\hbar \in \mathbb{R}_\Lambda$. This may seem naive at first sight, and in fact, under such replacement, the defining contour integral will be ill-defined for $\Lambda =1$. However, by replacing $\hbar$ by a general complex parameter $h$, and modifying the integral by slanting the contour $\Omega$ by an appropriate angle, i.e. considering a rotated contour $e^{{\rm i} \theta} \Omega$, it becomes possible to define new versions of the quantum dilogarithm function $\Phi^{\pm {\rm i}\hbar}$, for $\hbar \in \mathbb{R}$ (see \S\ref{subsec:trilogy}).

We note that the resulting functions $\Phi^{\pm {\rm i}\hbar}$, relevant for the case $\Lambda = 1$, can be expressed as ratios of honest compact quantum dilogarithms $\psi^{\exp(-\pi\hbar)}$ and $\psi^{\exp(-\pi/\hbar)}$
$$
	\Phi^{{\rm i}\hbar}(z) = \frac{ \psi^{\exp(-\pi\hbar)}(e^z)}{\psi^{\exp(-\pi /  \hbar)}(e^{z/({\rm i} \hbar)})}, \qquad
	\Phi^{-{\rm i}\hbar}(z) = \overline{\Phi^{{\rm i}\hbar}(\overline{z})}^{\,-1}.
$$
This suggests that  $\Phi^{\pm{\rm i}\hbar}$ can be seen as `modular double' versions of the compact quantum dilogarithm.

After a simple computation, one verifies that our proposed automorphism part operator ${\bf K}^\sharp_{T,T'}$ in eq.\eqref{eq:intro_our_K_sharp} is block-diagonal in the sense that ${\bf K}^\sharp_{T,T'} = {\rm Id}_{\mathbb{C}^2} \otimes \check{{\bf K}}^\sharp_{T,T'}$, where on the right-hand-side $\check{{\bf K}}^\sharp_{T,T'}$ denotes the unitary operator on the `single' Hilbert space $\mathring{\mathscr{H}}_T$
$$
	\check{{\bf K}}^\sharp_{T,T'} = \left\{
	\begin{array}{ll}
		\Phi^\hbar({\bf x}_k + \hbar {\bf y}_k) (\Phi^\hbar({\bf x}_k - \hbar {\bf y}_k))^{-1}                                 & \mbox{if $\Lambda=-1$}, \\
		\Phi^{{\rm i}\hbar}({\bf x}_k + {\rm i} \hbar {\bf y}_k) \, \Phi^{-{\rm i} \hbar}({\bf x}_k - {\rm i} \hbar {\bf y}_k) & \mbox{if $\Lambda=1$},  \\
		(1+e^{{\bf x}_k})^{{\bf y}_k/(\pi {\rm i})}                                                                            & \mbox{if $\Lambda=0$}.
	\end{array}
	\right.
$$

We note that for the anti-de Sitter case $\Lambda=-1$, this answer exactly coincides with Fock-Goncharov's intertwiner \cite{FG09a}. On the other hand, for the de Sitter and Minkowski cases $\Lambda=1,0$, the answer gives new unitary operators written respectively in terms of compact quantum dilogarithms and a certain linearized version of the quantum dilogarithm. Note also that for $\Lambda=0$, the intertwining operator does not involve $\hbar$, and although a priori the answer does not seem to be aligned with the answers for $\Lambda=-1,1$, it does indeed come from a same contour integral; see eq.\eqref{eq:F0_contour_integral}. We develop necessary functional equations for the functions $\Phi^{\pm {\rm i}\hbar}$ and $F_0(x,y)=(1+e^{x})^{y/(\pi {\rm i})}$. The desired intertwining equations and consistency equations are proved for the dS case $\Lambda=1$ mainly through analytic continuation of the known results for the AdS case $\Lambda=-1$ \cite{FG09a}, and for the flat case $\Lambda=0$ by direct proofs. This concludes our solution to (Q5) of the quantization problem for $\mathscr{X}_{|\varepsilon|}(\mathbb{R}_\Lambda^+)$, the set of $\mathbb{R}_\Lambda^+$-points of the cluster $\mathscr{X}$-variety $\mathscr{X}_{|\varepsilon|}$, for a general exchange matrix $\varepsilon$.

Coming back to our original motivation, recall that the moduli space $\mathcal{GH}_\Lambda(S\times \mathbb{R})$ is identified with a symplectic leaf of $\mathscr{X}_{|\varepsilon|}(\mathbb{R}_\Lambda^+)$ defined by the equations in eq.\eqref{eq:intro_constraint}, when $\varepsilon$ comes from an ideal triangulation $T$ of $S$. The corresponding constraint operators $\pi_T(\prod_{i\in T} \wh{Z}_i^{\theta_i})$ strongly commute with all other operators $\pi_T(u)$ for $u \in \mathbb{L}^\hbar_T$, so one obtains a quantization for the symplectic leaf $\mathcal{GH}_\Lambda(S\times \mathbb{R})$ by an irreducible representation, either through the simultaneous spectral decomposition of the constraint operators, or more explicitly through the Shale-Weil type construction; see \S\ref{subsec:a_deformation_quantization_of_moduli_spaces_of_3d_gravity} and \cite{Kim21b} for more details, which can be applied also to any exchange matrix $\varepsilon$.

\subsection{Consequences}

In the present paper we provide a precise formulation of the quantization problem of the moduli space $\mathcal{GH}_\Lambda(S\times \mathbb{R})$ of 3d gravity, for punctured surfaces $S$ and for each value of the cosmological constant $\Lambda \in \{-1,0,1\}$, and we establish a complete solution in terms of $\mathbb{R}_\Lambda$-versions of the theory of general quantum cluster $\mathscr{X}$-varieties and their representations.
Accordingly, we introduce a special class of classical observables (the ring of $\mathbb{R}_\Lambda$-valued regular functions/universally Laurent polynomials) on $\mathcal{GH}_\Lambda(S\times \mathbb{R}) = \mathscr{X}_{|\varepsilon|}(\mathbb{R}^+_\Lambda)$ and describe their non-commutative deformations and $*$-representations as quantum operators on an appropriate Hilbert space.
We emphasize that, as a symplectic manifold, $\mathcal{GH}_\Lambda(S\times \mathbb{R})$ is isomorphic to $T^*\mathcal{T}(S)$ for all values of $\Lambda$, so that all of the geometric information about the underlying Einstein 3-manifolds is encoded in the choice of observables.
Viewing the moduli spaces $\mathcal{GH}_\Lambda(S\times \mathbb{R})$ as cluster $\mathscr{X}$-varieties has the immediate benefit of providing a natural class of observables to be quantized.

Our constructions are also inherently invariant under the action of the mapping class group. Each element of ${\rm MCG}(S)$ is realized as the composition of flips and, upon quantization, we obtain $\Lambda$-dependent famililes of projective unitary representations of ${\rm MCG}(S)$.
We refer the readers to \cite{Kim_survey} which is a partial survey of the present paper, providing detailed steps to construct the ${\rm MCG}(S)$ representations using the results of the present paper.
For $\Lambda=-1$, these representations coincide with Chekhov-Fock-Goncharov's representations \cite{FC99,FG09a} based on the non-compact quantum dilogarithm function $\Phi^\hbar$ of Faddeev-Kashaev \cite{FK94}. For $\Lambda=1$, the representations provide a new family of projective unitary representations of the mapping class groups, in terms of the modular double versions $\Phi^{\pm {\rm i} \hbar}$ of the compact quantum dilogarithm function.
For $\Lambda=0$ the representation does not form a family because it does not involve the quantum parameter $\hbar$. Nonetheless, it is indeed a quantum representation and it also provides a new projective unitary representation of the mapping class group.
It would be interesting to compare our results with yet another quantum representation recently constructed by Goncharov and Shen \cite{GS19}, as both our results and theirs can be interpreted as letting the Planck constant $\hbar$ to have values other than real numbers.

Our constructions suggest natural generalizations of results and conjectures from the theory of quantum Teichm\"uller spaces and quantum cluster varieties. For example, the connection between quantum group representation theory and quantum Teichm\"uller theory \cite{FK12} generalizes to three dimensions \cite{Kim} using the $\Lambda$-dependent famililes of quantum dilogarithms $\Phi^{\pm\ell\hbar}$. Also, the ${\rm SL}_2(\mathbb{R})$ modular functor conjecture \cite{FG09a,Tes07}, on the relation among the quantum representations for different surfaces, immediately generalizes to an ${\rm SL}_2(\mathbb{R}_\Lambda)$-version which might eventually lead to a 3d topological quantum field theory relevant to 3d quantum gravity.

Finally, it would also be interesting to investigate possible applications of our methods to the quantization of the moduli space of asymptotically anti-de Sitter 3-manifolds \cite{BH86}. In this context, the appropriate boundary conditions are defined asymptotically, fixing the induced conformal structure on the boundary but allowing for additional degrees of freedom. The corresponding moduli space is then an infinite dimensional manifold, closely related to the universal Teichm\"uller space of quasiconformal deformations of the unit disk \cite{SK13}. One may thus expect that a generalization of our constructions, along the lines of Penner's and Fock-Goncharov's theory of universal Teichm\"uller space \cite{Pen93,FG06}, could provide a quantum theory of asymptotically anti-de Sitter 3d gravity. This is the most interesting situation, relating to 3d multi-black hole solutions \cite{BTZ92,ABB+98,Bar08b}, to the holographic principle and the AdS/CFT conjecture \cite{Mal98,Wit98} and ultimately to the Monster vertex operator algebra \cite{Wit07,FP20}.

\section{The moduli spaces of 3d gravity}
\label{sec:prototypical_examples}

\subsection{Maximal globally hyperbolic Einstein 3-manifolds}
\label{subsec:moduli_spaces_3d_gravity}

On 3-dimensional manifolds, the Einstein equation $\mathrm{Ric}-\frac{1}{2}R g = \Lambda g$, for a pseudo-Riemannian metric $g$ and a cosmological constant $\Lambda\in\R$, imposes a much stronger condition than on higher dimensional manifolds. As the Ricci tensor determines the full Riemann curvature tensor on manifolds of dimension 3, the solutions to the Einstein equation must all have constant sectional curvature $\Lambda$, and are therefore locally isometric to a homogeneous model geometry $\mathbb{X}_\Lambda = G_\Lambda/H$.
In Lorentzian signature, the isometry group $G_\Lambda$ can be identified with the projective special linear group $\mathrm{PSL}_2(\mathbb{R}_\Lambda)$ over the {\it ring of generalized complex numbers}
$$\mathbb{R}_\Lambda = \mathbb{R}[\ell]/(\ell^2+\Lambda)$$
and therefore depends on the value of the cosmological constant $\Lambda$, while the isotropy group $H$ is always isomorphic to the projective special linear group $\mathrm{PSL}_2(\mathbb{R})$ over the real numbers $\mathbb{R}$, isomorphic to the Lorentz group in dimension 3:
\begin{align*}
	G_\Lambda = \mathrm{PSL}_2(\mathbb{R}_\Lambda) =
	\begin{cases}
		\mathrm{PSL}_2(\mathbb{R})\times \mathrm{PSL}_2(\mathbb{R}),   & \Lambda < 0,
		\cr
		\mathrm{PSL}_2(\mathbb{R})\ltimes \mathfrak{sl}_2(\mathbb{R}), & \Lambda = 0,
		\cr
		\mathrm{PSL}_2(\mathbb{C}),                                    & \Lambda > 0,
	\end{cases}
	 &  &
	H = \mathrm{PSL}_2(\mathbb{R}).
\end{align*}
Changes in the magnitude of $\Lambda$ correspond to rescalings of the corresponding model Lorentzian metrics so, in essence, there are only three distinct local geometric models. These are the {\it anti-de Sitter space} $\mathbb{X}_{-1} = \mathrm{AdS}^3$ for $\Lambda = -1$, the {\it Minkowski space} $\mathbb{X}_{0} = \mathrm{Mink}^3$ for $\Lambda = 0$, and the {\it de Sitter space} $\mathbb{X}_{1} = \mathrm{dS}^3$ for $\Lambda = 1$.

A full classification of 3-dimensional Lorentzian Einstein manifolds can be achieved under mild assumptions on the causality of $M$ \cite{Mes07,Sca99,Bar05,BB09a}. We recall that a Lorentzian 3-manifold $(M,g)$ is called {\it globally hyperbolic} if it contains an embedded surface $S$ which intersects every inextendible timelike curve exactly once, see e.g. \cite{BE81,ONe83,HE73}; in particular, $M$ is homeomorphic to the product $S\times\mathbb{R}$. A globally hyperbolic 3-manifold is called {\it maximal} if every isometric embedding $(M,g) \to (M',g')$ into another globally hyperbolic 3-manifold $(M',g')$ is a global isometry. Finally, a maximal globally hyperbolic Lorentzian manifold $(M,g)$ is said to be {\it Cauchy-complete} if the induced Riemannian metric on the submanifold $S$ is geodesically complete. When the Cauchy surface $S$ is non-compact, we also need to impose boundary conditions for the maximal globally hyperbolic Einstein metrics. Here we will only consider surfaces $S$ of finite type and we will require that the holonomy around each of the boundary components be in a parabolic conjugacy class.
\begin{definition}
	We denote by $\mathcal{GH}_\Lambda(S\times\mathbb{R})$ the moduli space of maximal globally hyperbolic, Cauchy complete, Lorentizian metrics on $S\times\mathbb{R}$ with constant sectional curvature $\Lambda$ and parabolic boundary holonomies, considered up to isotopy.
\end{definition}

\subsection{Grafting parametrization and symplectic structures}

The moduli space $\mathcal{GH}_\Lambda(S\times\mathbb{R})$ can be parametrized, via a Lorentzian version of Thurston's grafting construction, as the bundle $\mathcal{ML}(S)$ of measured geodesic laminations over the Teichm\"uller space $\mathcal{T}(S)$ of the Cauchy surface $S$.
We recall that the {\it Teichm\"uller space} of an orientable surface $S$ is the space of {\it complete hyperbolic metrics} $h$ on $S$, considered modulo isotopy. Here, if the surface $S$ is non-compact, we assume that the boundary components are given by parabolic cusps; in other words, we consider only the case of finite area complete hyperbolic metrics $h$ on $S$. Given a hyperbolic metric $h$ on $S$, a {\it measured geodesic lamination} $\lambda$ is defined as a closed subset of $S$ foliated by complete geodesics of $h$, together with a positive Borel measure on transverse arcs and satisfying suitable properties; see e.g. \cite{Thu86,Bon96} for details. For non-compact hyperbolic surfaces $(S,h)$, we will also assume that the measured geodesic laminations are all compactly supported; in other words, each of their geodesic leaves must be bounded away from the cusps of $(S,h)$.
Measured geodesic laminations can be used to deform hyperbolic metrics via a cutting and gluing construction called an {\it earthquake}. Given a hyperbolic surface $(S,h)$ and a measured geodesic lamination $\lambda = (c,w)$ supported on a simple closed geodesic $c$ with a positice real weight $w$, one can then construct another hyperbolic surface $(S,h' = \mathrm{Eq}(h,\lambda))$ by cutting $S$ along $c$ and gluing the pieces back together after applying a hyperbolic translation of length $w$ to the right component. Such a deformation immediately generalizes to measured geodesic laminations supported on multicurves, that is disjoint union of simple closed geodesics, and from those to more general measured geodesic laminations via a limiting argument with respect to a suitable topology on $\mathcal{ML}(S)$; see \cite{Thu86,Bon96}. The resulting map $\mathrm{Eq}:\mathcal{ML}(S)\to\mathcal{T}(S)$ gives a bijection when restricted to each fiber of $\mathcal{ML}(S)$, so that every pair of hyperbolic metrics $h$ and $h'$ are related via earthquake along a unique measured geodesic lamination $\lambda\in\mathcal{ML}_h(S)$ \cite{Thu86,Ker83}.

The correspondence between the moduli space $\mathcal{GH}_\Lambda(S\times \mathbb{R})$ and the bundle $\mathcal{ML}(S)$ is obtained via a similar construction called {\it grafting}. Given a hyperbolic metric $h$ on a surface $S$, there is a unique `Fuchsian' maximal globally hyperbolic Einstein metric on $S\times \mathbb{R}$, foliated by hyperbolic Cauchy surfaces proportional to $(S,h)$. Given a measured geodesic lamination $\lambda = (c,w)$ supported on a simple closed geodesic $c$, one can then construct a deformation of this Fuchsian metric as follows: one cuts the 3-manifold $S\times\mathbb{R}$ along the timelike surface $c\times\mathbb{R}$ and replaces it by the product $c\times [0,w]\times\mathbb{R}$, with an appropriate extension of the induced metric on $c\times\mathbb{R}$. The resulting metric is globally hyperbolic, but might no longer be maximal. Nonetheless, it always admits a unique maximal extension, thus determing a point $g = \mathrm{Gr}_\Lambda(h,\lambda)\in \mathcal{GH}_\Lambda(S\times \mathbb{R})$. The construction for general measured geodesic laminations is again obtained via a limiting argument. Now, however, the resulting {\it grafting map} $\mathrm{Gr}_\Lambda:\mathcal{ML}(S)\to\mathcal{GH}_\Lambda(S\times\mathbb{R})$ is shown to be bijective: every maximal globally hyperbolic Einstein 3-manifold has a regular cosmological time function, whose level sets can be used to recover the original hyperbolic metric and the measured geodesic lamination \cite{Bon05,BB09a}.

The bundle of measured geodesic laminations $\mathcal{ML}(S)$ can thus be seen as a universal parameter space for maximal globally hyperbolic Einstein spacetimes, independently of the value of the cosmological constant $\Lambda\in\mathbb{R}$. On the other hand, the specific geometry of the Fuchsian metric and of its grafting deformation do depend explicitly on $\Lambda$. In particular, for distinct values of $\Lambda$, the construction above also provides $\Lambda$-dependent parametrizations of the moduli space $\mathcal{GH}_\Lambda(S\times\mathbb{R})$, obtained in terms of the geometry of special embedded surfaces on the maximal globally hyperbolic Einstein spacetimes, or on certain dual spaces determined via projective duality. These are given explicitly by: the product of two copies of the Teichm\"uller space $\mathcal{T}(S)\times\mathcal{T}(S)$ for $\Lambda = -1$, obtained by the left and right earthquake maps from $\mathcal{ML}(S)$; by the tangent bundle of the Teichm\"uller space $T\mathcal{T}(S)$ for $\Lambda = 0$, obtained by the infinitesimal earthquake map; and by the moduli space $\mathcal{CP}(S)$ of complex projective structures (with parabolic cusps) on $S$ modulo isotopy for $\Lambda = 1$, obtained by the Thurston grafting map from $\mathcal{ML}(S)$. Here, we will denote the grafting map by
\begin{align*}
	\mathrm{Gr}_\Lambda: \mathcal{ML}(S) \to \mathcal{GH}_\Lambda(S\times\mathbb{R}) =
	\begin{cases}
		\mathcal{T}(S)\times\mathcal{T}(S), & \Lambda =-1,
		\cr
		T\mathcal{T}(S),                    & \Lambda =0,
		\cr
		\mathcal{CP}(S),                    & \Lambda =1.
	\end{cases}
\end{align*}

Note that the $\Lambda$-dependent versions of the grafting map can also be given a unified interpretation as $\mathbb{R}_\Lambda$-complex earthquakes, obtained via analytic continuation of the measures on geodesic laminations with fixed support \cite{McM98,Meu06,Meu07}. In particular, the moduli space $\mathcal{GH}_\Lambda(S\times\mathbb{R})$ can be viewed as $\mathbb{R}_\Lambda$-complexification of the Teichm\"uller space. More precisely, $\mathcal{GH}_\Lambda(S\times\mathbb{R})$ admits an integrable almost product structure for $\Lambda=-1$, $\mathcal{GH}_\Lambda(S\times\mathbb{R})$ an integrable almost tangent structure for $\Lambda=0$ and an integrable almost complex structure for $\Lambda=1$, together with a totally real analytic embedding of $\mathcal{T}(S)$, given by the diagonal embedding into $\mathcal{T}(S)\times\mathcal{T}(S)$ for $\Lambda = -1$, the zero section of $T\mathcal{T}(S)$ for $\Lambda = 0$, and the Fuchsian embedding into $\mathcal{CP}(S)$ for $\Lambda = 1$, respectively.

\vs

Turning towards the description of symplectic structure on $\mathcal{GH}_\Lambda(S\times\mathbb{R})$, let us start by recalling that the Teichm\"uller space $\mathcal{T}(S)$ is endowed with a natural symplectic structure, the so-called Weil-Petersson symplectic structure \cite{Pet40,Wei58,Wol83}. The holonomy representation of hyperbolic structures on a 2-dimensional surface $S$ determines a local diffeomorphism
$$\mathrm{hol}:\mathcal{T}(S)\to \mathrm{Rep}(\pi_1(S),\mathrm{PSL}_2(\mathbb{R}))=\mathrm{Hom}(\pi_1(S),\mathrm{PSL}_2(\mathbb{R}))/\mathrm{PSL}_2(\mathbb{R}),$$
surjective onto a connected component of the $\mathrm{PSL}_2(\mathbb{R})$-representation variety of $S$ with maximal Euler number. In terms of this map, the Weil-Petersson symplectic structure on $\mathcal{T}(S)$ is identified with the restriction of the Atiyah-Bott-Goldman symplectic form \cite{AB83,Gol84} on $\mathrm{Rep}(\pi_1(S),\mathrm{PSL}_2(\mathbb{R}))$, defined by the group cohomology cup product on each $H^1(\pi_1(S),\mathfrak{sl}_2(\mathbb{R})_\rho) = T_\rho\mathrm{Rep}(\pi_1(S),\mathrm{PSL}_2(\mathbb{R}))$, with coefficients paired by the Killing form of $\mathfrak{sl}_2(\mathbb{R})$.
Similarly, we can define a natural symplectic structure on the moduli space of maximal globally hyperbolic Einstein metrics $\mathcal{GH}_\Lambda(S\times\mathbb{R})$. The holonomy representation of maximal globally hyperbolic Einstein metrics also determines a local diffeomorphism
$$\mathrm{hol}:\mathcal{GH}_\Lambda(S\times\mathbb{R})\to\mathrm{Rep}(\pi_1(S),\mathrm{PSL}_2(\mathbb{R}_\Lambda)),$$
and we can consider the pull-back of the Atiyah-Bott-Goldman symplectic form on the representation variety $\mathrm{Rep}(\pi_1(S),\mathrm{PSL}_2(\mathbb{R}_\Lambda))$. Here, however, there are distinct possible choices for this symplectic form, since the Lie algebra $\mathfrak{sl}_2(\mathbb{R}_\Lambda)$ admits a 1-parameter family of non-degenerate $\mathrm{PSL}_2(\mathbb{R}_\Lambda)$-invariant symmetric bilinear forms, up to constant rescaling, determined essentially by the real and the imaginary parts of the $\mathfrak{sl}_2(\mathbb{R}_\Lambda)$-Killing form.
The symplectic structure on $\mathcal{GH}_\Lambda(S\times\mathbb{R})$ relevant to 3d gravity is defined by the imaginary part of the Killing form, so it will be refered to as the {\it gravitational symplectic structure}. In terms of the $\Lambda$-dependent parametrizations of $\mathcal{GH}_\Lambda(S\times\mathbb{R})$ described above, this is equivalent to: the difference of Weil-Petersson symplectic forms on $\mathcal{T}(S)\times\mathcal{T}(S)$ for $\Lambda = -1$; the canonical cotangent bundle symplectic form on $T^*\mathcal{T}(S)$ for $\Lambda = 0$, under the identification between $T^*\mathcal{T}(S)$ and $T\mathcal{T}(S)$ induced by the Weil-Petersson symplectic form; and the imaginary part of the holomorphic Goldman symplectic structure on $\mathcal{CP}(S)$ for $\Lambda = 1$. Remarkably, all of these symplectic forms turn out to be equivalent, in the sense that the composition $\mathrm{Gr}_{\Lambda'}\circ\mathrm{Gr}_{\Lambda}^{-1}:\mathcal{GH}_\Lambda(S\times\mathbb{R})\to \mathcal{GH}_{\Lambda'}(S\times\mathbb{R})$ is a symplectomorphism for all values of $\Lambda$ and $\Lambda'$; see \cite{KS09} for $\Lambda = 1$ and \cite{SS18} for $\Lambda = -1$. The case $\Lambda = 0$, is not fully available in the current literature, but can be treated in a similar way.

Another important feature of the moduli space $\mathcal{GH}_\Lambda(S\times\mathbb{R})$ that generalizes from the classical Teichm\"uller space $\mathcal{T}(S)$ is the presence of a natural action of the mapping class group of $S$, ${\rm MCG}(S) = \mathrm{Diff}^+(S) / \mathrm{Diff}^+(S)_0$. This arises from extensions of diffeomorphisms of $S$ to diffeomorphisms of $S\times\mathbb{R}$ preserving the cosmological time function. In particular, it is compatible with the grafting parametrization, in the sense that the map $\mathrm{Gr}_\Lambda$ intertwines the mapping class group actions on $\mathcal{GH}_\Lambda(S\times\mathbb{R})$ and $\mathcal{ML}(S)$. As a consequence, ${\rm MCG}(S)$ preserves the gravitational symplectic structure on $\mathcal{GH}_\Lambda(S\times\mathbb{R})$.

\subsection{Generalized shear coordinates}
\label{subsec:generalized_shear_coordinates}

In \cite{MS16}, earthquakes and grafting were used to extend Thurston's shear coordinates on the Teichm\"uller space $\mathcal{T}(S)$ to $\mathbb{R}_\Lambda$-valued functions on the moduli spaces $\mathcal{GH}_\Lambda(S\times\mathbb{R})$. One starts with the choice of an {\it ideal triangulation} $T$ of the punctured surface $S$, which is a maximal collection of isotopy classes of simple unoriented paths called {\it ideal arcs} running between the punctures and dividing $S$ into {\it ideal triangles}. Here, we only allow ideal triangulations $T$ whose ideal triangles are bounded by three distinct ideal arcs.
The {\it Thurston shear coordinates} on the Teichm\"uller space $\mathcal{T}(S)$ are real analytic hyperbolic invariants associated to the ideal arcs in an ideal triangulation $T$. For hyperbolic surfaces of genus $g$ with $n$ punctures, satisfying $2g-2+n>0$, these form a global constrained coordinate system on $\mathcal{T}(S)$ with $6g-6+3n$ positive coordinate functions $X_i:\mathcal{T}(S)\to \mathbb{R}^+$, one per ideal each ideal arc $i$ of $T$, satisfying $n$ constraints $X_p = 1$, one per each puncture $p$ of $S$. Given an ideal arc $i\in T$ and a hyperbolic metric $h\in\mathcal{T}(S)$, one has unique ideal quadrilateral determined by the pair of adjacent ideal triangles incident to $i$. The shear coordinate $X_i(h)$ of the metric $h$ along the arc $i$ is defined as a cross-ratio of the ideal points of any lift of this quadrilateral to the universal cover $\mathbb{H}^2$ of $(S,h)$
\begin{align*}
	X_i(h) = -\frac{p_1-p_2}{p_1-p_4}\frac{p_3-p_4}{p_3-p_2}.
	 &  &  &  &
	\raisebox{-0.4\height}{\scalebox{.9}{
\begingroup%
  \makeatletter%
  \providecommand\color[2][]{%
    \errmessage{(Inkscape) Color is used for the text in Inkscape, but the package 'color.sty' is not loaded}%
    \renewcommand\color[2][]{}%
  }%
  \providecommand\transparent[1]{%
    \errmessage{(Inkscape) Transparency is used (non-zero) for the text in Inkscape, but the package 'transparent.sty' is not loaded}%
    \renewcommand\transparent[1]{}%
  }%
  \providecommand\rotatebox[2]{#2}%
  \newcommand*\fsize{\dimexpr\f@size pt\relax}%
  \newcommand*\lineheight[1]{\fontsize{\fsize}{#1\fsize}\selectfont}%
  \ifx\svgwidth\undefined%
    \setlength{\unitlength}{141.73228346bp}%
    \ifx\svgscale\undefined%
      \relax%
    \else%
      \setlength{\unitlength}{\unitlength * \real{\svgscale}}%
    \fi%
  \else%
    \setlength{\unitlength}{\svgwidth}%
  \fi%
  \global\let\svgwidth\undefined%
  \global\let\svgscale\undefined%
  \makeatother%
  \begin{picture}(1,0.6)%
    \lineheight{1}%
    \setlength\tabcolsep{0pt}%
    \put(0,0){\includegraphics[width=\unitlength,page=1]{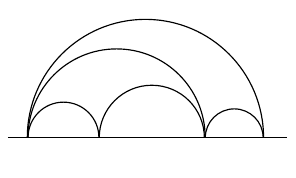}}%
    \put(0.56880848,0.41319576){\makebox(0,0)[lt]{\lineheight{1.25}\smash{\begin{tabular}[t]{l}$i$\end{tabular}}}}%
    \put(0.06575217,0.08370083){\makebox(0,0)[lt]{\lineheight{1.25}\smash{\begin{tabular}[t]{l}$p_1$\end{tabular}}}}%
    \put(0.30916872,0.08370083){\makebox(0,0)[lt]{\lineheight{1.25}\smash{\begin{tabular}[t]{l}$p_2$\end{tabular}}}}%
    \put(0.66900238,0.08370083){\makebox(0,0)[lt]{\lineheight{1.25}\smash{\begin{tabular}[t]{l}$p_3$\end{tabular}}}}%
    \put(0.87008591,0.08370083){\makebox(0,0)[lt]{\lineheight{1.25}\smash{\begin{tabular}[t]{l}$p_4$\end{tabular}}}}%
  \end{picture}%
\endgroup%
}}
\end{align*}
Given a puncture $p$ of $S$, the constraint results from the imposed boundary conditions at $p$ and can be described explicitly as
\begin{align*}
	X_p(h) = \prod_{i\in T}(X_i(h))^{\theta_{ip}} = 1,
\end{align*}
where $\theta_{ip}$ denotes the valence of the arc $i$ at the puncture $p$.
The {\it generalized shear coordinates} on $\mathcal{GH}_\Lambda(S\times\mathbb{R})$ can be obtained via an $\mathbb{R}_\Lambda$-analytic continuation of the Thurston shear coordinates with respect to the totally real embedding $\mathcal{T}(S)\to\mathcal{GH}_\Lambda(S\times\mathbb{R})$. For $\Lambda = 1$, the embedding is given by the Fuchsian section $\mathcal{T}(S) \to \mathcal{CP}(S)$ and the generalized shear coordinates are defined in an open neighborhood of $\mathcal{T}(S)$ by the usual theory of holomorphic continuation of real analytic functions. For $\Lambda = -1$ and $\Lambda = 0$, the totally real embedding is given, respectively, by the diagonal embedding $\mathcal{T}(S)\to \mathcal{T}(S)\times\mathcal{T}(S)$ and by the zero section $\mathcal{T}(S) \to T\mathcal{T}(S)$. The $\mathbb{R}_{\Lambda}$-continuation $Z:\mathcal{GH}_\Lambda(S\times\mathbb{R})\to \mathbb{R}_\Lambda$ of a real analytic function $X:\mathcal{T}(S)\to \mathbb{R}$ is then given, respectively, by
\begin{align*}
	\begin{cases}
		Z(h_+,h_-) = \frac{1+\ell}{2}X(h_+) + \frac{1-\ell}{2}X(h_-), & \Lambda = -1,
		\cr
		Z(h,\xi) = X(h) + \ell d_hX(\xi),                             & \Lambda = 0,
	\end{cases}
\end{align*}
where $(h_+,h_-)\in \mathcal{T}(S)\times\mathcal{T}(S)$ and $(h,\xi)\in T\mathcal{T}(S)$.
For each ideal arc $i\in T$, the $\mathbb{R}_\Lambda$-continuation of $X_i:\mathcal{T}(S)\to\mathbb{R}^+$ gives rise to an $\mathbb{R}_\Lambda$-valued coordinate function $Z_i:\mathcal{GH}_\Lambda(S\times\mathbb{R})\to \mathbb{R}_\Lambda^+$, satisfying a constraint for each puncture $p$ of $S$
\begin{align*}
	Z_p = \prod_{i\in T}(Z_i)^{\theta_{ip}} = 1.
\end{align*}
These are global constrained coordinates for $\Lambda = -1, 0$, while for $\Lambda = 1$ they at least cover an open neighborhood of the Fuchsian locus in $\mathcal{CP}(S)$.

Coordinate expressions for the symplectic structure or, more precisely, for the corresponding Poisson structure, were obtained in \cite{Wol83, FC99, Foc97} for $\mathcal{T}(S)$ and in \cite{MS16} for $\mathcal{GH}_\Lambda(S\times\mathbb{R})$. Given an ideal triangulation $T$ on $S$, the Poisson brackets are given by
\begin{align*}
	\{ X_i, X_j \} = \varepsilon_{ij} \, X_i X_j,
	 &  &
	\{ Z_i, Z_j \} = \ell \, \varepsilon_{ij} \, Z_i Z_j,
	 &  &
	\forall i,j \in T,
\end{align*}
where $\varepsilon_{ij} \in \mathbb{Z}$ encodes the combinatorics of $T$ as
\begin{align}
	\label{eq:varepsilon_for_triangulation}
	\begin{array}{ll}
		\varepsilon_{ij} & = a_{ij} - a_{ji},
		\\
		a_{ij}           & = \mbox{ the number of corners of triangles of $T$  }
		\cr
		                 & \quad\;\; \mbox{delimited by $i$ on the right and $j$ on the left };
	\end{array}
\end{align}
We call $\varepsilon_T = (\varepsilon_{ij})_{i,j\in T}$ the {\it exchange matrix} of $T$.

An important feature of the Thurston shear coordinates and of the generalized shear coordinates is their behavior under changes of ideal triangulations. Given a pair of ideal triangulations $T$ and $T'$ on $S$, let $X_i$ and $X'_i$ denote the corresponding Thurston shear coordinates on $\mathcal{T}(S)$ and let $Z_i$ and $Z'_i$ denote the corresponding generalized shear coordinates on $\mathcal{GH}_\Lambda(S\times\mathbb{R})$. It is well known that there exists a sequence of intermediate ideal triangulations obtained recursively by applying elementary mutations $\mu_k$, also known as flips, which change an ideal triangulation at a single ideal arc $k$ at a time.
\begin{center}
	\scalebox{.9}{
\begingroup%
  \makeatletter%
  \providecommand\color[2][]{%
    \errmessage{(Inkscape) Color is used for the text in Inkscape, but the package 'color.sty' is not loaded}%
    \renewcommand\color[2][]{}%
  }%
  \providecommand\transparent[1]{%
    \errmessage{(Inkscape) Transparency is used (non-zero) for the text in Inkscape, but the package 'transparent.sty' is not loaded}%
    \renewcommand\transparent[1]{}%
  }%
  \providecommand\rotatebox[2]{#2}%
  \newcommand*\fsize{\dimexpr\f@size pt\relax}%
  \newcommand*\lineheight[1]{\fontsize{\fsize}{#1\fsize}\selectfont}%
  \ifx\svgwidth\undefined%
    \setlength{\unitlength}{193.55124646bp}%
    \ifx\svgscale\undefined%
      \relax%
    \else%
      \setlength{\unitlength}{\unitlength * \real{\svgscale}}%
    \fi%
  \else%
    \setlength{\unitlength}{\svgwidth}%
  \fi%
  \global\let\svgwidth\undefined%
  \global\let\svgscale\undefined%
  \makeatother%
  \begin{picture}(1,0.328341)%
    \lineheight{1}%
    \setlength\tabcolsep{0pt}%
    \put(0,0){\includegraphics[width=\unitlength,page=1]{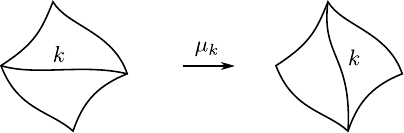}}%
  \end{picture}%
\endgroup%
}
\end{center}
When $T$ and $T' = \mu_k(T)$ are related by a mutation along $k\in T$, the underlying exchange matrices are related by
\begin{align}
	\nonumber
	\varepsilon'_{ij} = \left\{
	\begin{array}{ll}
		-\varepsilon_{ij}                                                                                        & \mbox{if $k\in \{i,j\}$},     \\
		\varepsilon_{ij} + \frac{1}{2}(\varepsilon_{ik}|\varepsilon_{kj}| + |\varepsilon_{ik}| \varepsilon_{kj}) & \mbox{if $k \notin \{i,j\}$}.
	\end{array}
	\right.
\end{align}
and the corresponding coordinate transformation between the Thurston shear coordinates $X_i$ and $X'_i$ is given by the {\it cluster $\mathscr{X}$-mutation formula}
\begin{align}
	X_i' =
	\left\{
	\begin{array}{ll}
		X_k^{-1},                                                     & \mbox{if $i =k$},    \\
		X_i(1+X_k^{-{\rm sgn}(\varepsilon_{ik})})^{-\varepsilon_{ik}} & \mbox{if $i\neq k$},
	\end{array}
	\right.
\end{align}
while the coordinate transformation between the generalized shear coordinates $Z_i$ and $Z'_i$ is given by a similar formula as in eq.\eqref{eq:intro_Z_transformation}
\begin{align}
	\nonumber
	Z_i' = \left\{
	\begin{array}{ll}
		Z_k^{-1}                                                      & \mbox{if $i=k$},     \\
		Z_i(1+Z_k^{-{\rm sgn}(\varepsilon_{ik})})^{-\varepsilon_{ik}} & \mbox{if $i\neq k$}.
	\end{array}
	\right.
\end{align}
One can easily verify with the above formulas that the coordinate transformations corresponding to a change of ideal triangulations preserve the Poisson brackets, in the sense that $\{Z_i, Z_j\} = \ell \varepsilon_{ij} Z_i Z_j$ implies $\{Z_i', Z_j'\} = \ell \varepsilon_{ij}' Z_i' Z_j'$. This suggests, in particular, that the shear coordinates on $\mathcal{T}(S)$ and the generalized shear coordinates on $\mathcal{GH}_\Lambda(S\times \mathbb{R})$ can be seen as certain versions of cluster $\mathscr{X}$-variables of cluster $\mathscr{X}$-varieties, as defined by Fock and Goncharov \cite{FG06,FG09b}. This viewpoint is the crucial starting point of our formulation of the quantization of the spaces $\mathcal{T}(S)$ and $\mathcal{GH}_\Lambda(S\times \mathbb{R})$, and will be described in detail in the following section.

\section{Generalized complex points of a cluster variety}

\subsection{Cluster $\mathscr{X}$-varieties}
\label{subsec:cluster_X-varieties}

Let us first recall the notion of Fock-Goncharov's cluster $\mathscr{X}$-variety \cite{FG06,FG09b}. Let $I$ be a finite index set. A {\it cluster $\mathscr{X}$-seed} $\Gamma = (\varepsilon, \{X_i\}_{i\in I})$ consists of an {\it exchange matrix} $\varepsilon = (\varepsilon_{ij})_{i,j\in I}$ which is a skew-symmetric $I\times I$ matrix with integer entries, and an assignment of a cluster $\mathscr{X}$-variable $X_i$ to each $i \in I$. Here, $X_i$'s are formal commuting variables; more precisely, $\{X_i\}_{i \in I}$ is required to form a transcendence basis over $\mathbb{Q}$ of an ambient field $\mathcal{F} = \mathbb{Q}(\{X_i\}_{i\in I})$. Often, $\varepsilon$ is identified with a quiver whose set of vertices is identified with $I$ and whose signed adjacency matrix is $\varepsilon$, where the variable $X_i$ is understood as being attached to the vertex $i$.
To a seed $\Gamma$ we associate an affine scheme $\mathscr{X}_\Gamma$ (over $\mathbb{Q}$), called the seed $\mathscr{X}$-torus, as the split algebraic torus $(\mathbb{G}_m)^I = ( \mathbb{G}_m(\mathbb{Q}) )^I$, whose coordinate functions are identified with $X_i$. In other words, $\mathscr{X}_\Gamma$ is viewed as ${\rm Spec}(R_\Gamma)$, where $R_\Gamma$ is the Laurent polynomial ring $\mathbb{Q}[\{X_i^{\pm 1} \, | \, i \in I\}]$. The torus $\mathscr{X}_\Gamma$ is also equipped with a Poisson structure given by
$$
	\{X_i,X_j\} = \varepsilon_{ij}X_iX_j,\qquad \forall i,j\in I.
$$

The {\it cluster mutation} $\mu_k$ at $k\in I$ is a procedure producing a new seed $\Gamma' = \mu_k(\Gamma) =  (\varepsilon',\{X_i'\}_{i\in I})$ out of a seed $\Gamma = (\varepsilon, \{X_i\}_{i\in I})$. The new exchange matrix $\varepsilon'$ is defined by the {\it quiver mutation formula}
\begin{align}
	\label{eq:varepsilon_prime_formula}
	\varepsilon'_{ij} = \left\{
	\begin{array}{ll}
		- \varepsilon_{ij},                                                                                      & \mbox{if $k\in \{i,j\}$},   \\
		\varepsilon_{ij} + \frac{1}{2}(\varepsilon_{ik}|\varepsilon_{kj}| + |\varepsilon_{ik}|\varepsilon_{kj}), & \mbox{if $k\notin\{i,j\}$.}
	\end{array}
	\right.
\end{align}
We consider a birational map
$$
	\mu_k : \mathscr{X}_\Gamma \dashrightarrow \mathscr{X}_{\Gamma'}
$$
between the seed $\mathscr{X}$-tori, given in terms of the $\mathscr{X}$-variables by the cluster $\mathscr{X}$-mutation formula
\begin{align}
	\label{eq:mu_k_star}
	\mu_k^* X_i' =
	\left\{
	\begin{array}{ll}
		X_k^{-1},                                                     & \mbox{if $i =k$},    \\
		X_i(1+X_k^{-{\rm sgn}(\varepsilon_{ik})})^{-\varepsilon_{ik}} & \mbox{if $i\neq k$}.
	\end{array}
	\right.
\end{align}
Importantly, this birational map preserves the corresponding Poisson structures in the sense that
$$
	\{\mu_k^*X_i',\mu_k^*X_j'\} = \mu_k^* \{X_i',X_j'\},\qquad \forall i,j\in I.
$$

Another way of producing a new seed is a seed automorphism $P_\sigma$ associated to a permutation $\sigma$ of $I$. More precisely, we set $P_\sigma(\Gamma) = \Gamma' = (\varepsilon',\{X_i'\}_{i\in I})$, where $\varepsilon'$ is given by
$$
	\varepsilon'_{\sigma(i) \sigma(j)} = \varepsilon_{ij}, \qquad \forall i,j \in I,
$$
and the corresponding isomorphism
$$
	P_\sigma : \mathscr{X}_\Gamma \to \mathscr{X}_{\Gamma'}
$$
between the tori is given by
$$
	P_\sigma^* X'_{\sigma(i)} = X_i, \qquad \forall i \in I.
$$

\vspace{2mm}

Starting with an initial seed $\Gamma = (\varepsilon,\{X_i\}_{i\in I})$ one then considers the equivalence class $|\Gamma|$ of all possible seeds obtained by a finite (possibly empty) sequence of mutations and seed automorphisms,  and defines a scheme $\mathscr{X}_{|\Gamma|}$, the {\it cluster $\mathscr{X}$-variety} associated to $|\Gamma|$, by gluing all of the $\mathscr{X}$-tori $\mathscr{X}_\Gamma$ for $\Gamma\in |\Gamma|$ along the birational gluing maps $\mu_{\Gamma,\Gamma'}$. As the structure of this scheme $\mathscr{X}_{|\Gamma|}$ depends essentially only on the quiver-mutation-equivalence class of the underlying  exchange matrix of the initial seed, it is also often denoted by $\mathscr{X}_{|\varepsilon|}$.

\subsection{The set of $\mathbb{R}_\Lambda$-points of a cluster $\mathscr{X}$-variety}
\label{subsec:R_Lambda}

For any field $\mathbb{F}$, one may consider the set $\mathscr{X}_{|\Gamma|}(\mathbb{F})$ of $\mathbb{F}$-points of the cluster $\mathscr{X}$-variety associated to $|\Gamma|$. In the theory of cluster varieties, one more often considers the set $\mathscr{X}_{|\Gamma|}(\mathbb{P})$ of $\mathbb{P}$-points, for a {\it semi-field} $\mathbb{P}$. In particular, when evaluated at the semi-field $\mathbb{R}^+$ of positive real numbers, the resulting set $\mathscr{X}_{|\Gamma|}(\mathbb{R}^+)$ is a smooth (real) Poisson manifold, which was quantized in \cite{FG09a}. In the present paper, we consider the set $\mathscr{X}_{|\Gamma|}(\mathbb{R}_\Lambda)$ of $\mathbb{R}_\Lambda$-points.

The following are basic observations on the structure of the ring $\mathbb{R}_\Lambda$.
\begin{definition}
	For each element $x+\ell y \in \mathbb{R}_\Lambda$, with $x,y\in\mathbb{R}$, the {\it $\Lambda$-real part} and the {\it $\Lambda$-imaginary part} of $x+\ell y$ are defined as
	\begin{align*}
		{\rm Re}_\Lambda(x+\ell y) := x,
		 &  &
		{\rm Im}_\Lambda(x+\ell y) := y.
	\end{align*}
	The {\it $\Lambda$-conjugate} of $x+ \ell y$ is defined to be $x - \ell y$.
\end{definition}
\begin{lemma}[matrix realization of $\mathbb{R}_\Lambda$]
	We have injective $\mathbb{R}$-algebra homomorphisms
	\begin{align}\label{eq:original_embedding_map}
		\mathbb{R}_\Lambda \to {\rm Mat}_{2\times 2}(\mathbb{R}),
		 &  &
		x+\ell y \mapsto \mattwo{x}{-\Lambda y}{y}{x}.
	\end{align}
\end{lemma}
For $\Lambda=-1$, these matrices can be simultaneously diagonalized over $\mathbb{R}$, and for $\Lambda=1$, this can be done over $\mathbb{C}$. Applying these diagonalizations, one obtains the $\mathbb{R}$-algebra embeddings:
\begin{align}
	\nonumber
	D_{-1} & ~:~ \mathbb{R}_{-1} \to {\rm Mat}_{2\times 2}(\mathbb{R}), &   &
	x+\ell y \mapsto \mattwo{x+y}{0}{0}{x-y},
	       &                                                            &
	\mbox{(for $\Lambda=-1$)}                                                 \\
	\nonumber
	D_1    & ~:~ \mathbb{R}_1 \to {\rm Mat}_{2\times 2}(\mathbb{C}),
	       &                                                            &
	x+\ell y \mapsto \mattwo{x+{\rm i} y}{0}{0}{x-{\rm i}y}.
	       &                                                            &
	\mbox{(for $\Lambda=1$)}
\end{align}
The image of $D_{-1}$ is $\{{\rm diag}(a,b) : a,b\in \mathbb{R} \}$ for $\Lambda=-1$, and that of $D_1$ is $\{{\rm diag}(a,\overline{a}) : a \in\mathbb{C}\}$ for $\Lambda=1$. Thus:
\begin{corollary}
	\label{cor:ring_realizations}
	We have $\mathbb{R}$-algebra isomorphisms
	\begin{align}
		\nonumber
		\Lambda=-1 & ~:~ \mathbb{R}_{-1} \to \mathbb{R} \times \mathbb{R}, \\
		\nonumber
		\Lambda=1  & ~:~ \mathbb{R}_1 \to \mathbb{C}.
	\end{align}
\end{corollary}
For $\Lambda=0$, we also denote the original embedding map (eq.\eqref{eq:original_embedding_map}) by
\begin{align*}
	D_0 ~:~ \mathbb{R}_0 \to {\rm Mat}_{2\times 2}(\mathbb{R}),
	 &  &
	x+\ell y \mapsto \mattwo{x}{0}{y}{x}.
	 &  &
	(\mbox{for $\Lambda=0$})
\end{align*}
Note that this cannot be diagonalized and the image is $\{\smallmattwo{a}{0}{b}{a} : a,b\in \mathbb{R}\}$.

\vspace{2mm}

Turning our attention back to $\mathscr{X}_{|\Gamma|}(\mathbb{R}_\Lambda)$, note that for any given seed $\Gamma\in|\Gamma|$ the $\mathscr{X}$-torus $\mathscr{X}_\Gamma(\mathbb{R}_\Lambda)$ is equal to $(\mathbb{R}_\Lambda^\times)^I$ as a set, where $\mathbb{R}_\Lambda^\times$ denotes the set of all units of $\mathbb{R}_\Lambda$. In particular, the cluster $\mathscr{X}$-variety $\mathscr{X}_{|\Gamma|}(\mathbb{R}_\Lambda)$ is obtained by gluing a collection of $(\mathbb{R}_\Lambda^\times)^I$ along mutation maps $\mu_k$ and index permutation maps $P_\sigma$. In other words, we can view each $\mathscr{X}_\Gamma(\mathbb{R}_\Lambda) = (\mathbb{R}_\Lambda^\times)^I$ as a chart of $\mathscr{X}_{|\Gamma|}(\mathbb{R}_\Lambda)$ associated to a seed $\Gamma$. We note here however that when two charts are glued, not all points of each chart are necessarily glued together. In fact, we will only be interested in the points of $\mathscr{X}_{|\Gamma|}(\mathbb{R}_\Lambda)$ whose $\mathbb{R}_\Lambda$-coordinates belong to the subset
$$
	\R_\Lambda^+ = \exp(\R_\Lambda)
$$
defined as the image of the exponential map on $\R_\Lambda$.

\begin{definition}[positive points]
	The subset of $\mathscr{X}_{|\Gamma|}(\mathbb{R}_\Lambda)$ obtained by gluing $\mathscr{X}_\Gamma(\mathbb{R}_\Lambda^+) := (\mathbb{R}_\Lambda^+)^I \subset (\mathbb{R}_\Lambda^\times)^I = \mathscr{X}_\Gamma(\mathbb{R}_\Lambda)$ for all $\Gamma \in |\Gamma|$ is denoted by $\mathscr{X}_{|\Gamma|}(\mathbb{R}_\Lambda^+)$.
\end{definition}
This set $\mathscr{X}_{|\Gamma|}(\mathbb{R}_\Lambda^+)$, which is an analog of Fock-Goncharov's $\mathscr{X}_{|\Gamma|}(\mathbb{R}^+)$ in \cite{FG06}, shall be the actual geometric object to be quantized in this paper, and we investigate it in more detail in the next subsection.

\subsection{Geometric structures on the set of $\mathbb{R}_\Lambda^+$-points}
\label{subsec:geometric_structures_on_the_set_of_R_Lambda_positive_points}

In the present subsection we study $\mathscr{X}_{|\Gamma|}(\mathbb{R}_\Lambda^+)$ in more detail. We start by noting that the matrix realizations of $\R_\Lambda^+$ are given by
\begin{align*}
	D_{-1}(\mathbb{R}_{-1}^+) & = \big\{ {\rm diag}(a,b) : a,b\in \mathbb{R}^+ \big\},                           \\
	D_1(\mathbb{R}_1^+)       & = \big\{{\rm diag}(a,\overline{a}) : a\in \mathbb{C}^\times\big\},               \\
	D_0(\mathbb{R}_0^+)       & = \big\{ \smallmattwo{a}{0}{b}{a} : a\in \mathbb{R}^+, \, b\in \mathbb{R}\big\}.
\end{align*}
In particular, via the isomorphisms in Cor.\ref{cor:ring_realizations}, we have $\mathbb{R}_{-1}^+=\mathbb{R}^+\times \mathbb{R}^+\subset\mathbb{R} \times \mathbb{R}$ and $\mathbb{R}_1^+=\mathbb{C}^\times\subset\mathbb{C}$. It is then not too hard to see that:
\begin{lemma}
	For $\Lambda=-1,0$, the gluing maps between the charts $\mathscr{X}_\Gamma(\mathbb{R}^+_\Lambda)$ are bijective.
\end{lemma}

In particular, for $\Lambda=-1$, each chart $\mathscr{X}_\Gamma(\mathbb{R}^+_\Lambda)$ is actually a global chart, and we have identifications
$$
	\mathscr{X}_{|\Gamma|} (\mathbb{R}_{-1}^+) = (\mathbb{R}_{-1}^+)^I = (\mathbb{R}^+ \times \mathbb{R}^+)^I
$$
enumerated by the seeds $\Gamma\in|\Gamma|$. Moreover, since the transition maps between two such identifications can be shown to be smooth, we obtain a well-defined smooth manifold structure on $\mathscr{X}_{|\Gamma|}(\mathbb{R}_{-1}^+)$. We note that these identifications hint that the set $\mathscr{X}_{|\Gamma|}(\mathbb{R}_{-1}^+)$ is closely related to Fock-Goncharov's $\mathscr{X}_{|\Gamma|} \times \mathscr{X}_{|\Gamma|}^{\rm op}$, or to the symplectic double cluster $\mathscr{D}$-variety $\mathscr{D}_{|\Gamma|}$ \cite{FG09a}.
For $\Lambda=0$, we have similar identifications
$$
	\mathscr{X}_{|\Gamma|}(\mathbb{R}_0^+) = (\mathbb{R}_{0}^+)^I = (\mathbb{R}^+ \times \mathbb{R})^I,
$$
for each $\Gamma\in|\Gamma|$, again inducing a well-defined smooth manifold structure on $\mathscr{X}_{|\Gamma|}(\mathbb{R}_0^+)$.
For $\Lambda=1$, we can view $\mathscr{X}_{|\Gamma|}(\mathbb{R}_1^+)$ as a complex variety covered by toric charts.

\vs

For the sake of a uniform discussion, for each value of $\Lambda$, we can regard $\mathscr{X}_{|\Gamma|}(\mathbb{R}^+_\Lambda)$ as an {\it $\mathbb{R}_\Lambda$-variety}. The coordinate functions for each `$\mathbb{R}_\Lambda$-toric' chart $\mathscr{X}_\Gamma(\mathbb{R}_\Lambda^+) = (\mathbb{R}_\Lambda^+)^I$ are $\mathbb{R}_\Lambda$-valued functions and will be denoted by
$$
	Z_i ~:~ \mathscr{X}_\Gamma(\mathbb{R}_\Lambda^+) \longrightarrow \mathbb{R}^+_\Lambda \subset \mathbb{R}_\Lambda.
$$
So, when $\Gamma' = \mu_k(\Gamma)$, the two tori $\mathscr{X}_\Gamma(\mathbb{R}_\Lambda^+)$ and $\mathscr{X}_{\Gamma'}(\mathbb{R}_\Lambda^+)$ are glued by the birational map
$$
	\mu_k : \mathscr{X}_\Gamma(\mathbb{R}_\Lambda^+) \dashrightarrow \mathscr{X}_{\Gamma'}(\mathbb{R}_\Lambda^+)
$$
given by
\begin{align}
	\label{eq:Z_mutation_formula}
	\mu_k^* Z_i' = \left\{
	\begin{array}{ll}
		Z_k^{-1},                                                     & \mbox{if $i =k$},    \\
		Z_i(1+Z_k^{-{\rm sgn}(\varepsilon_{ik})})^{-\varepsilon_{ik}} & \mbox{if $i\neq k$},
	\end{array}
	\right.
\end{align}
Similarly, the index-permutation isomorphism $P_\sigma$ is given by $P_\sigma^*(Z'_{\sigma(i)}) = Z_i$, $\forall i$.

\vs

We now turn to a discussion about a natural Poisson structure on this $\mathbb{R}_\Lambda$-variety $\mathscr{X}_{|\Gamma|}(\mathbb{R}_\Lambda^+)$. This can be formulated in terms of a ring of $\mathbb{R}_\Lambda$-valued functions
$$
	{\rm Fun}(\mathscr{X}_{|\Gamma|}(\mathbb{R}_\Lambda^+) ; \mathbb{R}_\Lambda)
$$
satisfying some desirable analytic condition. For example, one may consider the set of $\mathbb{R}_\Lambda$-analytic functions on $\mathscr{X}_{|\Gamma|}(\mathbb{R}_\Lambda^+)$ as in \cite{MS16} or, more generally, the set of $C^\infty$-functions using the natural smooth structures of $\mathscr{X}_{|\Gamma|}(\mathbb{R}_\Lambda^+)$ and $\mathbb{R}_\Lambda$. Either way, this ring is an algebra over $\mathbb{R}_\Lambda$, and the notion of an $\mathbb{R}_\Lambda$-Poisson bracket on $\mathscr{X}_{|\Gamma|}(\mathbb{R}_\Lambda^+)$ can be defined as an $\mathbb{R}_\Lambda$-bilinear and skew-symmetric map
$$
	\{ \cdot, \cdot \} ~:~ {\rm Fun}(\mathscr{X}_{|\Gamma|}(\mathbb{R}_\Lambda^+) ; \mathbb{R}_\Lambda) \times {\rm Fun}(\mathscr{X}_{|\Gamma|}(\mathbb{R}_\Lambda^+) ; \mathbb{R}_\Lambda) \to {\rm Fun}(\mathscr{X}_{|\Gamma|}(\mathbb{R}_\Lambda^+) ; \mathbb{R}_\Lambda)
$$
satisfying the Jacobi identity as well as the Leibniz rule.

The Poisson structure we will consider here is given on each chart $\mathscr{X}_\Gamma(\mathbb{R}_\Lambda^+)$ of $\mathscr{X}_{|\Gamma|}(\mathbb{R}_\Lambda^+)$ as
\begin{align}
	\label{eq:R_Lambda-Poisson_bracket}
	\{ Z_i, Z_j \} = \ell \, \varepsilon_{ij} \, Z_i Z_j, \qquad \forall i,j \in I.
\end{align}
We note that by reading the $\Lambda$-real and the $\Lambda$-imaginary parts of the above equation, one could deduce more usual ($\mathbb{R}$-valued) Poisson bracket on ${\rm Fun}(\mathscr{X}_{|\Gamma|}(\mathbb{R}_\Lambda^+);\mathbb{R})$. For example, if we write $Z_i$ in the exponential form
\begin{align*}
	Z_i = \exp(z_i),
	 &  &
	z_i = x_i + \ell y_i,
\end{align*}
where $x_i = {\rm Re}_\Lambda(z_i)$ and $y_i = {\rm Im}_\Lambda(z_i)$, then the above Poisson bracket is shown to be compatible with the following Poisson bracket among $x_i$ and $y_i$
\begin{align}
	\label{eq:x_i_and_y_i_Poisson}
	\{x_i,x_j\} = 0 = \{y_i,y_j\},
	\qquad
	\{x_i, y_j \} = \textstyle \frac{1}{2} \varepsilon_{ij},
	\qquad
	\forall i,j \in I.
\end{align}

In what follows, in addition to the $\mathbb{R}_\Lambda$-valued coordinates $Z_i$, we will also need to deal with their $\Lambda$-conjugates. We find it convenient to write $Z_i^{(+)}$ for the coordinate $Z_i$, and $Z_i^{(-)}$ for its $\Lambda$-conjugate. The $\mathbb{R}_\Lambda$-Poisson bracket then satisfies
\begin{align}
	\label{eq:R_Lambda_Poisson_brackets}
	\{Z_i^{(+)},Z_j^{(+)}\}  = \ell \varepsilon_{ij} Z_i^{(+)} Z_j^{(+)}, \;
	\{Z_i^{(-)},Z_j^{(-)}\}  = - \ell \varepsilon_{ij} Z_i^{(-)} Z_j^{(-)},  \;
	\{Z_i^{(+)},Z_j^{(-)}\}  = 0.
\end{align}
The mutation formulas for $Z_i^{(-)}$ are given by $\Lambda$-conjugation of the formulas in eq.\eqref{eq:Z_mutation_formula}. Note that, writing $Z_i^{(+)} = \exp(x_i + \ell y_i)$ in exponential form, the $\Lambda$-conjugate coordinates are given by $Z_i^{(-)} = \exp(x_i - \ell y_i)$.

\begin{proposition}
	\label{prop:Poisson_bracket_compatible}
	The Poisson bracket defined on each chart $\mathscr{X}_\Gamma(\mathbb{R}_\Lambda^+)$ by eq.\eqref{eq:R_Lambda_Poisson_brackets} is compatible with the gluing maps $\mu_k$ and $P_\sigma$ between different charts.
\end{proposition}
The proof for the canonical Poisson bracket on Fock-Goncharov's usual cluster $\mathscr{X}$-variety works almost verbatim, with $X_i$'s replaced by $Z_i$'s; see \cite{MS16} for details in the case of punctured surfaces. As a result, we obtain a Poisson structure on the $\mathbb{R}_\Lambda$-variety $\mathscr{X}_{|\Gamma|}(\mathbb{R}_\Lambda^+)$, which will be used in our formulation of the quantization of $\mathscr{X}_{|\Gamma|}(\mathbb{R}_\Lambda^+)$.

\vs

Note that for each vector $\theta = (\theta_i)_{i\in I} \in \mathbb{Z}^I$  in the kernel of $\varepsilon$, i.e. such that $\sum_{j\in I} \varepsilon_{ij} \theta_j=0$, $\forall i \in I$, the element
$$
	Z^\theta := \prod_{i\in I} Z_i^{\theta_i}
$$
belongs to the Poisson kernel of $\mathscr{X}_{|\Gamma|}(\mathbb{R}_\Lambda^+)$. We define the {\it cusped symplectic leaf} as the following subset of $\mathscr{X}_{|\Gamma|}(\mathbb{R}_\Lambda^+)$
\begin{align*}
	\mathscr{X}_{|\Gamma|}(\mathbb{R}_\Lambda^+)_{\rm cusp} := \big\{ \mbox{points of $\mathscr{X}_{|\Gamma|}(\mathbb{R}_\Lambda^+)$ s.t. $Z^\theta=1$ holds for }
	\cr
	\mbox{every $\theta$ in the kernel of $\varepsilon$}\big\}.
\end{align*}
When the exchange matrix $\varepsilon$ comes from an ideal triangulation of a punctured surface $S$ in the sense of eq.\eqref{eq:varepsilon_for_triangulation}, such a subset recovers the moduli space of 3d gravity $\mathcal{GH}_\Lambda(S\times \mathbb{R})$ discussed in the previous section. More precisely, for $\Lambda=-1,0$ the space $\mathcal{GH}_\Lambda(S\times \mathbb{R})$ maps bijectively onto the leaf $\mathscr{X}_{|\Gamma|}(\mathbb{R}_\Lambda^+)_{\rm cusp}$, while for $\Lambda=1$ both $\mathcal{GH}_\Lambda(S\times \mathbb{R})$ and $\mathscr{X}_{|\Gamma|}(\mathbb{R}_\Lambda^+)_{\rm cusp}$ can be viewed as complexifications of the real locus $\mathscr{X}_{|\Gamma|}(\mathbb{R}^+)_{\rm cusp}\subset \mathscr{X}_{|\Gamma|}(\mathbb{R}_\Lambda^+)_{\rm cusp}$, which is known to recover the Teichm\"uller space $\mathcal{T}(S)$. It is an interesting open question to identify the precise relation between $\mathcal{GH}_{1}(S\times \mathbb{R})$ and $\mathscr{X}_{|\Gamma|}(\mathbb{R}_{1}^+)_{\rm cusp}$ at a global level, i.e. to characterize the points of $\mathscr{X}_{|\Gamma|}(\mathbb{C}^\times)_{\rm cusp}\subset \mathscr{X}_{|\Gamma|}(\mathbb{C}^\times)$ that come from 3d spacetimes.

In the remainder of the present paper, we shall develop a theory for quantizing the space $\mathscr{X}_{|\Gamma|}(\mathbb{R}^+_\Lambda)$ and its cusped leaf $\mathscr{X}_{|\Gamma|}(\mathbb{R}^+_\Lambda)_{\rm cusp}$ (as well as other leaves) for a general seed $\Gamma$ using the language of the quantum theory of cluster varieties in the style of \cite{FG09a}; as a major consequence, this will provide a quantization of the moduli space $\mathcal{GH}_\Lambda(S\times \mathbb{R})$ of 3d gravity.

\section{The quantization problem}
\label{sec:a_deformation_quantization_problem}

In this section we formulate the quantization problem which we would like to solve.

\subsection{Poisson manifolds}

We begin by recalling a notion of {\it quantization} of a smooth real Poisson manifold $(M,\{\cdot,\cdot\})$. This consists of a separable complex Hilbert space $(\mathscr{H},\langle\cdot,\cdot\rangle)$, a subalgebra $\mathcal{A}$ of $C^\infty(M) = C^\infty(M;\mathbb{R})$ of classical observables to be quantized, together with a one-parameter family of maps
\begin{align}
	\label{eq:bf_Q_hbar}
	{\bf Q}^\hbar ~:~  \mathcal{A} \to \big\{\mbox{self-adjoint operators on $\mathscr{H}$}\big\}
\end{align}
depending real-analytically on a real quantum parameter $\hbar$ (the Planck constant), such that
\begin{enumerate}
	\item[\rm (DQ1)] ${\bf Q}^\hbar$ is $\mathbb{R}$-linear,

	\item[\rm (DQ2)] ${\bf Q}^\hbar(1) = {\rm Id}$,

	\item[\rm (DQ3)] $\forall f,g\in \mathcal{A}$, \,\, $[{\bf Q}^\hbar(f), {\bf Q}^\hbar(g)] = {\rm i} \hbar \, {\bf Q}^\hbar(\{f,g\}) + o(\hbar)$ as $\hbar\to 0$.
\end{enumerate}
Moreover, if some (discrete) group $G$ acts as Poisson automorphisms of $(M,\{\cdot,\cdot\})$ and of $\mathcal{A}$, it is natural to require that this $G$ action be quantized as well, in an equivariant manner with respect to ${\bf Q}^\hbar$; namely there should be a unitary representation of $G$ on $\mathscr{H}$
$$
	\rho^\hbar ~:~ G \to {\rm U}(\mathscr{H}) = \big\{\mbox{unitary operators on $\mathscr{H}$}\big\}
$$
such that
$$
	\rho^\hbar(\sigma) \circ {\bf Q}^\hbar(f)\circ \rho^\hbar(\sigma)^{-1} = {\bf Q}^\hbar( \sigma\cdot f), \qquad \forall f\in \mathcal{A}, \quad \forall \sigma \in G.
$$

Often, one also extends $C^\infty(M)$ to $C^\infty(M;\mathbb{C})$ equipped with the $*$-structure given by complex conjugation, and takes the classical (commutative) algebra $\mathcal{A}$ to be a $*$-subalgebra of $C^\infty(M;\mathbb{C})$. One then must require the quantization map ${\bf Q}^\hbar$ to preserve the $*$-structure and the operator adjoint, in an appropriate sense.

There are a few other subtleties in the definition above that we find worth mentioning here. First, each self-adjoint operator on $\mathscr{H}$ must be defined on some dense domain; such domain may differ from an operator to another, creating subtle issues in functional analysis. Another issue is related to the choice of an appropriate topology on the space of operators on $\mathscr{H}$; this is needed to make proper sense of the so-called semi-classical limit $\hbar\to 0$ in the item (DQ3). Yet another possible problem is the requirement of irreducibility of the quantum representation; this is not strictly necessary although it may be desired in particular situations.

In trying to tackle such subtleties, it is useful to decompose the quantization map ${\bf Q}^\hbar$ into algebraic and operator parts as follows. A {\it deformation quantization} map refers to a family of maps
$$
	\wh{Q}^\hbar ~:~ \mathcal{A} \to \mathcal{A}^\hbar,
$$
where $\mathcal{A}^\hbar$ is a family of non-commutative $*$-algebras over $\mathbb{C}$, with $\mathcal{A}^0 \cong \mathcal{A}$, that satisfies some algebraic analogues of (DQ1)--(DQ3). Here, one can make sense of (DQ3) by regarding $\hbar$ as a formal symbol and equipping the quantum algebra $\mathcal{A}^\hbar$ with a suitable topology, e.g. one may realize $\mathcal{A}^\hbar$ as a vector subspace of the vector space $\mathbb{C}[[\hbar]] \otimes_\mathbb{C} \mathcal{A}$, equipped with an appropriate topology and a non-commutative product structure. In this case, one would require $\mathcal{A}^\hbar /\hbar \mathcal{A}^\hbar \cong \mathcal{A}$, and would formulate (DQ3) in terms of the classicalization map $\mathcal{A}^\hbar \to \mathcal{A}^\hbar/\hbar \mathcal{A}^\hbar \cong \mathcal{A}$. The remaining is the operator part
$$
	\pi^\hbar~:~ \mathcal{A}^\hbar \to \big\{\mbox{densely-defined linear operators on $\mathscr{H}$}\big\}
$$
which is required to be a $*$-algebra homomorphism. That is, $\pi^\hbar$ is just a $*$-representation (irreducible or not) of the algebra $\mathcal{A}^\hbar$ on the Hilbert space $\mathscr{H}$. The issues with domains must still be dealt with in this setting, but it now becomes easier to formulate precise statements to be proved. In the end, the sought-for quantization map ${\bf Q}^\hbar$ as in eq.\eqref{eq:bf_Q_hbar} would be constructed as
$$
	{\bf Q}^\hbar = \pi^\hbar \circ \wh{Q}^\hbar.
$$

\subsection{Cluster $\mathscr{X}$-varieties at $\mathbb{R}^+$: the algebraic aspect}
\label{subsec:the_case_of_a_cluster_X-variety_at_R_positive_algebraic}

Here we review how the problem of quantization of the set $\mathscr{X}_{|\Gamma|}(\mathbb{R}^+)$ of positive real points of a cluster $\mathscr{X}$-variety is formulated \cite{FG09a,CKKO20}. Let $|\Gamma|$ denote an equivalence class of cluster $\mathscr{X}$-seeds, in the sense of \S\ref{subsec:cluster_X-varieties}. For each seed $\Gamma \in |\Gamma|$ is associated a positive real cluster $\mathscr{X}$-chart $\mathscr{X}_\Gamma(\mathbb{R}^+) = (\mathbb{R}^+)^I$ whose coordinate functions are denoted by $X_i$, $i \in I$, each of which is a smooth positive real valued function on $\mathscr{X}_\Gamma(\mathbb{R}^+)$. This chart is endowed with a Poisson structure, given in terms of the coordinate functions as $\{X_i,X_j \} = \varepsilon_{ij} X_i X_j$, $\forall i,j \in I$, with $\varepsilon$ the corresponding exchange matrix of $\Gamma$, and we consider the set of all functions on $\mathscr{X}_\Gamma(\mathbb{R}^+)$ that can be expressed as Laurent polynomials in $X_i$, $i \in I$, as a classical algebra of observables associated to $\Gamma$. More precisely, we define the Laurent polynomial algebra over the complex numbers generated by the variables $X_i$, $i\in I$,
$$
	\mathring{\mathcal{A}}_\Gamma := \mathbb{C}[\{X_i^{\pm 1} : i\in I\}]
$$
together with a natural $*$-structure given by $X_i^*=X_i$, $\forall i\in I$. For the corresponding non-commutative quantum algebra, we then consider the {\it quantum torus algebra} $\mathring{\mathcal{A}}^\hbar_\Gamma$, which is defined as the associative algebra over $\mathbb{C}$ with the following generators and relations:
\begin{align*}
	\mbox{generators} & ~:~ \wh{X}_i^{\pm 1}, \qquad i \in I,                                                       \\
	\mbox{relations}  & ~:~ \wh{X}_i \wh{X}_j = q^{2 \varepsilon_{ij}} \wh{X}_j \wh{X}_i, \qquad \forall i,j \in I.
\end{align*}
Here we omit the trivial relations $\wh{X}_i \wh{X}_i^{-1} = \wh{X}_i^{-1} \wh{X}_i = 1$ and set
$$
	q = e^{\pi {\rm i} \hbar},
$$
where $\hbar$ can be viewed as a real number parametrizing a family of algebras $\mathring{\mathcal{A}}^\hbar_\Gamma$. As shall be seen later, for our purposes it will be better to view $q$ as a formal symbol and $\mathring{\mathcal{A}}^\hbar_\Gamma$ as a $\mathbb{C}[q^{\pm 1}]$-algebra (or $\mathbb{C}(q)$-algebra) defined by generators and relations as above; when considering representations of $\mathring{\mathcal{A}}^\hbar_\Gamma$, we will require $q$ to be represented by the scalar $e^{\pi {\rm i} \hbar}$. The $*$-structure on $\mathring{\mathcal{A}}^\hbar_\Gamma$ is given on the quantum generators as
$$
	\wh{X}_i^* = \wh{X}_i, \quad \forall i \in I.
$$
To define a quantization of the cluster $\mathscr{X}$-chart $\mathscr{X}_\Gamma(\mathbb{R}^+)$, one then looks for a deformation quantization map $\mathring{\wh{Q}}{}^\hbar_\Gamma ~:~ \mathring{\mathcal{A}}_\Gamma \to \mathcal{A}^\hbar_\Gamma$ and for a representation $\pi_\Gamma$ of $\mathcal{A}^\hbar_\Gamma$ on a Hilbert space $\mathscr{H}_\Gamma$ that satisfy the desired conditions; one noteworthy point is that we require each generator $\wh{X}_i$ to be represented by a {\it positive} self-adjoint operator $\pi_\Gamma(\wh{X}_i)$. This quantization problem can be solved rather easily for each seed $\Gamma$, and indeed there are many solutions. However, in order to define a quantization of the whole cluster variety $\mathscr{X}_{|\Gamma|}(\mathbb{R}^+)$, one must ensure that the solutions for different seeds are compatible with one another in a certain precise sense. Only then one will achieve a quantization of the entire variety, not only of a single chart.

\vs

To formulate this compatibility, one first notes that the classical algebras for different seeds are related by a sequence of mutations and index permutations. That is, for each pair of seeds $\Gamma,\Gamma' \in |\Gamma|$, there is a uniquely determined $*$-isomorphism
$$
	\mathring{\mu}_{\Gamma,\Gamma'} ~:~ {\rm Frac}(\mathring{\mathcal{A}}_{\Gamma'}) \to {\rm Frac}(\mathring{\mathcal{A}}_{\Gamma})
$$
between the corresponding fields of fractions of the coordinate rings. These are defined by a composition of mutations $\mu_k$ and index permutations $P_\sigma$ as described in \S\ref{subsec:cluster_X-varieties}. On the quantum side, for each pair of seeds $\Gamma,\Gamma'$, there must also be a corresponding quantum mutation isomorphism
$$
	\mathring{\mu}^\hbar_{\Gamma,\Gamma'} ~:~ {\rm Frac}(\mathring{\mathcal{A}}^\hbar_{\Gamma'}) \to {\rm Frac}(\mathring{\mathcal{A}}^\hbar_{\Gamma}),
$$
which is a $*$-isomorphism between the corresponding skew-fields of fractions such that
\begin{enumerate}
	\item[\rm (QM1)] $\mathring{\mu}^\hbar_{\Gamma,\Gamma'}$ recovers $\mathring{\mu}_{\Gamma,\Gamma'}$ as $q\to 1$ (or, as $\hbar \to 0$),

	\item[\rm (QM2)] $\mathring{\mu}^\hbar_{\Gamma,\Gamma'} \circ \mathring{\mu}^\hbar_{\Gamma',\Gamma''}= \mathring{\mu}^\hbar_{\Gamma,\Gamma''}$ holds for each triple of seeds $\Gamma,\Gamma',\Gamma'' \in |\Gamma|$.
\end{enumerate}
Another stipulation is that the entire construction should be invariant (or equivariant) under the action of the cluster mapping class group, which consists of transformations of seeds that preserve the underlying exchange matrices. That is, we also ask for:
\begin{enumerate}
	\item[\rm (QM3)] The map $\mathring{\mu}^\hbar_{\Gamma,\Gamma'}$ depends only on the underlying exchange matrices $\varepsilon,\varepsilon'$ of the seeds $\Gamma,\Gamma'$.
\end{enumerate}
Note that this condition makes sense because the algebra $\mathring{\mathcal{A}}^\hbar_\Gamma$ is constructed in such a way that it only depends on the underlying exchange matrix $\varepsilon$. Another important remark here is that the skew-fields of fractions above are indeed well defined because the algebras $\mathring{\mathcal{A}}^\hbar_\Gamma$ satisfy the Ore condition \cite{Coh95}.

\begin{proposition}[\cite{FG09a,BZ05,FC99,Kas98}]\label{prop:quantum_mutation}
	There exists a quantum mutation isomorphism $\mathring{\mu}^\hbar_{\Gamma,\Gamma'}$ satisfying (QM1)--(QM3).
\end{proposition}

When $\Gamma' = \mu_k(\Gamma)$, the isomorphism $\mathring{\mu}^\hbar_{\Gamma,\Gamma'}$ in Prop.\ref{prop:quantum_mutation} is described as the composition of $*$-isomorphisms
\begin{align}
	\label{eq:mu_hbar_decomposition}
	\mathring{\mu}^\hbar_{\Gamma,\Gamma'} = \mathring{\mu}^\sharp_{\Gamma,\Gamma'} \circ \mathring{\mu}'_{\Gamma,\Gamma'} ~:~ {\rm Frac}(\mathring{\mathcal{A}}^\hbar_{\Gamma'}) ~ \overset{\mathring{\mu}'_{\Gamma,\Gamma'}}{\longrightarrow} ~ {\rm Frac}(\mathring{\mathcal{A}}^\hbar_{\Gamma}) ~ \overset{\mathring{\mu}^\sharp_{\Gamma,\Gamma'}}{\longrightarrow} ~ {\rm Frac}(\mathring{\mathcal{A}}^\hbar_{\Gamma}),
\end{align}
with the {\it monomial transformation part} $\mathring{\mu}'_{\Gamma,\Gamma'}$ and the {\it automorphism part} $\mathring{\mu}^\sharp_{\Gamma,\Gamma'}$ being given on the generators by
\begin{align}
	\label{eq:mu_hbar_prime_k}
	\mathring{\mu}'_{\Gamma,\Gamma'}(\wh{X}_i')      & = \left\{
	\begin{array}{ll}
		\wh{X}_k^{-1}                                                                       & \mbox{if $i =k$},    \\
		q^{-\varepsilon_{ik} [\varepsilon_{ik}]_+} \wh{X}_i \wh{X}_k^{[\varepsilon_{ik}]_+} & \mbox{if $i\neq k$},
	\end{array}
	\right.                                                                                                                                                                                 \\
	\label{eq:mu_hbar_sharp_k}
	\mathring{\mu}^\sharp_{\Gamma,\Gamma'}(\wh{X}_i) & = \wh{X}_i \prod_{r=1}^{|\varepsilon_{ik}|} (1 + (q^{-{\rm sgn}(\varepsilon_{ik})})^{2r-1} \wh{X}_k)^{-{\rm sgn}(\varepsilon_{ik})},
\end{align}
where $\varepsilon$ is the exchange matrix of $\Gamma$ and $[\sim]_+$ denotes the positive part of a real number
$$
	\textstyle [a]_+ := \frac{1}{2}(a+|a|), \qquad \forall a\in \mathbb{R}.
$$
A better understanding of the above complicated-looking formulas for $\mathring{\mu}'_{\Gamma,\Gamma'}$ and $\mathring{\mu}^\sharp_{\Gamma,\Gamma'}$ will be reviewed later. When $\Gamma' = P_\sigma(\Gamma)$, the isomorphism $\mathring{\mu}^\hbar_{\Gamma,\Gamma'}$ is given as $\mathring{\mu}^\hbar_{\Gamma,\Gamma'}(\wh{X}'_{\sigma(i)}) = \wh{X}_i$. For a general pair of seeds $\Gamma,\Gamma'$, the isomorphism $\mathring{\mu}^\hbar_{\Gamma,\Gamma'}$ is given as the composition of the above two elementary kinds.

\vs

The isomorphisms $\mathring{\mu}_{\Gamma,\Gamma'}$ and $\mathring{\mu}^\hbar_{\Gamma,\Gamma'}$ should in principle allow us to compare the deformation quantization maps $\mathring{\wh{Q}} {}^\hbar_\Gamma : \mathring{\mathcal{A}}_\Gamma \to \mathring{\mathcal{A}}^\hbar_\Gamma$ associated to different seeds $\Gamma$ and to formulate the compatibility condition. However, the classical and quantum algebras $\mathring{\mathcal{A}}_\Gamma$ and $\mathring{\mathcal{A}}^\hbar_\Gamma$ are not in general preserved under such isomorphisms. We thus define more invariant algebras as follows. For each fixed seed $\Gamma$, an element of ${\rm Frac}(\mathring{\mathcal{A}}_\Gamma)$ is said to be {\it Laurent} for $\Gamma$ if it belongs to $\mathring{\mathcal{A}}_\Gamma$. Likewise, an element of ${\rm Frac}(\mathring{\mathcal{A}}^\hbar_\Gamma)$ is Laurent for $\Gamma$ if it belongs to $\mathring{\mathcal{A}}_\Gamma^\hbar$. We define the algebra of {\it universally Laurent} elements of ${\rm Frac}(\mathring{\mathcal{A}}_\Gamma)$ and those of ${\rm Frac}(\mathring{\mathcal{A}}^\hbar_\Gamma)$ respectively as
\begin{align}
	\nonumber
	\mathring{\mathbb{L}}_\Gamma       & := \bigcap_{\Gamma' \in |\Gamma|} \mathring{\mu}_{\Gamma,\Gamma'}(\mathring{\mathcal{A}}_{\Gamma'}) ~\subset~ \mathring{\mathcal{A}}_\Gamma ~ \subset ~ {\rm Frac}(\mathring{\mathcal{A}}_\Gamma),                         \\
	\nonumber
	\mathring{\mathbb{L}}^\hbar_\Gamma & := \bigcap_{\Gamma' \in |\Gamma|} \mathring{\mu}^\hbar_{\Gamma,\Gamma'}(\mathring{\mathcal{A}}^\hbar_{\Gamma'}) ~\subset~ \mathring{\mathcal{A}}^\hbar_\Gamma ~ \subset ~ {\rm Frac}(\mathring{\mathcal{A}}^\hbar_\Gamma).
\end{align}
Note that the algebras $\mathring{\mathbb{L}}_\Gamma$ for different $\Gamma$'s are now canonically identified via the maps $\mathring{\mu}_{\Gamma,\Gamma'}$, while the algebras $\mathring{\mathbb{L}}^\hbar_{\Gamma}$'s are canonically identified via $\mathring{\mu}^\hbar_{\Gamma,\Gamma'}$. In particular we can indeed compare the deformation quantization maps for different seeds when restricted to the universally Laurent algebras. For the compatibility condition, we now require the quantization maps
$$
	\mathring{\wh{Q}} {}^\hbar_\Gamma ~:~ \mathring{\mathbb{L}}_\Gamma \to \mathring{\mathbb{L}}^\hbar_\Gamma
$$
to make the following diagram commute for each pair of seeds $\Gamma,\Gamma'\in|\Gamma|$
$$
	\xymatrix@C+10mm@R-2mm{
	\mathring{\mathbb{L}}_\Gamma \ar[r]^-{\mathring{\wh{Q}} {}^\hbar_\Gamma} & \mathring{\mathbb{L}}_\Gamma^\hbar \\
	\mathring{\mathbb{L}}_{\Gamma'} \ar[r]^-{\mathring{\wh{Q}} {}^\hbar_{\Gamma'}} \ar[u]^{\mathring{\mu}_{\Gamma,\Gamma'}} & \mathring{\mathbb{L}}_{\Gamma'}^\hbar  \ar[u]_{\mathring{\mu}^\hbar_{\Gamma,\Gamma'}},
	}
$$
i.e.
$$
	\mathring{\wh{Q}} {}^\hbar_\Gamma(\mathring{\mu}_{\Gamma,\Gamma'}(u)) = \mathring{\mu}^\hbar_{\Gamma,\Gamma'}(\mathring{\wh{Q}} {}^\hbar_{\Gamma'}(u)), \quad \forall u \in \mathring{\mathbb{L}}_{\Gamma'}.
$$
Then, if one identifies each $\mathring{\mathbb{L}}_\Gamma$ as the ring $\mathscr{O}(\mathscr{X}_{|\Gamma|})$ of globally regular functions on the variety $\mathscr{X}_{|\Gamma|}$, and each $\mathring{\mathbb{L}}^\hbar_\Gamma$ as the ring $\mathscr{O}^\hbar(\mathscr{X}_{|\Gamma|})$ of globally regular quantum functions, the quantization maps $\mathring{\wh{Q}} {}^\hbar_\Gamma$ yield a well-defined deformation quantization map
\begin{align}
	\label{eq:mathring_wh_Q}
	\mathring{\wh{Q}} {}^\hbar_{|\Gamma|} ~:~ \mathscr{O}(\mathscr{X}_{|\Gamma|}) \to \mathscr{O}^\hbar(\mathscr{X}_{|\Gamma|}),
\end{align}
independently of the choice of seeds.
Constructing such a deformation quantization map $\mathring{\wh{Q}} {}^\hbar_{|\Gamma|}$ is quite a non-trivial task, and was not done in \cite{FG09a}. In the present paper we exhibit a solution by assembling various results on {\it duality maps}, which are related to special bases of $\mathscr{O}(\mathscr{X}_{|\Gamma|})$ and $\mathscr{O}^\hbar(\mathscr{X}_{|\Gamma|})$, and which constitute one of the prominent topics in the theory of cluster varieties. This idea is hinted briefly in \cite{CKKO20}, and is described in more detail in \S\ref{subsec:algebraic_deformation_quantization_through_duality_maps} of the present paper.

\subsection{Cluster $\mathscr{X}$-varieties at $\mathbb{R}^+$: the operator aspect}
\label{subsec:the_case_of_a_cluster_X-variety_at_R_positive_operator}

It still remains to formulate the operator part, where we mostly follow \cite{FG09a}. First, we seek to begin with a representation of $\mathring{\mathbb{L}}^\hbar_\Gamma$, defined on a dense subspace $\mathring{\mathscr{D}}_\Gamma$ of a Hilbert space $\mathring{\mathscr{H}}_\Gamma$, in the form of an algebra homomorphism
$$
	\mathring{\til{\pi}}_\Gamma = \mathring{\til{\pi}} {}^\hbar_\Gamma ~:~ \mathring{\mathbb{L}}^\hbar_\Gamma \to {\rm End}_\mathbb{C}(\mathring{\mathscr{D}}_\Gamma).
$$
In particular, for each element $u$ of $\mathring{\mathbb{L}}^\hbar_\Gamma$, is associated a linear operator $\mathring{\til{\pi}}_\Gamma (u) : \mathring{\mathscr{D}}_\Gamma \to \mathring{\mathscr{D}}_\Gamma$, so $\mathring{\til{\pi}}_\Gamma$ makes sense as a genuine representation on the space $\mathring{\mathscr{D}}_\Gamma$ in a usual linear algebraic sense. We require $\mathring{\til{\pi}}_\Gamma$ to preserve the $*$-structure in the sense that $\mathring{\til{\pi}}_\Gamma(u^*) \subseteq \mathring{\til{\pi}}_\Gamma(u)^*$, or $\mathring{\til{\pi}}_\Gamma(u^*) = \mathring{\til{\pi}}_\Gamma(u)^* \restriction{\mathring{\mathscr{D}}_\Gamma}$, where $\mathring{\til{\pi}}_\Gamma(u)^*$ means the operator adjoint of $\mathring{\til{\pi}}_\Gamma(u)$. The construction of a representation satisfying the above was given in \cite{FG09a} by first considering a representation of the quantum torus algebra $\mathring{\mathcal{A}}^\hbar_\Gamma$ on $\mathring{\mathscr{D}}_\Gamma$, and restricting it to $\mathring{\mathbb{L}}^\hbar_\Gamma$.

\vs

We then extend the operators $\mathring{\til{\pi}}_\Gamma(u)$, defined on $\mathring{\mathscr{D}}_\Gamma$, to their common maximal domain
$$
	\mathring{\mathscr{S}}_\Gamma := \bigcap_{u\in \mathring{\mathbb{L}}^\hbar_\Gamma} {\rm Dom}(\mathring{\til{\pi}}_\Gamma(u)^*)
$$
which Fock and Goncharov refer to as the {\it Schwartz space}; here, ${\rm Dom}(\mathring{\til{\pi}}_\Gamma(u)^*)$ denotes the domain of the operator adjoint of $\mathring{\til{\pi}}_\Gamma(u)$, i.e.
\begin{align*}
	{\rm Dom}(\mathring{\til{\pi}}_\Gamma(u)^*) = \Big\{ \xi \in \mathring{\mathscr{H}}_\Gamma \, \left| \, \mbox{the linear functional $\mathring{\mathscr{D}}_\Gamma \to \mathbb{C}$ given by} \right.
	\cr
	\mbox{$\eta \mapsto \langle \mathring{\til{\pi}}_\Gamma(u) \eta, \xi \rangle$ is bounded} \Big\}.
\end{align*}
In particular, by the Riesz representation theorem, for each $\xi \in {\rm Dom}(\mathring{\til{\pi}} {}^\hbar_\Gamma(u)^*)$ there exists a vector $\xi' \in \mathring{\mathscr{H}}_\Gamma$ such that $\langle \mathring{\til{\pi}}_\Gamma(u) \eta, \xi \rangle = \langle \eta, \xi' \rangle$ holds for all $\eta \in \mathring{\mathscr{D}}_\Gamma$, i.e. $\mathring{\til{\pi}}_\Gamma(u)^* \xi =\xi'$. This lets us extend the operators $\mathring{\til{\pi}}_\Gamma(u)$ to $\mathring{\mathscr{S}}_\Gamma$, which we denote by $\mathring{\pi}_\Gamma(u)$: for each $u\in \mathring{\mathbb{L}}_\Gamma$ and $\xi \in \mathring{\mathscr{S}}_\Gamma$, define $\mathring{\pi}_\Gamma(u) \xi \in \mathring{\mathscr{H}}_\Gamma$ as follows \cite{FG09a}
$$
	\mathring{\pi}_\Gamma(u) \xi := \mathring{\til{\pi}}_\Gamma(u^*)^* \xi.
$$
One of the crucial analytic arguments in \cite{FG09a} is to equip the Schwartz space $\mathring{\mathscr{S}}_\Gamma$ a Frech\'et topology given by the family of seminorms $\Vert\,\cdot\,\Vert_u$ on $\mathring{\mathscr{S}}_\Gamma$, enumerated by the elements $u$ of (a countable basis of) $\mathring{\mathbb{L}}^\hbar_\Gamma$, defined as $\Vert\xi\Vert_u := \Vert\mathring{\pi}_\Gamma(u)\xi\Vert$, where $\Vert\,\cdot\,\Vert$ is the usual Hilbert space norm. It is then shown that $\mathring{\mathscr{D}}_\Gamma$ is dense in $\mathring{\mathscr{S}}_\Gamma$ under this topology, which allows one to check various statements just for vectors in $\mathring{\mathscr{D}}_\Gamma$ in order to prove those for vectors in $\mathring{\mathscr{S}}_\Gamma$; we will also be using this strategy implicitly.

\vs

Finally, the representations $\mathring{\pi}_\Gamma$ on the Schwartz spaces $\mathring{\mathscr{S}}_\Gamma$ for different $\Gamma$'s must be compatible with each other in the following sense; for each pair of seeds $\Gamma,\Gamma' \in |\Gamma|$ there must be a unitary operator
$$
	\mathring{\bf K}_{\Gamma,\Gamma'} = \mathring{\bf K}^\hbar_{\Gamma,\Gamma'} ~:~ \mathring{\mathscr{H}}_{\Gamma'} \to \mathring{\mathscr{H}}_\Gamma
$$
called the {\it intertwiner} for the quantum coordinate change map $\mu^\hbar_{\Gamma,\Gamma'}$, such that

\begin{enumerate}
	\item[\rm (IT1)] $\mathring{\bf K}_{\Gamma,\Gamma'}(\mathring{\mathscr{S}}_{\Gamma'}) = \mathring{\mathscr{S}}_\Gamma$;

	\item[\rm (IT2)] $\mathring{\bf K}_{\Gamma,\Gamma'}$ intertwines the representations $\mathring{\pi}_{\Gamma'}$ and $\mathring{\pi}_{\Gamma}$ in the sense that the following diagram commutes for all $u \in \mathring{\mathbb{L}}^\hbar_{\Gamma'}$
	      $$
		      \xymatrix@C+10mm@R-2mm{
		      \mathring{\mathscr{S}}_\Gamma \ar[r]^-{\mathring{\pi}_\Gamma(\mathring{\mu}^\hbar_{\Gamma,\Gamma'}(u))} & \mathring{\mathscr{S}}_\Gamma \\
		      \mathring{\mathscr{S}}_{\Gamma'} \ar[r]^-{\mathring{\pi}_{\Gamma'}(u)} \ar[u]^{\mathring{\bf K}_{\Gamma,\Gamma'}} & \mathring{\mathscr{S}}_{\Gamma'} \ar[u]_{\mathring{\bf K}_{\Gamma,\Gamma'}} \
		      }
	      $$
	      i.e. the following intertwining equations are satisfied for all $u \in \mathring{\mathbb{L}}^\hbar_{\Gamma'}$
	      $$
		      \mathring{\bf K}_{\Gamma,\Gamma'} \circ \mathring{\pi}_{\Gamma'}(u) = \mathring{\pi}_\Gamma(\mathring{\mu}^\hbar_{\Gamma,\Gamma'}(u)) \circ \mathring{\bf K}_{\Gamma,\Gamma'};
	      $$

	\item[\rm (IT3)] for each triple of seeds $\Gamma,\Gamma',\Gamma'' \in |\Gamma|$, the consistency equations
	      $$
		      \mathring{\bf K}_{\Gamma,\Gamma'} \circ \mathring{\bf K}_{\Gamma',\Gamma''} = c_{T,T',T''}\mathring{\bf K}_{\Gamma,\Gamma''}
	      $$
	      hold up to multiplicative constants;

	\item[\rm (IT4)] $\mathring{\bf K}^\hbar_{\Gamma,\Gamma'}$ depends only on the underlying exchange matrices $\varepsilon,\varepsilon'$ of the seeds $\Gamma,\Gamma'$.
\end{enumerate}

These intertwiners allow us to identify the representations $\mathring{\pi}_\Gamma$ for different seeds $\Gamma$ in a unitary and consistent manner, yielding the sought-for quantization map ${\bf Q}^\hbar$ for the cluster $\mathscr{X}$-variety $\mathscr{X}_{|\Gamma|}$. Even without relating to the quantization map, constructing such a system of intertwiners can be interpreted as constructing a projective unitary representation of the groupoid of seeds in $|\Gamma|$. A morphism in this groupoid can be written as a pair of seeds $(\Gamma,\Gamma')$, and for a fixed exchange matrix $\varepsilon$, the set of all morphisms $(\Gamma,\Gamma')$ such that $\varepsilon$ is the underlying exchange matrix for both $\Gamma$ and $\Gamma'$ form a group called the {\it cluster mapping class group}. This recovers the usual mapping class group in case when $\varepsilon$ comes from an ideal triangulation of a punctured surface. The above quantum representation of the groupoid of seeds then leads to representations of the cluster mapping class groups, which have been viewed as one of the main results of the theory of quantum cluster varieties and quantum Teichm\"uller theory. The construction of such a representation is indeed the main achievement of \cite{FG09a}; this also hints that it is one of the most difficult and central parts of the entire quantization problem.
\begin{proposition}[\cite{FC99,Kas98,FG09a,Kim21b}]
	There exist representations $\mathring{\pi}_\Gamma$ and intertwiners $\mathring{\bf K}_{\Gamma,\Gamma'}$ satisfying (IT1)--(IT4).
\end{proposition}

The formulation in \cite{FG09a} is actually for the cluster $\mathscr{D}$-variety, which is the `symplectic double' of the cluster $\mathscr{X}$-variety; one can restrict the result of \cite{FG09a}  to the cluster $\mathscr{X}$-variety, which yields a reducible representation. For an irreducible representation of a quantum cluster $\mathscr{X}$-variety, see \cite{Kim21b}.

\subsection{Cluster $\mathscr{X}$-varieties at $\mathbb{R}_\Lambda^+$}
\label{subsec:the_case_of_a_cluster_X-variety_and_R_Lambda}

Now we formulate a version of the quantization of $\mathscr{X}_{|\Gamma|}(\mathbb{R}_\Lambda^+)$, which is the main subject of the present paper. First we complexify the ring of generalized complex numbers to $\mathbb{C} \otimes_\mathbb{R} \mathbb{R}_\Lambda = \mathbb{C}[\ell] / (\ell^2 + \Lambda)$. In order to incorporate the $*$-structure at the element $\ell$, we further extend ring in the case $\Lambda=0$.
\begin{definition}
	\label{def:C_Lambda}
	Define the complexified version of the ring of generalized complex numbers as the $\mathbb{C}$-algebra
	$$
		\mathbb{C}_\Lambda := \left\{
		\begin{array}{ll}
			\mathbb{C}[\ell]/(\ell^2+\Lambda)           & \mbox{if $\Lambda=-1,1$}, \\
			\mathbb{C}[\ell,\ell^*]/(\ell^2,(\ell^*)^2) & \mbox{if $\Lambda=0$}.
		\end{array}
		\right.
	$$
	The $*$-structure is given by the unique $\mathbb{C}$-conjugate-linear map $*:\mathbb{C}_\Lambda \to \mathbb{C}_\Lambda$, $u\mapsto u^*$, s.t.
	\begin{align*}
		\left\{
		\begin{array}{ll}
			* : \ell \mapsto -\Lambda\ell & \mbox{if $\Lambda=-1,1$}, \\
			* : \ell \mapsto \ell^*       & \mbox{if $\Lambda=0$}.
		\end{array}
		\right.
	\end{align*}
\end{definition}

\vs

For each seed $\Gamma \in |\Gamma|$, is associated a chart $\mathscr{X}_\Gamma(\mathbb{R}_\Lambda^+) = (\mathbb{R}_\Lambda^+)^I$, with the $\mathbb{R}_\Lambda$-valued coordinate functions $Z^{(+)}_i$, $Z^{(-)}_i$, $i \in I$. The ring $\mathcal{A}_\Gamma$ of classical observables for $\Gamma$ is taken to be the ring of Laurent polynomials in $Z_i^{(\epsilon)}$, for $i \in I$ and $\epsilon \in \{+,-\}$, which are now viewed as formal generators rather than actual functions. To make sense of the Poisson bracket on $\mathcal{A}_\Gamma$ using the formulas in eq.\eqref{eq:R_Lambda_Poisson_brackets}, one should in principle define it as a Laurent polynomial algebra over $\mathbb{C}_\Lambda$. However, when dealing with algebraic statements it will be sufficient to consider $\mathcal{A}_\Gamma$ as an algebra over $\mathbb{C}$. In fact, we would like to avoid using the ring $\mathbb{C}_\Lambda$ as much as possible due to the presence of zero divisors for $\Lambda=0,-1$, which complicates considerations about fields of fractions needed later. To be more precise:

\begin{definition}
	For $\Lambda=-1,1$, we define the classical observable algebra $\mathcal{A}_\Gamma$ as the (commutative) Laurent polynomial algebra over $\mathbb{C}$ with the set of generators $\big\{ Z_i^{(\epsilon)} ~:~ i \in I, \epsilon \in \{+,-\}\big\}$ and with the $*$-structure
	$$
		\left\{
		\begin{array}{ll}
			* : Z_i^{(\epsilon)} \mapsto Z_i^{(\epsilon)},  & \mbox{if $\Lambda=-1$}, \\
			* : Z_i^{(\epsilon)} \mapsto Z_i^{(-\epsilon)}, & \mbox{if $\Lambda=1$}.
		\end{array}
		\right.
	$$
	For $\Lambda=0$, we define $\mathcal{A}_\Gamma$ as the Laurent polynomial algebra over $\mathbb{C}$ with generators $\big\{ Z_i^{(\epsilon)}, ( Z_i^{(\epsilon)})^* ~:~ i \in I, \epsilon \in \{+,-\}\big\}$ equipped with the $*$-structure
	$$
		\left\{
		\begin{array}{ll}
			* : Z_i^{(\epsilon)} \mapsto (Z_i^{(\epsilon)})^*, & \mbox{if $\Lambda=0$}.
		\end{array}
		\right.
	$$
\end{definition}

The quantum algebra $\mathcal{A}_\Gamma^\hbar$ could then be defined as the free associative (non-commutative) $\mathbb{C}_\Lambda$-algebra generated by $(\wh{Z}_i^{(\epsilon)})^{\pm 1}$, $\epsilon \in \{+,-\}$, $i\in I$, modulo by relations such as $\wh{Z}_i^{(+)} \wh{Z}_j^{(+)} = q_\Lambda^{2 \varepsilon_{ij}} \wh{Z}_j^{(+)} \wh{Z}_i^{(+)}$, where $\varepsilon$ is the underlying exchange matrix for $\Gamma$ and
$$
	q_\Lambda := e^{\pi {\rm i} \ell \hbar} ~\in~\mathbb{C}_\Lambda.
$$
However, again to avoid using the ring $\mathbb{C}_\Lambda$ as much as possible, we first introduce the following $\mathbb{C}$-algebra.
\begin{definition}
	\label{def:C_hbar_Lambda}
	We define the quantum coefficient ring $\mathbb{C}^\hbar_\Lambda$ as the $\mathbb{C}$-algebra
	\begin{align*}
		\mathbb{C}^\hbar_\Lambda := \left\{
		\begin{array}{ll}
			\mathbb{C}[q_\Lambda^{\pm 1}]                        & \mbox{if $\Lambda=-1,1,$} \\
			\mathbb{C}[q_\Lambda^{\pm 1}, (q_\Lambda^*)^{\pm 1}] & \mbox{if $\Lambda=0$},
		\end{array}
		\right.
	\end{align*}
	with the $*$-structure
	$$
		\left\{
		\begin{array}{ll}
			* : q_\Lambda \mapsto q_\Lambda^{-1} & \mbox{if $\Lambda=-1$}, \\
			* : q_\Lambda \mapsto q_\Lambda      & \mbox{if $\Lambda=1$},  \\
			* : q_\Lambda \mapsto q_\Lambda^*    & \mbox{if $\Lambda=0$}.
		\end{array}
		\right.
	$$
	Here $q_\Lambda$ and $q_\Lambda^*$ are considered as formal symbols; indeed we also consider the superscritp $\hbar$ in $\mathbb{C}^\hbar_\Lambda$ to be a formal symbol indicating the quantum setting, rather than an actual real parameter.
\end{definition}
Note that the $*$-structure is compatible with the natural algebra map $\mathbb{C}^\hbar_\Lambda \to \mathbb{C}_\Lambda$ sending $q_\Lambda \mapsto e^{\pi {\rm i} \ell\hbar}$ (and $q_\Lambda^* \mapsto e^{-\pi{\rm i} \ell^*\hbar}$).
\begin{definition}
	For $\Lambda=-1,1$, we define the quantum observable algebra $\mathcal{A}^\hbar_\Gamma$ as the free associative algebra over $\mathbb{C}^\hbar_\Lambda$ generated by $\{ (Z_i^{(\epsilon)})^{\pm 1} ~:~ i\in I, \, \epsilon \in \{+,-\}\}$, modulo the relations
	$$
		\wh{Z}_i^{(+)} \wh{Z}_j^{(+)} = q_\Lambda^{2 \varepsilon_{ij}} \wh{Z}_j^{(+)} \wh{Z}_i^{(+)}, \;
		\wh{Z}_i^{(-)} \wh{Z}_j^{(-)} = q_\Lambda^{-2 \varepsilon_{ij}} \wh{Z}_j^{(-)} \wh{Z}_i^{(-)}, \;
		\wh{Z}_i^{(+)} \wh{Z}_j^{(-)} = \wh{Z}_j^{(-)} \wh{Z}_i^{(+)}.
	$$
	The $*$-structure is given by
	$$
		\left\{
		\begin{array}{ll}
			* : \wh{Z}_i^{(\epsilon)} \mapsto \wh{Z}_i^{(\epsilon)},  & \mbox{if $\Lambda=-1$}, \\
			* : \wh{Z}_i^{(\epsilon)} \mapsto \wh{Z}_i^{(-\epsilon)}, & \mbox{if $\Lambda=1$}.
		\end{array}
		\right.
	$$
	For $\Lambda=0$, we define $\mathcal{A}^\hbar_\Gamma$ as the free associative algebra over $\mathbb{C}^\hbar_\Lambda$ generated by $\big\{(Z_i^{(\epsilon)})^{\pm 1}, ((Z_i^{(\epsilon)})^*)^{\pm 1} ~:~ i \in I, \, \epsilon \in \{+,-\}\big\}$, modulo the above relations for $\wh{Z}_i^{(\epsilon)}$, together with the corresponding relations for $(\wh{Z}_i^{(\epsilon)})^*$ with $q_\Lambda$ replaced by $(q_\Lambda^*)^{-1}$, and equipped with the obvious $*$-structure.
\end{definition}

Since we view $\mathring{\mathcal{A}}^\hbar_\Gamma$ and $\mathcal{A}^\hbar_\Gamma$ as algebras over $\mathbb{C}[q^{\pm 1}]$ and $\mathbb{C}^\hbar_\Lambda$, we may also regard them as $\mathbb{C}$-algebras, where $q^{\pm 1}$, $q_\Lambda^{\pm 1}$ and $(q_\Lambda^*)^{\pm 1}$ are viewed as elements. Then, for each $\epsilon \in \{+,-\}$, we have an injective $\mathbb{C}$-algebra homomorphism
\begin{align}
	\label{eq:two_algebra_embeddings}
	\iota^{(\epsilon)}_\Gamma ~:~ \mathring{\mathcal{A}}^\hbar_\Gamma \to \mathcal{A}^\hbar_\Gamma, \qquad \wh{X}_i \mapsto \wh{Z}_i^{(\epsilon)}, \quad q \mapsto (q_\Lambda)^{\epsilon}.
\end{align}
The two images $\mathcal{A}^{\hbar(\epsilon)}_\Gamma := \iota^{(\epsilon)}_\Gamma(\mathring{\mathcal{A}}^\hbar_\Gamma)$ are isomorphic to $\mathring{\mathcal{A}}^\hbar_\Gamma$ as $\mathbb{C}$-algebras. For $\Lambda = -1,1$, the algebra $\mathcal{A}^\hbar_\Gamma$ is in fact equal to the tensor product of these two algebras. For $\Lambda=0$, this tensor product forms a subalgebra of $\mathcal{A}^\hbar_\Gamma$, which we denote by $(\mathcal{A}^\hbar_\Gamma)_0$; one can also compose these maps with the $*$-maps, whose images generate another subalgebra which we may denote by $(\mathcal{A}^\hbar_\Gamma)_0^*$. Via the embedding maps $\iota^{(\epsilon)}_\Gamma$, algebraic statements about $\mathring{\mathcal{A}}^\hbar_\Gamma$ carry over to the two subalgebras $\mathcal{A}^{\hbar(+)}_\Gamma$ and $\mathcal{A}^{\hbar(-)}_\Gamma$ of $\mathcal{A}^\hbar_\Gamma$.

\vs

Now, for each pair of seeds $\Gamma,\Gamma' \in |\Gamma|$, we must associate the classical coordinate change map
$$
	\mu_{\Gamma,\Gamma'} ~:~ {\rm Frac}(\mathcal{A}_{\Gamma'}) \to {\rm Frac}(\mathcal{A}_\Gamma),
$$
and the quantum coordinate change map
$$
	\mu^\hbar_{\Gamma,\Gamma'} ~:~ {\rm Frac}(\mathcal{A}^\hbar_{\Gamma'}) \to {\rm Frac}(\mathcal{A}^\hbar_\Gamma).
$$
The classical coordinate change formulas are given as in \S\ref{subsec:geometric_structures_on_the_set_of_R_Lambda_positive_points}; these also coincide with the formulas for $\mathring{\mu}_{\Gamma,\Gamma'}$ in \S\ref{subsec:the_case_of_a_cluster_X-variety_at_R_positive_algebraic} on the generators, where $Z_i^{(\epsilon)},Z_i' {}^{(\epsilon)}$ for each $\epsilon \in \{+,-\}$ play the role of $X_i$, $X_i'$. Accordingly, we define the quantum isomorphism $\mu^\hbar_{\Gamma,\Gamma'}$ as the $*$-isomorphism given by the same formulas as $\mathring{\mu}^\hbar_{\Gamma,\Gamma'}$ in \S\ref{subsec:the_case_of_a_cluster_X-variety_at_R_positive_algebraic} on the generators, for each $\epsilon \in \{+,-\}$, with $\wh{X}_i$ and $\wh{X}_i'$ replaced by $\wh{Z}_i^{(\epsilon)}$ and $\wh{Z}_i' {}^{(\epsilon)}$ respectively and the parameter $q$ replaced by $q_\Lambda^{\epsilon}$. So, in case $\Gamma' = \mu_k(\Gamma)$, we have
\begin{align}
	\label{eq:quantum_isomorphism}
	\mu^\hbar_{\Gamma,\Gamma'} = \mu^\sharp_{\Gamma,\Gamma'} \circ \mu'_{\Gamma,\Gamma'} ~:~ {\rm Frac}(\mathcal{A}^\hbar_{\Gamma'}) ~ \overset{\mu'_{\Gamma,\Gamma'}}{\longrightarrow} ~ {\rm Frac}(\mathcal{A}^\hbar_\Gamma) ~ \overset{\mu^\sharp_{\Gamma,\Gamma'}}{\longrightarrow} ~ {\rm Frac}(\mathcal{A}^\hbar_\Gamma),
\end{align}
where
\begin{align}
	\nonumber
	\mu'_{\Gamma,\Gamma'} (\wh{Z}_i'{}^{(\epsilon)})   & = \left\{
	\begin{array}{ll}
		(\wh{Z}_k^{(\epsilon)})^{-1}                                                                                                        & \mbox{if $i =k$},    \\
		q_\Lambda^{- \epsilon\, \varepsilon_{ik} [\varepsilon_{ik}]_+} \wh{Z}_i^{(\epsilon)} (\wh{Z}_k^{(\epsilon)})^{[\varepsilon_{ik}]_+} & \mbox{if $i\neq k$},
	\end{array}
	\right.                                                                                                                                                                                                                                 \\
	\label{eq:mu_hbar_sharp_k_Lambda}
	\mu^\sharp_{\Gamma,\Gamma'}(\wh{Z}_i^{(\epsilon)}) & = \wh{Z}_i^{(\epsilon)} \prod_{r=1}^{|\varepsilon_{ik}|} (1 + (q_\Lambda^{- \epsilon\, {\rm sgn}(\varepsilon_{ik})})^{2r-1} \wh{Z}_k^{(\epsilon)})^{-{\rm sgn}(\varepsilon_{ik})},
\end{align}
In case $\Gamma' = P_\sigma(\Gamma)$, we have $\mu^\hbar_{\Gamma,\Gamma'}(\wh{Z}_{\sigma(i)}' {}^{\hspace{-4mm}(\epsilon)}) = \wh{Z}_i^{(\epsilon)}$. For a general pair of seeds $\Gamma,\Gamma' \in |\Gamma|$, we define the quantum coordinate change maps $\mu^\hbar_{\Gamma,\Gamma'}$ as the composition of the above two elementary kinds. The sequence of mutations $\mu_k$ and seed automorphisms $P_\sigma$ connecting $\Gamma$ to $\Gamma'$ is not unique, so one has to check whether the resulting isomorphism $\mu^\hbar_{\Gamma,\Gamma'}$ does not depend on the choice of such a sequence. This has been shown in the usual cluster $\mathscr{X}$-variety case for the  quantum isomorphisms $\mathring{\mu}^\hbar_{\Gamma,\Gamma'}$ \cite{FG09a,BZ05}; the corresponding statement for $\mu^\hbar_{\Gamma,\Gamma'}$ follows by applying the algebra embeddings of eq.\eqref{eq:two_algebra_embeddings}. Thus we obtain:
\begin{proposition}
	The quantum coordinate change maps $\mu^\hbar_{\Gamma,\Gamma'}$ constructed above satisfy:
	\begin{enumerate}
		\item[\rm (QML1)] $\mu^\hbar_{\Gamma,\Gamma'}$ recovers $\mu_{\Gamma,\Gamma'}$ as $q_\Lambda \to 1$;

		\item[\rm (QML2)] $\mu^\hbar_{\Gamma,\Gamma'} \circ \mu^\hbar_{\Gamma',\Gamma''}= \mu^\hbar_{\Gamma,\Gamma''}$ holds for each triple of seeds $\Gamma,\Gamma',\Gamma'' \in |\Gamma|$;

		\item[\rm (QML3)] $\mu^\hbar_{\Gamma,\Gamma'}$ depends only on the underlying exchange matrices $\varepsilon,\varepsilon'$ of the seeds $\Gamma,\Gamma'$.
	\end{enumerate}

\end{proposition}

We now define the classical and quantum universally Laurent algebras as before:
\begin{align}
	\nonumber
	\mathbb{L}_\Gamma       & := \bigcap_{\Gamma' \in |\Gamma|} \mu_{\Gamma,\Gamma'}(\mathcal{A}_{\Gamma'}) ~\subset~ \mathcal{A}_\Gamma ~ \subset ~ {\rm Frac}(\mathcal{A}_\Gamma),                          \\
	\nonumber
	\mathbb{L}^\hbar_\Gamma & := \bigcap_{\Gamma' \in |\Gamma|} \mu^\hbar_{\Gamma,\Gamma' }(\mathcal{A}^\hbar_{\Gamma'}) ~\subset~ \mathcal{A}^\hbar_\Gamma ~ \subset ~ {\rm Frac}(\mathcal{A}^\hbar_\Gamma).
\end{align}
For $\Lambda=0$, we may also define the subalgebra versions:
$$
	(\mathbb{L}_\Gamma)_0 := \bigcap_{\Gamma' \in |\Gamma|} \mu_{\Gamma,\Gamma'}( (\mathcal{A}_{\Gamma'})_0 ), \qquad
	(\mathbb{L}^\hbar_\Gamma)_0 := \bigcap_{\Gamma' \in |\Gamma|} \mu^\hbar_{\Gamma,\Gamma'}( (\mathcal{A}^\hbar_{\Gamma'})_0).
$$
The first goal is to find a deformation quantization map
$$
	\wh{Q}_\Gamma^\hbar ~:~ \mathbb{L}_\Gamma \to \mathbb{L}^\hbar_\Gamma,
$$
for each $\Gamma$, that is compatible with $\mu_{\Gamma,\Gamma'}$ and $\mu^\hbar_{\Gamma,\Gamma'}$, in the sense that
$$
	\wh{Q}^\hbar_\Gamma(\mu_{\Gamma,\Gamma'}(u)) = \mu^\hbar_{\Gamma,\Gamma'}(\wh{Q}^\hbar_{\Gamma'}(u)), \qquad \forall u \in \mathbb{L}_{\Gamma'},\quad \forall \Gamma,\Gamma'.
$$
By identifying all $\mathbb{L}_\Gamma$ for different $\Gamma \in |\Gamma|$ to denote it by $\mathscr{O}_\Lambda(\mathscr{X}_{|\Gamma|})$, and likewise identifying all $\mathbb{L}^\hbar_\Gamma$ to denote it by $\mathscr{O}^\hbar_\Lambda(\mathscr{X}_{|\Gamma|})$, one could package the above deformation quantization maps as
$$
	\wh{Q}^\hbar_{|\Gamma|} ~:~ \mathscr{O}_\Lambda(\mathscr{X}_{|\Gamma|}) \to \mathscr{O}_\Lambda^\hbar(\mathscr{X}_{|\Gamma|}).
$$
We will see in detail in \S\ref{subsec:algebraic_deformation_quantization_through_duality_maps} how we obtained such a map. One thing to remark is that, by formulating the quantization problem of the $\mathbb{R}_\Lambda$-variety $\mathscr{X}_{|\Gamma|}(\mathbb{R}_\Lambda^+)$ this way, we are declaring what the ring of classical observables to be quantized should be. We chose it to be the ring of regular, or universally Laurent, functions, i.e. the functions that, for each seed $\Gamma \in |\Gamma|$, can be written as a Laurent polynomial in $\big\{Z_i^{(\epsilon)} : i \in I, \epsilon \in \{+,-\}\big\}$. That is, we decided to work under the algebro-geometric formulation in terms of the generators $Z_i^{(\epsilon)}$. However, going back to the initial setting, note that $Z_i^{(\epsilon)}$ is an $\mathbb{R}_\Lambda$-valued function, so it can't be viewed as a usual real-valued classical observable in general. An honest quantization should tell us how to quantize some class of real-valued classical observables. So, what would be such a class of real-valued functions? One could for example take the $\Lambda$-real and the $\Lambda$-imaginary parts of the $\mathbb{R}_\Lambda$-valued functions we quantized; the resulting quantization formula written in terms of these real-valued functions might be complicated, especially when $\Lambda=0$, which are probably difficult to obtain or even to guess without resorting to the $\mathbb{R}_\Lambda$-valued functions.

\vs

What remains in the process of quantization, which in fact constitutes the heart of the present paper, is the operator aspect. That is, for each classical observable $u \in \mathscr{O}_\Lambda(\mathscr{X}_{|\Gamma|})$ to be quantized, we should study how the quantum element $\wh{Q}^\hbar_{|\Gamma|}(u) \in \mathscr{O}^\hbar_\Lambda(\mathscr{X}_{|\Gamma|})$ would be represented as an operator on a Hilbert space. As before, a basic step is to study representations of the algebras $\mathbb{L}^\hbar_\Gamma$, or of $\mathcal{A}^\hbar_\Gamma$, on a Hilbert space $\mathscr{H}_\Gamma$, for each seed $\Gamma$. Here arises a crucial difference from the previous case for the usual cluster $\mathscr{X}$-variety, because of the element $\ell$, or $q_\Lambda = e^{\pi {\rm i} \ell\hbar}$; this will be dealt with in the next section in detail. Likewise as before, first we consider a representation
$$
	\til{\pi}_\Gamma ~:~ \mathbb{L}^\hbar_\Gamma \to {\rm End}(\mathscr{D}_\Gamma)
$$
on a nice dense subspace $\mathscr{D}_\Gamma$ of $\mathscr{H}_\Gamma$, then define the Schwartz space as
$$
	\mathscr{S}_\Gamma := \bigcap_{u \in \mathbb{L}^\hbar_\Gamma}{\rm Dom}(\til{\pi}_\Gamma(u)^*)
$$
where $\til{\pi}_\Gamma(u)^*$ means the operator adjoint of $\til{\pi}_\Gamma(u)$. Then one naturally obtains a representation $\pi_\Gamma$ of $\mathbb{L}^\hbar_\Gamma$ on the Schwartz space $\mathscr{S}_\Gamma$. Next, in order to ensure that these representations $\pi_\Gamma$ for different seeds $\Gamma$ are compatible with each other, for each pair of seeds $\Gamma,\Gamma' \in |\Gamma|$, one seeks to construct an intertwining operator for the quantum coordinate change map $\mu^\hbar_{\Gamma,\Gamma'}$
$$
	{\bf K}_{\Gamma,\Gamma'} = {\bf K}^\hbar_{\Gamma,\Gamma'} ~:~ \mathscr{H}_{\Gamma'} \to \mathscr{H}_\Gamma
$$
which is preferably unitary, such that
\begin{enumerate}
	\item[\rm (ITL1)] ${\bf K}_{\Gamma,\Gamma'}(\mathscr{S}_{\Gamma'}) = \mathscr{S}_\Gamma$;

	\item[\rm (ITL2)] the following intertwining equations are satisfied for all $u \in \mathbb{L}^\hbar_{\Gamma'}$
	      $$
		      {\bf K}_{\Gamma,\Gamma'} \circ \pi_{\Gamma'}(u) = \pi_\Gamma (\mu^\hbar_{\Gamma,\Gamma'}(u)) \circ {\bf K}_{\Gamma,\Gamma'};
	      $$

	\item[\rm (ITL3)] for each triple of seeds $\Gamma,\Gamma',\Gamma'' \in |\Gamma|$, the consistency equations
	      $$
		      {\bf K}_{\Gamma,\Gamma'} \circ {\bf K}_{\Gamma',\Gamma''} = c_{T,T',T''}{\bf K}_{\Gamma,\Gamma''}
	      $$
	      hold up to multiplicative constants;

	\item[\rm (ITL4)] ${\bf K}_{\Gamma,\Gamma'}$ depends only on the underlying exchange matrices $\varepsilon,\varepsilon'$ of the seeds $\Gamma,\Gamma'$.
\end{enumerate}

\subsection{A summary of the quantization problem} We first summarize the situation for the quantization problem for the usual cluster $\mathscr{X}$-variety $\mathscr{X}_{|\Gamma|}(\mathbb{R}^+)$ at $\mathbb{R}^+$.
\begin{enumerate}
	\item[\rm (QPu1)] For each seed $\Gamma\in|\Gamma|$, construct a $*$-algebra $\mathring{\mathcal{A}}^\hbar_\Gamma$ over $\mathbb{C}[q^{\pm 1}]$ that deforms the classical coordinate algebra $\mathring{\mathcal{A}}_\Gamma$, which is a Laurent polynomial algebra;

	\item[\rm (QPu2)] For each pair of seeds $\Gamma,\Gamma' \in |\Gamma|$, construct a quantum coordinate change isomorphism $\mathring{\mu}^\hbar_{\Gamma,\Gamma'} : {\rm Frac}(\mathring{\mathcal{A}}^\hbar_{\Gamma'}) \to {\rm Frac}(\mathring{\mathcal{A}}^\hbar_\Gamma)$ satisfying (QM1)--(QM3);

	\item[\rm (QPu3)] For each seed $\Gamma \in |\Gamma|$, construct a deformation quantization map $\mathring{\wh{Q}} {}^\hbar_\Gamma : \mathring{\mathbb{L}}_\Gamma \to \mathring{\mathbb{L}}^\hbar_\Gamma$ that is compatible with the isomorphisms $\mathring{\mu}_{\Gamma,\Gamma'}$ and $\mathring{\mu}^\hbar_{\Gamma,\Gamma'}$, in the sense that $\mathring{\wh{Q}} {}^\hbar_\Gamma \circ \mathring{\mu}_{\Gamma,\Gamma'}= \mathring{\mu}^\hbar_{\Gamma,\Gamma'} \circ \mathring{\wh{Q}} {}^\hbar_{\Gamma'}$;

	\item[\rm (QPu4)] For each seed $\Gamma \in |\Gamma|$, construct a $*$-representation of the algebra $\mathring{\mathcal{A}}^\hbar_\Gamma$ on a dense subspace $\mathring{\mathscr{D}}_\Gamma$ of a Hilbert space $\mathring{\mathscr{H}}_\Gamma$ so that each generator of $\mathring{\mathcal{A}}^\hbar_\Gamma$ is represented by (essentially) self-adjoint positive operator, and a $*$-representation of the universally Laurent algebra $\mathring{\mathbb{L}}^\hbar_\Gamma \subset \mathring{\mathcal{A}}^\hbar_\Gamma$ on a Schwartz subspace $\mathring{\mathscr{S}}_\Gamma$ of $\mathring{\mathscr{H}}_\Gamma$;

	\item[\rm (QPu5)] For each pair of seeds $\Gamma,\Gamma' \in |\Gamma|$, construct a unitary intertwining operator $\mathring{\bf K}_{\Gamma,\Gamma'} : \mathscr{H}_{\Gamma'} \to \mathscr{H}_\Gamma$ (representing the transformation of seeds $\Gamma \leadsto \Gamma'$ such as mutations and seed automorphisms) satisfying (IT1)--(IT4).

\end{enumerate}
Note that (QPu1)--(QPu2) constitutes the problem of constructing a quantum cluster $\mathscr{X}$-variety, (QPu3) that of constructing a deformation quantization map, and (QPu4)--(QPu5) that of constructing a representation of the constructed quantum cluster $\mathscr{X}$-variety that is equivariant under the cluster mapping class group. The parts (QPu1)--(QPu2) are (QPu4)--(QPu5) are resolved in \cite{FG09a}, based on earlier works including \cite{BZ05,FC99,Kas98,Gon08}; we recalled this solution to (QPu1)--(QPu2) in \S\ref{subsec:the_case_of_a_cluster_X-variety_at_R_positive_algebraic}, and the solution to (QPu4)--(QPu5) will be reviewed in the next section. The part (QPu3) can be solved, as mentioned in \S\ref{subsec:the_case_of_a_cluster_X-variety_at_R_positive_algebraic}, using the results in the literature on classical and quantum duality maps; see \S\ref{subsec:algebraic_deformation_quantization_through_duality_maps} for details. So, one can say that the above deformation quantization problem (QPu1)--(QPu5) for the cluster $\mathscr{X}$-variety $\mathscr{X}_{|\Gamma|}(\mathbb{R}^+)$ at $\mathbb{R}^+$ has been completely solved more or less in the literature.

\vs

Now we formulate the quantization problem for the cluster variety $\mathscr{X}_{|\Gamma|}(\mathbb{R}_\Lambda^+)$ at $\mathbb{R}_\Lambda^+$, which we have been vaguely referring to as a (cluster) $\mathbb{R}_\Lambda$-variety. Keep in mind the dependence on $\Lambda \in \{-1,1,0\}$.
\begin{enumerate}
	\item[\rm (QPL1)] For each seed $\Gamma\in|\Gamma|$, construct a $\mathbb{C}[q^{\pm 1}_\Lambda]$-algebra $\mathcal{A}^\hbar_\Gamma$ with a suitable $*$-structure that deforms the classical coordinate algebra $\mathcal{A}_\Gamma$, which is a Laurent polynomial algebra;

	\item[\rm (QPL2)] For each pair of seeds $\Gamma,\Gamma \in |\Gamma|$, construct a quantum coordinate change isomorphism $\mu^\hbar_{\Gamma,\Gamma'} : {\rm Frac}(\mathcal{A}^\hbar_{\Gamma'}) \to {\rm Frac}(\mathcal{A}^\hbar_\Gamma)$ satisfying (QML1)--(QML3);

	\item[\rm (QPL3)] For each seed $\Gamma \in |\Gamma|$, construct a deformation quantization map $\wh{Q}^\hbar_\Gamma : \mathbb{L}_\Gamma \to \mathbb{L}^\hbar_\Gamma$ that is compatible with the isomorphisms $\mu_{\Gamma,\Gamma'}$ and $\mu^\hbar_{\Gamma,\Gamma'}$, in the sense that $\wh{Q}^\hbar_\Gamma \circ \mu_{\Gamma,\Gamma'}= \mu^\hbar_{\Gamma,\Gamma'} \circ \wh{Q}^\hbar_{\Gamma'}$;

	\item[\rm (QPL4)] For each seed $\Gamma \in |\Gamma|$, construct a $*$-representation of the universally Laurent algebra $\mathbb{L}^\hbar_\Gamma \subset \mathcal{A}^\hbar_\Gamma$ on a Schwartz subspace $\mathscr{S}_\Gamma$ of a Hilbert space $\mathscr{H}_\Gamma$;

	\item[\rm (QPL5)] For each pair of seeds $\Gamma,\Gamma' \in |\Gamma|$, construct an intertwining operator ${\bf K}_{\Gamma,\Gamma' } : \mathscr{H}_{\Gamma'} \to \mathscr{H}_\Gamma$ satisfying (ITL1)--(ITL4).

\end{enumerate}
A solution to this quantization problem can be viewed as the principal result of the present paper, which will be described in detail in section \S\ref{sec:main}.

\begin{theorem}[main result]
	There exists a solution to the above quantization problem (QPL1)--(QPL5) for the cluster variety $\mathscr{X}_{|\Gamma|}(\mathbb{R}_\Lambda^+)$ at $\mathbb{R}_\Lambda^+$.
\end{theorem}

As mentioned, the above formulation of the problem is in fact suggesting which classical algebra to quantize. By the very formulation, one can apply the algebra maps in eq.\eqref{eq:two_algebra_embeddings} to the solutions for (QPu1)--(QPu3) to obtain solutions for (QPL1)--(QPL3). In a sense, we formulated the algebraic part of the quantization problem in such a way that the solution for the usual cluster $\mathscr{X}$-variety at $\mathbb{R}^+$ carries over. The more non-trivial part consists of the remaining (QPL4)--(QPL5), i.e. the operator aspect, and we present our solution in section \S\ref{sec:main}.

\section{Fock-Goncharov representations of quantum $\mathscr{X}$-varieties at $\mathbb{R}^+$}
\label{sec:FG_representations}

In this section we review in more detail previously known results in the literature which constitute a solution to the parts (QPu4)--(QPu5) of the quantization problem of the cluster variety $\mathscr{X}_{|\Gamma|}(\mathbb{R}^+)$ at $\mathbb{R}^+$. As mentioned, these parts can be viewed as the problem of constructing a representation of a quantum cluster $\mathscr{X}$-variety, and what we will recall in this section is the solution obtained by Fock and Goncharov \cite{FG09a}, which generalize the earlier works \cite{FC99,Kas98}. The contents of this section is not merely for a review, but will be crucially used in our solution to (QPL4)--(QPL5) in the next section.

\subsection{A positive representation for each seed}
\label{subsec:a_positive_representation_for_each_seed}

We deal with (QPu4) first, i.e. for each seed $\Gamma \in |\Gamma|$, we must construct a $*$-representation of the algebra $\mathring{\mathcal{A}}^\hbar_\Gamma$ and the universally Laurent subalgebra $\mathring{\mathbb{L}}^\hbar_\Gamma$ on a Hilbert space $\mathring{\mathscr{H}}_\Gamma$, or more precisely on the corresponding Schwartz subspace $\mathring{\mathscr{S}}_\Gamma$.

\vs
Consider the Hilbert space
\begin{align}
	\label{eq:H_Gamma}
	\mathring{\mathscr{H}}_\Gamma := L^2(\mathbb{R}^I, \underset{i \in I}{\wedge} dt_i).
\end{align}
Mimicking and somewhat generalizing \cite{FG09a}, we begin with the following nice dense subspace $\mathring{\mathscr{D}}_\Gamma$ of $\mathring{\mathscr{H}}_\Gamma$, see e.g. in \cite[eq.(3.4)]{Kim21a},
\begin{align}
	\label{eq:D_Gamma}
	\mathring{\mathscr{D}}_\Gamma := {\rm span}_\mathbb{C} \left\{ e^{\langle t, M\cdot t + v\rangle } \cdot P(t) \left| \begin{array}{lll} \mbox{$M\in {\rm Mat}_{I\times I}(\mathbb{C})$ with negative-definite real part}, \\ \mbox{$ v = (v_i)_{i\in I} \in \mathbb{C}^I$,} \\ \mbox{$P(t)$ a polynomial in $t_i$'s, $i\in I$, over $\mathbb{C}$} \end{array} \right. \right\}
\end{align}
where $t = (t_i)_{i\in I} \in \mathbb{R}^I$ denote the variables in $\mathbb R^I$ and $\langle t, v\rangle = \sum_{i\in I} t_iv_i$.
As a basic tool we will use the standard position and momentum operators $\{ t_i, {\rm i} \frac{\partial}{\partial t_i} : i \in I\}$; any $\mathbb{R}$-linear combination of these makes sense as a symmetric linear operator $\mathring{\mathscr{D}}_\Gamma \to \mathring{\mathscr{D}}_\Gamma$, acting on the elements of $\mathring{\mathscr{D}}_\Gamma$ just as the notation suggests (i.e. by multiplying and by differentiating), and it is well known that it is essentially self-adjoint on $\mathring{\mathscr{D}}_\Gamma$; see e.g. \cite{Hal13}. We define
\begin{align}
	\label{eq:x_and_y_operators}
	{\bf x}_i := -\pi {\rm i} \frac{\partial}{\partial t_i}, \qquad
	{\bf y}_i := \sum_{j \in I}   \varepsilon_{ij} t_j, \qquad \forall i \in I
\end{align}
which are examples of such operators. By a slight abuse of notation, by the symbols ${\bf x}_i$ and ${\bf y}_i$ we mean the unique self-adjoint operators extending the operators defined on $\mathring{\mathscr{D}}_\Gamma$ by the above formulas. One can observe that they satisfy the Heisenberg commutation relations
$$
	[{\bf x}_i, {\bf x}_j] = 0, \qquad
	[{\bf y}_i, {\bf y}_j] = 0, \qquad
	[{\bf x}_i, {\bf y}_j] = \pi {\rm i} \, \varepsilon_{ij} \cdot {\rm Id}, \qquad \forall i,j \in I,
$$
e.g. as operators $\mathring{\mathscr{D}}_\Gamma\to \mathring{\mathscr{D}}_\Gamma$. Moreover, the {\it Weyl-relation-version} of these relations hold, where, for self-adjoint operators $A,B$ on a Hilbert space, the Weyl-relation-version of the Heisenberg commutation relation $[A,B] = {\rm i} c \cdot {\rm Id}$ (for some $c\in \mathbb{R}$) refers to the family of equalities
$$
	e^{{\rm i} \alpha A} e^{{\rm i} \beta B} = e^{-\alpha \beta {\rm i} c} e^{{\rm i} \beta B} e^{{\rm i} \alpha A}, \quad \forall \alpha,\beta\in \mathbb{R},
$$
of unitary operators $e^{{\rm i} \alpha A}$ and $e^{{\rm i} \beta B}$, which are defined by the functional calculus of $A$ and $B$ (see e.g. \cite{Hal13}). The operators ${\bf x}_i$ and ${\bf y}_i$ can be thought of as quantum versions of the functions $x_i$ and $y_i$ appearing in eq.\eqref{eq:x_i_and_y_i_Poisson}; this viewpoint will become more relevant in the next section. In fact, in view of eq.\eqref{eq:x_and_y_operators}, these operators can be seen as a representation of Fock-Goncharov's symplectic double cluster $\mathscr{D}$-variety \cite{FG09a}. The relation between our constructions and the cluster $\mathscr{D}$-varieties is left for future research.

\vs

We define the linear quantum coordinate operators (however confusing the notation might be)
\begin{align}
	\label{eq:x_hbar}
	\mathring{\bf x}_i := {\bf x}_i + \hbar {\bf y}_i, \qquad \forall i \in I,
\end{align}
so that they satisfy the Weyl-relation-version of the Heisenberg commutation relations
$$
	[\mathring{\bf x}_i, \mathring{\bf x}_j] = 2\pi {\rm i} \hbar\, \varepsilon_{ij} \cdot {\rm Id}, \qquad \forall i,j \in I.
$$
We also define
$$
	\mathring{\til{\pi}}_\Gamma(\wh{X}_i) := \exp(\mathring{\bf x}_i)\restriction \mathring{\mathscr{D}}_\Gamma ~:~ \mathring{\mathscr{D}}_\Gamma \to \mathring{\mathscr{D}}_\Gamma,
$$
so that $\mathring{\til{\pi}}_\Gamma(\wh{X}_i) \mathring{\til{\pi}}_\Gamma(\wh{X}_j) = e^{2\pi {\rm i} \hbar \varepsilon_{ij}} \mathring{\til{\pi}}_\Gamma(\wh{X}_j) \mathring{\til{\pi}}_\Gamma(\wh{X}_i)$ holds on $\mathring{\mathscr{D}}_\Gamma$. Then there is a unique algebra homomorphism $\mathring{\til{\pi}}_\Gamma : \mathring{\mathcal{A}}^\hbar_\Gamma \to {\rm End}(\mathring{\mathscr{D}}_\Gamma)$ whose values at the generators are given as above, and one can define the Schwartz space $\mathring{\mathscr{S}}_\Gamma \subset \mathring{\mathscr{H}}_\Gamma$ and the representation $\mathring{\pi}_\Gamma$ of $\mathring{\mathbb{L}}^\hbar_\Gamma$ on $\mathring{\mathscr{S}}_\Gamma$ as described in \S\ref{subsec:the_case_of_a_cluster_X-variety_at_R_positive_operator}.
\begin{proposition}[\cite{FG09a}]
	\label{prop:D_dense}
	$\mathring{\mathscr{D}}_\Gamma$ is dense in $\mathring{\mathscr{S}}_\Gamma$ in the Frech\'et topology described in \S\ref{subsec:the_case_of_a_cluster_X-variety_at_R_positive_operator}.
\end{proposition}

This completes (QPu4). We note that the main point of (QPu4) is to make sure that one chooses a suitable representation $\pi_\Gamma$ for each seed $\Gamma$ so that the more important intertwiner problem (QPu5) can be solved. In a sense, (QPu4) and (QPu5) should be thought of as a single problem tied together.

\vs

We note that what Fock and Goncharov quantized in \cite{FG09a} is not just one copy of cluster $\mathscr{X}$-variety $\mathscr{X}_{|\Gamma|}(\mathbb{R}^+)$ at $\mathbb{R}^+$, but the symplectic double cluster $\mathscr{D}$-variety at $\mathbb{R}^+$, which `contains' two opposite copies of the cluster $\mathscr{X}$-variety. The linear quantum operators for the opposite copy are given by
$$
	\mathring{\til{{\bf x}}}_i := {\bf x}_i  - \hbar {\bf y}_i,
$$
which satisfy $[\mathring{\til{\bf x}}_i, \mathring{\til{\bf x}}_j] = - 2\pi {\rm i} \hbar \, \varepsilon_{ij} \cdot {\rm Id}$ and $[\mathring{\bf x}_i, \mathring{\til{\bf x}}_j]=0$. We will not review this symplectic double story in the present paper, but will have to use these opposite operators when describing Fock-Goncharov's results.

\subsection{The non-compact quantum dilogarithm function}
\label{subsec:Phi_hbar}

What plays a central role in the solution to (QPu5) of \cite{FG09a,FC99,Kas98} is the special function called the {\it quantum dilogarithm}, which was studied in a modern form by Faddeev and Kashaev \cite{FK94}, but which in fact had appeared in different guises in previous works, going back to Barnes 100 years ago \cite{Bar01}. The simplest version of the quantum dilogarithm is the following function
$$
	\psi^{\bf q} (z) = \prod_{n=1}^\infty (1+{\bf q}^{2n-1}z)^{-1}
$$
defined on the complex plane, where ${\bf q}$ is a nonzero complex parameter. When $|{\bf q}|<1$, this is a well-defined meromorphic function, sometimes called the {\it compact} quantum dilogarithm, and one of its characteristic property is the difference equation
$$
	\psi^{\bf q}({\bf q}^2 z)= (1+{\bf q}z) \psi^{\bf q}(z),
$$
which resembles the defining equation of the classical Gamma function. One sometimes view ${\bf q}$ as a formal symbol, and understand the above as formal power series in ${\bf q}$ (and $z$). Taking advantage of the above difference equation, the automorphism part $\mathring{\mu}^\sharp_{\Gamma,\Gamma'}$ of the quantum mutation map for the quantum cluster $\mathscr{X}$-variety which is defined by the somewhat enigmatic-looking formula in eq.\eqref{eq:mu_hbar_sharp_k} can be understood as (formal) conjugation by $\psi^{\bf q}(\wh{X}_k)$.

\vs

Meanwhile, when $U,V$ are elements of an algebra or operators satisfying $UV = {\bf q}^2 VU$, it is known that the pentagon equation
\begin{align}
	\label{eq:psi_q_pentagon}
	\psi^{\bf q}(U) \psi^{\bf q}(V) = \psi^{\bf q}(V) \psi^{\bf q}({\bf q} VU) \psi^{\bf q}(U)
\end{align}
holds. We will be dealing with some other versions of this pentagon equation in a rigorous way.

\vs

For the known construction of the quantum cluster $\mathscr{X}$-variety \cite{FG09a}, already the compact quantum dilogarithm $\psi^{\bf q}$ is useful to understand the algebraic quantum coordinate change maps as mentioned above, at least heuristically. However, in this setting, $\bf q$ must equal $q = e^{\pi {\rm i} \hbar}$ where $\hbar \in \mathbb{R}$, so that $|{\bf q}|=1$; in particular $\psi^{\bf q}$ is not well-defined as it stands. What is actually used in \cite{FG09a}, especially for the intertwining operators representing the quantum mutations, is the so-called {\it non-compact} quantum dilogarithm function $\Phi^\hbar(z)$ defined by the contour integral formula
$$
	\Phi^\hbar(z) = \exp\left( - \frac{1}{4} \int_\Omega \frac{e^{-{\rm i} p z}}{\sinh(\pi p) \sinh(\pi \hbar p)} \frac{dp}{p} \right).
$$
Here, $z$ is a complex number living in the strip $|\Im z|<\pi(1+\hbar)$ (we assume $\hbar>0$ for simplicity) and $\Omega$ is the contour in $\mathbb{C}$ following the real line and avoiding the origin along a small half circle from above. This integral formula can be found already in the work of Barnes \cite{Bar01}, hence we refer to it as the {\it Barnes integral}. It has been proved that this formula defines a non-vanishing holomorphic function on the strip. We recollect some useful properties of $\Phi^\hbar$.

\begin{proposition}[properties of the non-compact quantum dilogarithm; see e.g. \cite{FK94,FG09a} and references therein]
	\label{prop:properties_of_Phi_hbar}
	The function $\Phi^\hbar(z)$ on the strip $|{\rm Im}\,z| < \pi(1+\hbar)$ analytically continues to the meromorphic function $\Phi^\hbar(z)$ on the complex plane, which satisfies the following properties.
	\begin{itemize}
		\item[\rm (1)] The zeros and poles are at
		      \begin{align*}
			      \mbox{the set of zeros} & = \big\{(2n+1) \pi {\rm i} + (2m+1) \pi {\rm i} \hbar \, | \, n,m\in \mathbb{Z}_{\ge 0} \big\},  \\
			      \mbox{the set of poles} & = \big\{-(2n+1) \pi {\rm i} - (2m+1) \pi {\rm i} \hbar \, | \, n,m\in \mathbb{Z}_{\ge 0} \big\}.
		      \end{align*}
		      These zeros and poles are simple if and only if $\hbar \notin \mathbb{Q}$.

		\item[\rm (2)] (difference equations) Each of the functional relations
		      \begin{align*}
			      \left\{
			      \begin{array}{lcr}
				      \Phi^\hbar(z+2\pi {\rm i} \hbar) & = & (1+e^{\pi {\rm i} \hbar} \, e^z) \, \Phi^\hbar(z),        \\
				      \Phi^\hbar(z+2\pi {\rm i})       & = & (1+e^{\pi {\rm i}/\hbar} \, e^{z/\hbar}) \, \Phi^\hbar(z)
			      \end{array}
			      \right.
		      \end{align*}
		      holds, whenever the arguments of $\Phi^h$ are not poles.

		\item[\rm (3)] (involutivity) One has
		      $$
			      \Phi^\hbar(z) \, \Phi^\hbar(-z) = c_\hbar \, \exp\left( {z^2}/{(4\pi {\rm i} \hbar)} \right),
		      $$
		      whenever $z$ and $-z$ are not poles of $\Phi^\hbar$, where
		      $
			      c_\hbar := e^{-\frac{\pi {\rm i} }{12} (\hbar+\hbar^{-1})} \in {\rm U}(1) \subset \mathbb{C}^\times.
		      $

		\item[\rm (4)] (unitarity) One has
		      $$
			      \overline{\Phi^\hbar(z)} = \Phi^\hbar(\overline{z})^{-1}
		      $$
		      whenever $z$ and $\overline{z}$ are not poles.

	\end{itemize}

\end{proposition}

The solution to the sought-for intertwining operator $\mathring{{\bf K}}_{\Gamma,\Gamma'}$, which will be described in the next subsection, involves the application of the functional calculus of some self-adjoint operator to this function $\Phi^\hbar$. When proving the desired properties of such an operator, the following operator identity becomes crucial.

\begin{proposition}[the pentagon identity for the non-compact quantum dilogarithm $\Phi^\hbar$; {\cite{Wor00,FKV01,Gon08}}]
	\label{prop:pentagon_Phi_hbar}
	If $\hbar \in \mathbb{R}^+$, and if ${\bf x},{\bf y}$ are self-adjoint operators on a separable Hilbert space that satisfy the Weyl-relation-version of the Heisenberg commutation relation $[{\bf x}, {\bf y}] = 2\pi {\rm i} \hbar \cdot {\rm Id}$, then the following holds as the equality of unitary operators:
	$$
		\Phi^\hbar({\bf x}) \, \Phi^\hbar({\bf y})
		= \Phi^\hbar({\bf y}) \, \Phi^\hbar({\bf x} + {\bf y}) \, \Phi^\hbar({\bf x}).
	$$
\end{proposition}

\subsection{Mutation intertwiner for quantum cluster $\mathscr{X}$-variety at $\mathbb{R}^+$}
\label{subsec:FG_mutation_intertwiner}

We now describe Fock-Goncharov's intertwiner $\mathring{\bf K}_{\Gamma,\Gamma'} : \mathring{\mathscr{H}}_{\Gamma'} \to \mathring{\mathscr{H}}_\Gamma$ \cite{FG09a} associated to each pair of seeds $\Gamma,\Gamma' \in |\Gamma|$. The most important case is $\Gamma' = \mu_k(\Gamma)$, i.e. $\Gamma'$ and $\Gamma$ are related by a single mutation $\mu_k$. In this case the intertwiner is given in terms of a decomposition
$$
	\xymatrix@R+5mm{
	\mathring{\bf K}_{\Gamma,\Gamma'} = \mathring{\bf K}^\sharp_{\Gamma,\Gamma'} \circ \mathring{\bf K}'_{\Gamma,\Gamma'} ~:~ \mathring{\mathscr{H}}_{\Gamma'} \ar[r]^-{\mathring{\bf K}'_{\Gamma,\Gamma'}}
	& \mathring{\mathscr{H}}_{\Gamma} \ar[r]^-{\mathring{\bf K}^\sharp_{\Gamma,\Gamma'}}
	& \mathring{\mathscr{H}}_{\Gamma},
	}
$$
where the two parts $\mathring{\bf K}^\sharp_{\Gamma,\Gamma'}$ and $\mathring{\bf K}'_{\Gamma,\Gamma'}$ must satisfy the intertwining equations for the automorphism part $\mathring{\mu}^\sharp_{\Gamma,\Gamma'}$ and the monomial transformation part $\mathring{\mu}'_{\Gamma,\Gamma'}$ defined in eq.\eqref{eq:mu_hbar_sharp_k} and eq.\eqref{eq:mu_hbar_prime_k}. That is, we require the following two diagrams to commute:
$$
	\xymatrix@C+14mm{
	\mathring{\mathscr{H}}_{\Gamma} \ar[d]_{\mathring{\bf K}^\sharp_{\Gamma,\Gamma'}} \ar[r]^-{\mathring{\pi}_{\Gamma}(u)} & \mathring{\mathscr{H}}_{\Gamma} \ar[d]^{\mathring{\bf K}^\sharp_{\Gamma,\Gamma'}} \\
	\mathring{\mathscr{H}}_{\Gamma} \ar[r]^-{\mathring{\pi}_{\Gamma}(\mathring{\mu}^\sharp_{\Gamma,\Gamma'}(u))} & \mathring{\mathscr{H}}_{\Gamma}
	} \qquad\qquad
	\xymatrix@C+14mm{
	\mathring{\mathscr{H}}_{\Gamma'} \ar[r]^-{\mathring{\pi}_{\Gamma'}(u')} \ar[d]_{\mathring{\bf K}'_{\Gamma,\Gamma'}} & \mathring{\mathscr{H}}_{\Gamma'} \ar[d]^{\mathring{\bf K}'_{\Gamma,\Gamma}} \\
	\mathring{\mathscr{H}}_{\Gamma} \ar[r]^-{\mathring{\pi}_{\Gamma} (\mathring{\mu}'_{\Gamma,\Gamma'}(u'))} & \mathring{\mathscr{H}}_{\Gamma}
	}
$$
for all $u \in \mathring{\mu}'_{\Gamma,\Gamma'}(\mathring{\mathbb{L}}^\hbar_{\Gamma'})$ and $u' \in \mathring{\mathbb{L}}^\hbar_{\Gamma'}$.  The representation operators for the four horizontal arrows such as $\mathring{\pi}_{\Gamma}(u)$ are usually not defined on the whole Hilbert space $\mathring{\mathscr{H}}_{\Gamma}$ or $\mathring{\mathscr{H}}_{\Gamma'}$, so these spaces must be replaced by the corresponding Schwartz subspaces $\mathring{\mathscr{S}}_{\Gamma}$ or $\mathring{\mathscr{S}}_{\Gamma'}$. That is, we require the following intertwining equations to hold:
\begin{align}
	\label{eq:K_sharp_intertwining_equations}
	( \mathring{\bf K}^\sharp_{\Gamma,\Gamma'} \, \mathring{\pi}_{\Gamma'}( u) ) \, \eta & = ( \mathring{\pi}_{\Gamma}(\mathring{\mu}^\sharp_{\Gamma,\Gamma'}( u)) \, \mathring{\bf K}^\sharp_{\Gamma,\Gamma'} ) \, \eta, \quad \mbox{$\forall u \in \mathring{\mu}'_{\Gamma,\Gamma'}(\mathring{\mathbb{L}}^\hbar_{\Gamma'}),\;  \forall \eta \in \mathring{\bf K}'_{\Gamma,\Gamma'}(\mathring{\mathscr{S}}_{\Gamma'}$),} \\
	\label{eq:K_prime_intertwining_equations}
	( \mathring{\bf K}'_{\Gamma,\Gamma'} \, \mathring{\pi}_{\Gamma'}( u') ) \, \eta'     & = ( \mathring{\pi}_{\Gamma}(\mathring{\mu}'_{\Gamma,\Gamma'}( u')) \, \mathring{\bf K}'_{\Gamma,\Gamma'} ) \, \eta', \quad \mbox{$\forall u' \in \mathring{\mathbb{L}}^\hbar_{\Gamma'},\; \forall \eta' \in \mathring{\mathscr{S}}_{\Gamma'}$.}
\end{align}
In \cite{FG09a}, the intertwining equations are shown only for the vectors living in the nice subspaces $\mathring{\mathscr{D}}_\Gamma$ and $\mathring{\mathscr{D}}_{\Gamma'}$, for it suffices to do so in view of Prop.\ref{prop:D_dense}.

\vs

We first deal with $\mathring{\bf K}'_{\Gamma,\Gamma'} : \mathring{\mathscr{H}}_{\Gamma'} \to \mathring{\mathscr{H}}_{\Gamma}$, which is the easier of the two parts; define
$$
	\mathring{\bf K}'_{\Gamma,\Gamma'} ~:~ \mathring{\mathscr{H}}_{\Gamma'} = L^2(\mathbb{R}^I, \underset{i \in I}{\wedge} dt_i') \longrightarrow L^2(\mathbb{R}^I, \underset{i \in I}{\wedge} dt_i) = \mathring{\mathscr{H}}_\Gamma
$$
as the natural map induced by the map between the Euclidean spaces $\chi_{\Gamma,\Gamma'} : \mathbb{R}^I \to \mathbb{R}^I$ (unprimed to primed) whose pullback sends each coordinate function as
$$
	\chi_{\Gamma,\Gamma'}^* t_i' =  \left\{
	\begin{array}{ll}
		t_i                                            & \mbox{if $i\neq k$,} \\
		-t_k + \sum_{j\in I} [-\varepsilon_{kj}]_+ t_j & \mbox{if $i=k$},
	\end{array}
	\right.
$$
where $\varepsilon$ denotes the underlying exchange matrix for $\Gamma$.
Operators on $L^2(\mathbb{R}^N)$ induced by linear maps on $\mathbb{R}^N$ are studied systematically in \cite{Kim21a}, precisely to the extent that we need here. In particular, the resulting operator $\mathring{\bf K}'_{\Gamma,\Gamma'}$ is unitary, and the conjugation actions of $\mathring{\bf K}'_{\Gamma,\Gamma'}$ on the position and momentum operators are computed.
\begin{lemma}[follows from {\cite[Lem.3.18]{Kim21a}}]
	When $\Gamma' = \mu_k(\Gamma)$, for each $i\in I$, one has
	\begin{align*}
		\mathring{\bf K}'_{\Gamma,\Gamma'} \, t_i \, (\mathring{\bf K}'_{\Gamma,\Gamma'})^{-1}                                                & = \left\{
		\begin{array}{ll}
			t_i                                            & \mbox{if $i\neq k$}, \\
			-t_k + \sum_{j\in I} [-\varepsilon_{kj}]_+ t_j & \mbox{if $i=k$},
		\end{array}
		\right.                                                                                                                                           \\
		\textstyle \mathring{\bf K}'_{\Gamma,\Gamma'} \, ({\rm i} \frac{\partial}{\partial t_i}) \, (\mathring{\bf K}'_{\Gamma,\Gamma'})^{-1} & = \left\{
		\begin{array}{ll}
			{\rm i} \frac{\partial}{\partial t_i} + [-\varepsilon_{ki}]_+ ({\rm i} \frac{\partial}{\partial t_k}) & \mbox{if $i\neq k$}, \\
			- {\rm i} \frac{\partial}{\partial t_k}                                                               & \mbox{if $i = k$.}
		\end{array}
		\right.
	\end{align*}
\end{lemma}
It is straightforward (although not completely obvious) to deduce the sought-for intertwining equations for $\mathring{\bf K}'_{\Gamma,\Gamma'}$, as hinted in \cite{FG09a} and checked in detail in \cite{Kim21a}. The first step is:
\begin{corollary}[{\cite[Lem.4.9]{Kim21a}}]
	\label{cor:conjugation_action_of_K_prime_on_x_and_y}
	When $\Gamma' = \mu_k(\Gamma)$, for each $i\in I$, one has
	\begin{align*}
		\mathring{\bf K}'_{\Gamma,\Gamma'} \, {\bf x}_i' \, (\mathring{\bf K}'_{\Gamma,\Gamma'})^{-1}            & = \left\{
		\begin{array}{ll}
			{\bf x}_i + [\varepsilon_{ik}]_+ \, {\bf x}_k & \mbox{if $i\neq k$}, \\
			- {\bf x}_k                                   & \mbox{if $i=k$},
		\end{array}
		\right.                                                                                                              \\
		\textstyle \mathring{\bf K}'_{\Gamma,\Gamma'} \, {\bf y}_i' \, (\mathring{\bf K}'_{\Gamma,\Gamma'})^{-1} & = \left\{
		\begin{array}{ll}
			{\bf y}_i + [\varepsilon_{ik}]_+ \, {\bf y}_k & \mbox{if $i\neq k$}, \\
			- {\bf y}_k                                   & \mbox{if $i = k$,}
		\end{array}
		\right.
	\end{align*}
	where the primed operators ${\bf x}'_i$ and ${\bf y}'_i$ are for $\Gamma'$, while the non-primed operators ${\bf x}_i$ and ${\bf y}_i$ are for $\Gamma$.
\end{corollary}
The conjugation action of $\mathring{\bf K}'_{\Gamma,\Gamma'}$ on the linear quantum coordinate operators in eq.\eqref{eq:x_hbar} is:
\begin{corollary}
	\label{cor:K_prime_intertwining_equations}
	When $\Gamma'=\mu_k(\Gamma)$, for each $i\in I$, one has
	$$
		\mathring{\bf K}'_{\Gamma,\Gamma'} \, \mathring{\bf x}_i' \, (\mathring{\bf K}'_{\Gamma,\Gamma'} )^{-1} = \left\{
		\begin{array}{ll}
			- \mathring{\bf x}_k                                             & \mbox{if $i=k$},     \\
			\mathring{\bf x}_i + [\varepsilon_{ik}]_+ \, \mathring{\bf x}_k, & \mbox{if $i\neq k$.}
		\end{array}
		\right.
	$$
	Exponentiating yields
	$
		\mathring{\bf K}'_{\Gamma,\Gamma'} \, \mathring{\pi}_{\Gamma'}( \wh{X}_i) \, (\mathring{\bf K}'_{\Gamma,\Gamma'} )^{-1} = \mathring{\pi}_{\Gamma}(\mathring{\mu}'_{\Gamma,\Gamma'}( \wh{X}_i) ).
	$
\end{corollary}

From this it follows:
\begin{proposition}[part of {\cite[Thm.5.6]{FG09a}}]
	When $\Gamma'=\mu_k(\Gamma)$, the above $\mathring{\bf K}'_{\Gamma,\Gamma'}$ satisfies the sought-for intertwining equations in eq.\eqref{eq:K_prime_intertwining_equations}.
\end{proposition}
The remaining automorphism part operator is constructed as
$$
	\mathring{\bf K}^\sharp_{\Gamma,\Gamma'} := \Phi^\hbar(\mathring{\bf x}_k) (\Phi^\hbar(\mathring{\til{\bf x}}_k))^{-1} ~:~ \mathscr{H}_\Gamma = L^2(\mathbb{R}^I, \underset{i \in I}{\wedge} dt_i) \to L^2(\mathbb{R}^I, \underset{i \in I}{\wedge} dt_i) = \mathscr{H}_\Gamma,
$$
given by applying the functional calculus of the self-adjoint quantum coordinate operators $\mathring{\bf x}_k$ and $\mathring{\til{\bf x}}_k$ (relevant to $\Gamma$) to the non-compact quantum dilogarithm function $\Phi^\hbar$. By Prop.\ref{prop:properties_of_Phi_hbar}(4), it follows that $\mathring{\bf K}^\sharp_{\Gamma,\Gamma'}$ is unitary. One may expect that the intertwining equations would hold, in view of the difference equations as in Prop.\ref{prop:properties_of_Phi_hbar}(2); in fact, we need an operator version of the difference equations, which requires some careful analytical arguments, which are established in \cite{FG09a}, resulting in the following statement.
\begin{proposition}[part of {\cite[Thm.5.6]{FG09a}}]
	When $\Gamma'=\mu_k(\Gamma)$, the above $\mathring{\bf K}^\sharp_{\Gamma,\Gamma'}$ satisfies the sought-for intertwining equations in eq.\eqref{eq:K_sharp_intertwining_equations}.
\end{proposition}
Thus indeed $\mathring{\bf K}_{\Gamma,\Gamma'} = \mathring{\bf K}^\sharp_{\Gamma,\Gamma'} \circ \mathring{\bf K}'_{\Gamma,\Gamma'}$ satisfies the intertwining equations, i.e. (IT2) of \S\ref{subsec:the_case_of_a_cluster_X-variety_at_R_positive_operator}; then one can deduce (IT1) too. The property (IT4) is obvious from the construction.

\vs

The other kind of an elementary pair of seeds $\Gamma,\Gamma'$ is when $\Gamma' = P_\sigma(\Gamma)$ for a permutation $\sigma$ of the index set $I$. In this case, the intertwiner $\mathring{\bf K}_{\Gamma,\Gamma'} : \mathring{\mathscr{H}}_{\Gamma'} = L^2(\mathbb{R}^I, \wedge_i dt_i') \to L^2(\mathbb{R}^I, \wedge_i dt_i) = \mathring{\mathscr{H}}_\Gamma$ is given in a straightforward manner; namely, the operator induced by the index permutation map $\mathbb{R}^I \to \mathbb{R}^I$ associated to $\sigma:I \to I$. Then it is very easy to see (IT1) and (IT2).

\vs

For a general pair of seeds $\Gamma,\Gamma'$, one would express $\Gamma'$ as the result of applying a sequence of mutations $\mu_k$ and seed automorphisms $P_\sigma$ to $\Gamma$; to each such elementary transformation of seeds is associated the above intertwiner, which we compose to construct $\mathring{\bf K}_{\Gamma,\Gamma'}$. More precisely, find a sequence of seeds $\Gamma = \Gamma^{(0)}$, $\Gamma^{(1)}$, \ldots, $\Gamma^{(n)} = \Gamma'$ so that each adjacent pair $\Gamma^{(r)}, \Gamma^{(r+1)}$ is either a mutation or a seed automorphism, then define
$$
	\mathring{\bf K}_{\Gamma,\Gamma'} := \mathring{\bf K}_{\Gamma^{(0)},\Gamma^{(1)}} \circ \mathring{\bf K}_{\Gamma^{(1)},\Gamma^{(2)}} \circ \cdots \circ \mathring{\bf K}_{\Gamma^{(n-1)},\Gamma^{(n)}}.
$$
In order for this to not depend on the choice of such a sequence and be well-defined at least up to multiplicative constants, and also for these operators to satisfy the consistency equations as in (IT3), it suffices to make sure that the intertwiners we constructed for the mutations and seed automorphisms satisfy all relations satisfied by mutations and seed automorphisms at the seed level.

\vs

The first step would thus be to classify all relations satisfied by $\mu_k$ and $P_\sigma$ at the seed level. This is a fundamental problem in the theory of cluster algebras and varieties, and is still not resolved. There are two possible remedies, both taken by \cite{FG09a}. One is to deal with only the known relations, and be content with having such a partial proof for the consistency. The other is to modify the representations $\pi_\Gamma$ for each seed $\Gamma$ so that they are strongly irreducible in a certain sense, and use the Schur's-Lemma-type argument to assert the uniqueness of intertwiners. Here we will take the first method at the moment, to check only the known relations as follows.
\begin{lemma}[known simple relations of mutations and seed automorphisms: seed level]
	\label{lem:known_simple_relations}
	One has:
	\begin{enumerate}
		\item[\rm (R1)] (twice-flip, or $A_1$ identity) One has $\mu_k \mu_k = {\rm Id}$ when applied to any seed $\Gamma$, i.e. $\mu_k(\mu_k(\Gamma)) = \Gamma$ for any seed $\Gamma$.

		\item[\rm (R2)] (quadrilateral, or $A_1\times A_1$ identity) One has $\mu_i \mu_j \mu_i \mu_j = {\rm Id}$ when applied to a seed $\Gamma$ with $\varepsilon_{ij}=0$.

		\item[\rm (R3)] (pentagon, or $A_2$ identity) One has $\mu_i \mu_j \mu_i \mu_j \mu_i P_{(ij)} = {\rm Id}$ when applied to a seed $\Gamma$ with $\varepsilon_{ij}=\pm 1$.

		\item[\rm (R4)] (permutation identity) One has $P_{(\sigma_1\sigma_2)^{-1}} P_{\sigma_1} P_{\sigma_2} = {\rm Id}$ and $P_{\rm Id} = {\rm Id}$ when applied to any seed.

		\item[\rm (R5)] (index permutation identity) One has $\mu_{\sigma(i)} P_\sigma \mu_i P_{\sigma^{-1}} = {\rm Id}$ when applied to any seed.
	\end{enumerate}
\end{lemma}
Fortunately, for the following main example of the initial seed for the present paper, the above relations are known to generate all relations.
\begin{proposition}[\cite{FST08,Lab09}]
	\label{prop:FST}
	If the initial seed $\Gamma$ comes from an ideal triangulation of a punctured surface $S$, then the relations in the Lem.\ref{lem:known_simple_relations} generate all relations satisfied by mutations and seed automorphisms among the seeds in $|\Gamma|$ that come from ideal triangulations of $S$.
\end{proposition}
We recall from \S\ref{subsec:generalized_shear_coordinates} that by an ideal triangulation we mean one that does not have any self-folded ideal triangle. Note that in a statement like Prop.\ref{prop:FST}, one must be a bit careful even for such a seed $\Gamma$ which comes from an ideal triangulation, for there are seeds in the equivalence class $|\Gamma|$ which do not come from ideal triangulations, and it is known that there exists a relation not generated by the above ones if we consider all seeds in $|\Gamma|$; see \cite{FST08,KY20}.

\begin{proposition}[simple consistency equations for the elementary intertwiners for known relations; {\cite[Thm.5.4]{FG09a},\cite{KN11},\cite{Kim21a},\cite{Kim21b}}]
	\label{prop:consistency_equations_for_elementary_intertwiners}
	For any seed-level relation of mutations and seed automorphisms appearing in Lem.\ref{lem:known_simple_relations}, the operator identity for the corresponding intertwiners $\mathring{\bf K}_{\Gamma,\Gamma'}$ constructed in the previous subsection holds up to multiplicative constants.
\end{proposition}

For example, when $\Gamma' = \mu_k(\Gamma)$, we have $\Gamma = \mu_k(\Gamma')$, and the following equality, for (R1), of unitary operators holds up to constant:
\begin{align}
	\nonumber
	\mathring{\bf K}_{\Gamma,\Gamma'} \circ \mathring{\bf K}_{\Gamma',\Gamma} = c_{\Gamma,\Gamma'}{\rm Id}.
\end{align}
A proof of this relation uses Prop.\ref{prop:properties_of_Phi_hbar}(3). A proof of the operator identity corresponding to the most important relation (R3) uses Prop.\ref{prop:pentagon_Phi_hbar}.

\vs

This settles the part (QPu5) of the quantization problem to a certain extent, which is especially satisfactory for the initial seed coming from a triangulable punctured surface, which in turn is related to the classical Teichm\"uller spaces. For the cluster $\mathscr{X}$-variety $\mathscr{X}_{|\Gamma|}(\mathbb{R}_\Lambda^+)$ at $\mathbb{R}_\Lambda^+$, such seeds are of our main interest, being related to the moduli spaces of 3d spacetimes.

\section{A quantization of the cluster $\mathscr{X}$-varieties at $\mathbb{R}^+_\Lambda$}
\label{sec:main}

In this section we present our solution to the quantization problem of the $\mathbb{R}_\Lambda$-variety $\mathscr{X}_{|\Gamma|}(\mathbb{R}_\Lambda^+)$ posed in the previous section. The parts (QPL1)--(QPL2) are already solved in \S\ref{sec:a_deformation_quantization_problem}, so we deal with the remaining parts (QPL3)--(QPL5).

\subsection{A deformation quantization through doubled duality maps}
\label{subsec:algebraic_deformation_quantization_through_duality_maps}

As promised in \S\ref{subsec:the_case_of_a_cluster_X-variety_at_R_positive_algebraic}, we first describe a solution to (QPu3), which is to construct a deformation quantization map $\mathring{\wh{Q}} {}^\hbar_{|\Gamma|} ~:~ \mathscr{O}(\mathscr{X}_{|\Gamma|}) \to \mathscr{O}^\hbar(\mathscr{X}_{|\Gamma|})$, as in eq.\eqref{eq:mathring_wh_Q}. As mentioned, we will use the results on the so-called duality maps from the theory of cluster varieties. One version is a map
$$
	\mathring{\mathbb{I}}_{|\Gamma|} ~:~ \mathscr{A}_{|\Gamma^\vee|}(\mathbb{Z}^t) \to \mathscr{O}(\mathscr{X}_{|\Gamma|})
$$
satisfying some favorable properties, where $\mathscr{A}_{|\Gamma^\vee|}(\mathbb{Z}^t)$ stands for the set of tropical integer points of the cluster $\mathscr{A}$-variety $\mathscr{A}_{|\Gamma^\vee|}$ associated to the mutation-equivalence class $|\Gamma^\vee|$ of the `cluster $\mathscr{A}$-seed' $\Gamma^\vee$, whose underlying exchange matrix is same as that of $\Gamma$. One of the desired properties is that the image of $\mathring{\mathbb{I}}_{|\Gamma|}$ forms a basis of the algebra of regular functions $\mathscr{O}(\mathscr{X}_{|\Gamma|})$. The existence of such duality maps for general initial seed $\Gamma$ has been expected by the Fock-Goncharov duality conjectures \cite{FG06,FG09b}, and was proved for a large class of seeds by Gross, Hacking, Keel and Kontsevich \cite{GHKK18}: namely, for all seeds whose underlying exchange matrices satisfy some combinatorial condition, such as the existence of a maximal green sequence. The solution uses a tool called a consistent scattering diagram, and except for few examples, an explicit enough construction of a consistent scattering diagram is not known, although the existence is proved; this makes the solution of \cite{GHKK18} not constructive in general. Davison and Mandel \cite{DM21} later constructed a quantum version of that duality map
$$
	\mathring{\mathbb{I}}^\hbar_{|\Gamma|} ~:~  \mathscr{A}_{|\Gamma^\vee|}(\mathbb{Z}^t) \to \mathscr{O}^\hbar(\mathscr{X}_{|\Gamma|}).
$$
So, the sought-for deformation quantization map
$$
	\mathring{\wh{Q}} {}^\hbar_{|\Gamma|} ~:~ \mathscr{O}(\mathscr{X}_{|\Gamma|}) \to \mathscr{O}^\hbar(\mathscr{X}_{|\Gamma|})
$$
can then be defined as the unique linear map that sends each basis vector $\mathring{\mathbb{I}}_{|\Gamma|}(l) \in \mathscr{O}(\mathscr{X}_{|\Gamma|})$ to the corresponding basis vector $\mathring{\mathbb{I}}^\hbar_{|\Gamma|}(l) \in \mathscr{O}^\hbar(\mathscr{X}_{|\Gamma|})$, for $l \in \mathscr{A}_{|\Gamma^\vee|}(\mathbb{Z}^t)$. This yields a solution to the deformation quantization problem (QPu3), for any initial seed satisfying the above mentioned combinatorial condition. We remark that the above construction of $\mathring{\wh{Q}} {}^\hbar_{|\Gamma|}$ using duality maps has not been emphasized in the literature, although all the necessary ingredients were already known.

\vs

The above mentioned particular duality maps of \cite{GHKK18} and \cite{DM21} are not the only possible duality maps with certain desired properties in general; see e.g. \cite{Qin19}. In principle, if one can find different answers for the duality maps, they may lead to a different deformation quantization map. In particular, for the case when the initial seed $\Gamma$ comes from an ideal triangulation of  a punctured surface, which is related to the classical Teichm\"uller theory and also to the moduli spaces of 3d spacetimes, the latter being the objects of our main interest, we suggest to use other known duality maps: namely, Fock-Goncharov's geometric solution to $\mathring{\mathbb{I}}_{|\Gamma|}$ \cite{FG06}, and the corresponding quantum version $\mathring{\mathbb{I}}^\hbar_{|\Gamma|}$ constructed by Allegretti and the first author \cite{AK17}. Unlike the constructions of \cite{GHKK18,DM21}, these latter two constructions \cite{FG06,AK17} are completely constructive and explicit, heavily relying on the topology and geometry of the relevant punctured surface. Recently, Mandel and Qin \cite{MQ} proved that the former two constructions coincide with the latter two, for $\Gamma$ coming from a punctured surface.

\vs

Now we apply the algebra maps in eq.\eqref{eq:two_algebra_embeddings} to the above solution of (QPu3) in order to construct a solution to (QPL3), for the case of $\mathscr{X}_{|\Gamma|}(\mathbb{R}_\Lambda^+)$.  First, the classical duality map $\mathring{\mathbb{I}}_{|\Gamma|}$ for the usual cluster $\mathscr{X}$-variety $\mathscr{X}_{|\Gamma|}$ can be understood as a package of duality maps
$$
	\mathring{\mathbb{I}}_\Gamma ~:~ \mathscr{A}_{|\Gamma^\vee|}(\mathbb{Z}^t) \to \mathring{\mathbb{L}}_\Gamma
$$
compatible with the coordinate change maps $\mathring{\mu}_{\Gamma,\Gamma'}$, in the sense that $\mathring{\mathbb{I}}_\Gamma = \mathring{\mu}_{\Gamma,\Gamma'} \circ \mathring{\mathbb{I}}_{\Gamma'}$. So, for each seed $\Gamma \in |\Gamma|$, per each $l \in \mathscr{A}_{|\Gamma^\vee|}(\mathbb{Z}^t)$ is associated a basis vector $\mathring{\mathbb{I}}_\Gamma(l)$ of $\mathring{\mathbb{L}}_\Gamma$; likewise for the quantum case, so we have quantum duality maps
$$
	\mathring{\mathbb{I}}^\hbar_\Gamma :  \mathscr{A}_{|\Gamma^\vee|}(\mathbb{Z}^t) \to \mathring{\mathbb{L}}^\hbar_\Gamma.
$$
For the $\mathbb{R}_\Lambda$-side, as seen in \S\ref{subsec:the_case_of_a_cluster_X-variety_and_R_Lambda}, the quantum algebra $\mathcal{A}^\hbar_\Gamma$ or its subalgebra $(\mathcal{A}^\hbar_\Gamma)_0$ for each seed $\Gamma \in |\Gamma|$ is isomorphic to the tensor product
$$
	\mathcal{A}^{\hbar(+)}_\Gamma \otimes \mathcal{A}^{\hbar(-)}_\Gamma \cong \left\{
	\begin{array}{ll}
		\mathcal{A}^\hbar_\Gamma     & \mbox{if $\Lambda=-1,1$}, \\
		(\mathcal{A}^\hbar_\Gamma)_0 & \mbox{if $\Lambda=0$},
	\end{array}
	\right.
$$
where each subalgebra $\mathcal{A}^{\hbar(\epsilon)}_\Gamma$ is generated by $\big\{ (Z_i^{(\epsilon)})^{\pm 1} : i \in I\big\}$, isomorphic to $\mathring{\mathcal{A}}^\hbar_\Gamma$ via the isomorphism $\iota^{(\epsilon)}_\Gamma$ in eq.\eqref{eq:two_algebra_embeddings}. For each $\epsilon \in \{+,-\}$, the quantum coordinate change map $\mu^\hbar_{\Gamma,\Gamma'} : {\rm Frac}(\mathcal{A}^\hbar_{\Gamma'}) \to {\rm Frac}(\mathcal{A}^\hbar_\Gamma)$ sends elements of ${\rm Frac}(\mathcal{A}^{\hbar(\epsilon)}_{\Gamma'})$ to those of ${\rm Frac}(\mathcal{A}^{\hbar(\epsilon)}_\Gamma)$; thus in fact it can be decomposed as
$$
	\mu^\hbar_{\Gamma,\Gamma'} = \mu^{\hbar(+)}_{\Gamma,\Gamma'} \otimes \mu^{\hbar(-)}_{\Gamma,\Gamma'}
$$
with
$$
	\mu^{\hbar(\epsilon)}_{\Gamma,\Gamma'} ~:~ {\rm Frac}(\mathcal{A}^{\hbar(\epsilon)}_{\Gamma'}) \to {\rm Frac}(\mathcal{A}^{\hbar(\epsilon)}_\Gamma)
$$
given by the same formula as $\mu^\hbar_{\Gamma,\Gamma'}$ on the generators. This yields the decomposition
$$
	\mathbb{L}^{\hbar(+)}_\Gamma \otimes \mathbb{L}^{\hbar(-)}_\Gamma = \left\{
	\begin{array}{ll}
		\mathbb{L}^\hbar_\Gamma     & \mbox{if $\Lambda=-1,1$}, \\
		(\mathbb{L}^\hbar_\Gamma)_0 & \mbox{if $\Lambda=-0$},
	\end{array}
	\right.
$$
where, for each $\epsilon \in \{+,-\}$, the restriction of the map $\iota^{(\epsilon)}_\Gamma : \mathring{\mathcal{A}}^\hbar_\Gamma \to \mathcal{A}^\hbar_\Gamma$ induces the isomorphism
$$
	\iota^{(\epsilon)}_\Gamma ~:~ \mathring{\mathbb{L}}^\hbar_\Gamma \to \mathbb{L}^{\hbar(\epsilon)}_\Gamma.
$$

\vs

Such decomposition statements hold for the classical setting too, via the $\mathbb{C}$-algebra embeddings
\begin{align*}
	\iota^{(\epsilon)}_\Gamma ~:~ \mathring{\mathcal{A}}_\Gamma \to \mathcal{A}_\Gamma,
	 &  &
	X_i \mapsto Z_i^{(\epsilon)};
\end{align*}
likewise, we get
$$
	\mathbb{L}_\Gamma^{(+)} \otimes \mathbb{L}^{(-)}_\Gamma = \left\{
	\begin{array}{ll}
		\mathbb{L}_\Gamma     & \mbox{if $\Lambda=-1,1$,} \\
		(\mathbb{L}_\Gamma)_0 & \mbox{if $\Lambda=0$}
	\end{array}
	\right.
$$
with the isomorphisms
$$
	\iota^{(\epsilon)}_\Gamma ~:~ \mathbb{L}_\Gamma \to \mathbb{L}^{(\epsilon)}_\Gamma.
$$
As a result, $\mathbb{L}_\Gamma$ or $(\mathbb{L}_\Gamma)_0$ is isomorphic to $\mathring{\mathbb{L}}_\Gamma \otimes \mathring{\mathbb{L}}_\Gamma$, hence admits a basis enumerated by the product of two copies of $\mathscr{A}_{|\Gamma^\vee|}(\mathbb{Z}^t)$. For each $\Gamma \in |\Gamma|$ and $\epsilon \in \{+,-\}$, define the classical half duality map
$$
	\mathbb{I}^{(\epsilon)}_\Gamma := \iota^{(\epsilon)}_\Gamma \circ \mathring{\mathbb{I}}_\Gamma ~:~ \mathscr{A}_{|\Gamma^\vee|}(\mathbb{Z}^t) \to \mathbb{L}^{(\epsilon)}_\Gamma,
$$
and the classical full duality map
$$
	\mathbb{I}_\Gamma ~:~  \mathscr{A}_{|\Gamma^\vee|}(\mathbb{Z}^t) \times \mathscr{A}_{|\Gamma^\vee|}(\mathbb{Z}^t)  \to \mathbb{L}^{(+)}_\Gamma \otimes \mathbb{L}^{(-)}_\Gamma = \left\{
	\begin{array}{ll}
		\mathbb{L}_\Gamma     & \mbox{if $\Lambda=-1,1$} \\
		(\mathbb{L}_\Gamma)_0 & \mbox{if $\Lambda=0$}
	\end{array}
	\right.
$$
as
$$
	\mathbb{I}_\Gamma (l_1,l_2) := \mathbb{I}^{(+)}_\Gamma (l_1) \otimes \mathbb{I}^{(-)}_\Gamma (l_2).
$$
The image of $\mathbb{I}_\Gamma$ then forms a basis of $\mathbb{L}_\Gamma$ or $(\mathbb{L}_\Gamma)_0$, and is compatible with the coordinate change maps, in the sense that
$$
	\mathbb{I}_\Gamma = \mu_{\Gamma,\Gamma'} \circ \mathbb{I}_{\Gamma'},
$$
hence forming the single-packaged classical duality map
$$
	\mathbb{I}_{|\Gamma|} ~:~ \mathscr{A}_{|\Gamma^\vee|}(\mathbb{Z}^t) \times \mathscr{A}_{|\Gamma^\vee|}(\mathbb{Z}^t) \to \mathscr{O}_\Lambda(\mathscr{X}_{|\Gamma|}).
$$
Likewise for the quantum case, define the quantum half duality maps
$$
	\mathbb{I}^{\hbar(\epsilon)}_\Gamma := \iota^{(\epsilon)}_\Gamma \circ \mathring{\mathbb{I}}^\hbar_\Gamma ~:~  \mathscr{A}_{|\Gamma^\vee|}(\mathbb{Z}^t)  \to \mathbb{L}^{\hbar(\epsilon)}_\Gamma
$$
and the quantum full duality map
$$
	\mathbb{I}^\hbar_\Gamma ~:~ \mathscr{A}_{|\Gamma^\vee|}(\mathbb{Z}^t) \times \mathscr{A}_{|\Gamma^\vee|}(\mathbb{Z}^t)  \to \mathbb{L}^{\hbar(+)}_\Gamma \otimes \mathbb{L}^{\hbar(-)}_\Gamma =  \left\{
	\begin{array}{ll}
		\mathbb{L}^\hbar_\Gamma     & \mbox{if $\Lambda=-1,1$}, \\
		(\mathbb{L}^\hbar_\Gamma)_0 & \mbox{if $\Lambda=-0$},
	\end{array}
	\right.
$$
as
$$
	\mathbb{I}^\hbar_\Gamma(l_1,l_2) := \mathbb{I}^{\hbar(+)}_\Gamma(l_1) \otimes \mathbb{I}^{\hbar(-)}_\Gamma(l_2).
$$
Finally, the deformation quantization map $\wh{Q}^\hbar_\Gamma : \mathbb{L}_\Gamma \to \mathbb{L}^\hbar_\Gamma$, or $(\wh{Q}^\hbar_\Gamma)_0 : (\mathbb{L}_\Gamma)_0 \to (\mathbb{L}^\hbar_\Gamma)_0$ for $\Lambda=0$, is constructed as the unique linear map sending each basis vector $\mathbb{I}_\Gamma(l_1,l_2)$ to the corresponding quantum element $\mathbb{I}^\hbar_\Gamma(l_1,l_2)$, for $l_1,l_2 \in \mathscr{A}_{|\Gamma^\vee|}(\mathbb{Z}^t)$. Moreover, for $\Lambda=0$, one can compose the $*$-maps to construct a quantization map between the full algebras $\mathbb{L}_\Gamma$ and $\mathbb{L}^\hbar_\Gamma$. This is our solution to the deformation quantization problem (QPL4) of the $\mathbb{R}_\Lambda$-variety $\mathscr{X}_{|\Gamma|}(\mathbb{R}_\Lambda^+)$ for a general initial seed $\Gamma$.

\subsection{Representations on doubled Hilbert spaces}
\label{subsec:representations_on_doubled_Hilbert_spaces}

Now we touch upon the operator aspect of the quantization problem for the $\mathbb{R}_\Lambda$-variety $\mathscr{X}_{|\Gamma|}(\mathbb{R}_\Lambda^+)$; here we first deal with (QPL3). We aim to consider and construct $*$-representations of algebras over $\mathbb{C}_\Lambda^\hbar$ (Def.\ref{def:C_hbar_Lambda}), and as mentioned, one crucial point is on how to represent the elements of $\mathbb{C}_\Lambda^\hbar$. We suggest that $\mathbb{C}_\Lambda^\hbar$ should be represented in an irreducible or indecomposable manner as much as possible, and also in a uniform way for all three values of $\Lambda$. As mentioned, we realize the elements of $\mathbb{C}^\hbar_\Lambda$ as elements of $\mathbb{C}_\Lambda$ (Def.\ref{def:C_Lambda}) by the natural $\mathbb{C}$-algebra map $\mathbb{C}^\hbar_\Lambda \to \mathbb{C}_\Lambda$ sending $q_\Lambda$ to $e^{\pi {\rm i} \ell\hbar}$ and $q_\Lambda^*$ to $e^{\pi {\rm i} \ell^*\hbar}$. So it suffices to determine how the elements of $\mathbb{C}_\Lambda$ are represented.

\vs

We shall require that $\mathbb{C}_\Lambda$, viewed as a $\mathbb{C}$-algebra, should be represented by the following standard complex $*$-representation on $\mathbb{C}^2$ equipped with the standard Hermitian inner product, which is a complexification of the map in eq.\eqref{eq:original_embedding_map}:
$$
	\pi ~:~ \mathbb{C}_\Lambda \to {\rm End}(\mathbb{C}^2), \qquad x + \ell y \mapsto \mattwo{x}{-\Lambda y}{y}{x}, \quad \forall x,y\in \mathbb{C};
$$
in particular, $\pi(\ell) = \smallmattwo{0}{-\Lambda}{1}{0}$, and $\pi(\ell^*) = \pi(\ell)^* = \smallmattwo{0}{1}{-\Lambda}{0}$.

\vs

For each seed $\Gamma \in |\Gamma|$, we propose to quantize classical real-valued functions on $\mathscr{X}_{|\Gamma|}(\mathbb{R}_\Lambda^+)$ by self-adjoint operators on the Hilbert space $\mathring{\mathscr{H}}_\Gamma = L^2(\mathbb{R}^I, \wedge_{i\in I} dt_i)$ in eq.\eqref{eq:H_Gamma} which we used in the quantization of the cluster variety $\mathscr{X}_{|\Gamma|}(\mathbb{R}^+)$ at $\mathbb{R}^+$, and then to quantize $\mathbb{R}_\Lambda$-valued functions on $\mathscr{X}_{|\Gamma|}(\mathbb{R}_\Lambda^+)$ by operators on the Hilbert space
$$
	\mathscr{H}_{\Gamma} := \mathbb{C}^2 \otimes \mathring{\mathscr{H}}_\Gamma = \mathscr{H}^{(+)}_{\Gamma} \oplus \mathscr{H}^{(-)}_{\Gamma},
$$
where each of $\mathscr{H}^{(+)}_{\Gamma}$ and $\mathscr{H}^{(-)}_{\Gamma}$ is isomorphic to $\mathring{\mathscr{H}}_\Gamma$ as Hilbert spaces. Here $\mathscr{H}_\Gamma$ can be thought of as a double of the Hilbert space $\mathring{\mathscr{H}}_\Gamma$, or a {\it doubled Hilbert space}. We take $\mathscr{H}_\Gamma$ as the quantum Hilbert space for the seed $\Gamma$ for the quantization problem for the $\mathbb{R}_\Lambda$-variety $\mathscr{X}_{|\Gamma|}(\mathbb{R}_\Lambda^+)$ as formulated in \S\ref{sec:a_deformation_quantization_problem}.

\vs

A linear operator $A : \mathscr{H}_{\Gamma} \to \mathscr{H}_{\Gamma}$ could be expressed as an operator-valued $2$-by-$2$ matrix
$$
	A = \mattwo{A^{(++)}}{A^{(-+)}}{A^{(+-)}}{A^{(--)}}
$$
with some operators
$$
	A^{(\epsilon \epsilon')} ~:~ \mathscr{H}_{\Gamma}^{(\epsilon)} \to \mathscr{H}_{\Gamma}^{(\epsilon')}.
$$

For a vector subspace $\mathring{V}_\Gamma$ of $\mathscr{H}_\Gamma$, we denote by $\mathbb{C}^2 \otimes \mathring{V}_\Gamma$ the corresponding subspace of $\mathbb{C}^2\otimes \mathring{\mathscr{H}}_\Gamma = \mathscr{H}_\Gamma$. For a linear operator $\mathring{B}$ on $\mathring{\mathscr{H}}_\Gamma$, we denote the corresponding doubled operator on $\mathscr{H}_\Gamma$ by ${\rm Id}_{\mathbb{C}^2} \otimes \mathring{B}$, which is given in the operator-valued matrix format as
$$
	{\rm Id}_{\mathbb{C}^2} \otimes \mathring{B} = \mattwo{\mathring{B}}{0}{0}{\mathring{B}}.
$$
In case $\mathring{B}$ is defined only on a domain ${\rm Dom}(\mathring{B}) = \mathring{V}_\Gamma$ which is a subspace of $\mathring{\mathscr{H}}_\Gamma$, then the operator ${\rm Id}_{\mathbb{C}^2}\otimes \mathring{B}$ is defined by the above formula on the domain ${\rm Dom}({\rm Id}_{\mathbb{C}^2} \otimes \mathring{B}) := \mathbb{C}^2 \otimes \mathring{V}_\Gamma = \mathring{V}_\Gamma \oplus \mathring{V}_\Gamma$. When we refer to an {\it operator} $\mathring{B}$ on $\mathring{\mathscr{H}}_\Gamma$, we would mean a densely defined operator on $\mathring{\mathscr{H}}_\Gamma$, which consists of a dense subspace ${\rm Dom}(\mathring{B})$ of $\mathring{\mathscr{H}}_\Gamma$ and a linear map $\mathring{B} : {\rm Dom}(\mathring{B}) \to \mathring{\mathscr{H}}_\Gamma$. Thus we need to be careful when dealing with the domains.

\vs

We let each element $u$ of $\mathbb{C}_\Lambda$ act on  $\mathscr{H}_\Gamma$ as the operator $\pi(u) \otimes {\rm Id}$, which is a bounded operator (hence is everywhere defined). Then the complex numbers $u\in \mathbb{C} \subset \mathbb{C}_\Lambda$ act on $\mathscr{H}_\Gamma$ as the usual scalar $u$, so the only important element to consider is $\ell \in \mathbb{C}_\Lambda$, which is represented by the operator
$$
	\wh{\ell} := \pi(\ell) \otimes {\rm Id} = \mattwo{0}{-\Lambda}{1}{0} \otimes {\rm Id} = \mattwo{0}{-\Lambda \cdot {\rm Id}}{{\rm Id}}{0}.
$$
We note that the operator $\wh{\ell}$ is a (bounded) normal operator in the cases $\Lambda = -1,1$, while it is not when $\Lambda=0$. For any densely defined operator $\mathring{B}$ on $\mathscr{H}_\Gamma$, one has the equality
$$
	({\rm Id}_{\mathbb{C}^2}\otimes \mathring{B}) \circ \wh{\ell} = \wh{\ell} \circ ({\rm Id}_{\mathbb{C}^2}\otimes \mathring{B})
$$
as densely defined operators. We will denote this operator $({\rm Id}_{\mathbb{C}^2}\otimes \mathring{B}) \circ \wh{\ell} = \wh{\ell} \circ ({\rm Id}_{\mathbb{C}^2}\otimes \mathring{B})$ just by $\wh{\ell} \, ({\rm Id}_{\mathbb{C}^2}\otimes \mathring{B})$, without the composition symbol $\circ$.

\vs

For a $\mathbb{C}_\Lambda$-algebra, we will consider only those representations whose underlying Hilbert space is of the form $\mathscr{H}_\Gamma$ (associated to some Hilbert space $\mathring{\mathscr{H}}_\Gamma$) and the elements $u$ of $\mathbb{C}_\Lambda$ are represented by $\pi(u) \otimes {\rm Id}$ as above.

\vs

Now, for each $\epsilon \in \{+,-\}$, we define the operator
\begin{align}
	\label{eq:our_log_quantum_operator}
	{\bf z}^{(\epsilon)}_i := ({\rm Id}_{\mathbb{C}^2}\otimes {\bf x}_i) + \epsilon \, \wh{\ell} \, \hbar \, ({\rm Id}_{\mathbb{C}^2} \otimes {\bf y}_i) = \mattwo{{\bf x}_i}{- \epsilon  \Lambda \hbar {\bf y}_i}{\epsilon \hbar {\bf y}_i}{{\bf x}_i}
\end{align}
on the doubled Hilbert space $\mathscr{H}_\Gamma = \mathbb{C}^2 \otimes \mathring{\mathscr{H}}_\Gamma$, where the self-adjoint operators ${\bf x}_i$ and ${\bf y}_i$ on $\mathring{\mathscr{H}}_\Gamma$ are as defined in eq.\eqref{eq:x_and_y_operators} of \S\ref{subsec:a_positive_representation_for_each_seed}. Then, one has the commutation relations
$$
	[{\bf z}^{(+)}_i, {\bf z}^{(+)}_j] = 2\pi {\rm i} \hbar \wh{\ell} \varepsilon_{ij}, \qquad
	[{\bf z}^{(-)}_i, {\bf z}^{(-)}_j] = - 2\pi {\rm i} \hbar \wh{\ell} \varepsilon_{ij}, \qquad
	[{\bf z}^{(+)}_i, {\bf z}^{(-)}_j] = 0,
$$
e.g. as operators $\mathscr{D}_\Gamma \to \mathscr{D}_\Gamma$, where
$$
	\mathscr{D}_\Gamma := \mathbb{C}^2 \otimes \mathring{\mathscr{D}}_\Gamma,
$$
and $\mathring{\mathscr{D}}_\Gamma$ is as in eq.\eqref{eq:D_Gamma}.
One can also show that when $\Lambda = -1,1$, the above ${\bf z}^{(\epsilon)}_i$ yields a normal operator; we do not prove this fact here and leave it to readers, as our proofs will not explicitly depend on it.

We now develop a version of functional calculus of some normal operators on the doubled Hilbert space $\mathscr{H}_\Gamma = \mathscr{H}_\Gamma^{(+)} \oplus \mathscr{H}_\Gamma^{(-)}$. Rigorous proofs can be found involving only usual functional calculus of self-adjoint operators. However, we will often give a formulation using normal operators, for such a description can be more intuitive.

\begin{definition}
	\label{def:admissible_function}
	A function $f : \mathbb{R}_\Lambda \to \mathbb{C}_\Lambda$ is {\it admissible} if its matrix form $D_\Lambda \circ f\circ D_\Lambda^{-1} : D_\Lambda(x+\ell y) \mapsto D_\Lambda(f(x+\ell y))$ is given as follows (see \S\ref{subsec:R_Lambda} for the definition of $D_\Lambda$):
	\begin{enumerate}
		\item[\rm (1)] for $\Lambda =-1$, there exist functions $f^{(+)},f^{(-)} : \mathbb{R} \to \mathbb{C}$ such that
		      \begin{align*}
			      D_\Lambda \circ f\circ D_\Lambda^{-1}  ~:~
			      \begin{pmatrix}
				      x+y & 0 \cr 0 & x-y
			      \end{pmatrix}
			      \mapsto
			      \begin{pmatrix}
				      f^{(+)}(x+y) & 0 \cr 0 & f^{(-)}(x-y)
			      \end{pmatrix};
		      \end{align*}

		\item[\rm (2)] for $\Lambda =1$, there exist functions $f^{(+)},f^{(-)} : \mathbb{C} \to \mathbb{C}$ such that
		      \begin{align*}
			      D_\Lambda \circ f\circ D_\Lambda^{-1}  ~:~
			      \begin{pmatrix}
				      x+{\rm i} y & 0 \cr 0 & x-{\rm i}y
			      \end{pmatrix}
			      \mapsto
			      \begin{pmatrix}
				      f^{(+)}(x+{\rm i}y) & 0 \cr 0 & f^{(-)}(x-{\rm i}y)
			      \end{pmatrix};
		      \end{align*}

		\item[\rm (3)] for $\Lambda =0$, there exists a differentiable function $f_0 : \mathbb{R} \to \mathbb{C}$ such that
		      \begin{align*}
			      D_\Lambda \circ f\circ D_\Lambda^{-1}  ~:~ \mattwo{x}{0}{y}{x} \mapsto \mattwo{f_0(x)}{0}{f_0'(x) y}{f_0(x)};
		      \end{align*}
	\end{enumerate}
\end{definition}
One prototypical example of an admissible function $f:\mathbb{R}_\Lambda \to \mathbb{C}_\Lambda$ is a function coming from an analytic function $g:\mathbb{R} \to \mathbb{C}$ or $g:\mathbb{C} \to \mathbb{C}$, such as $g(z) = \exp(z)$; in this case $f^{(+)}=f^{(-)}=f_0=g$; see e.g. \cite{MS16}. We define an $\mathbb{R}_\Lambda$-version of functional calculus as follows.
\begin{definition}[$\mathbb{R}_\Lambda$ functional calculus]
	Let ${\bf x}, {\bf y}$ be self-adjoint operators on a complex Hilbert space $\mathring{\mathscr{H}}$ that strongly commute with each other. Consider the normal operator ${\bf z} = ({\rm Id}_{\mathbb{C}^2} \otimes {\bf x}) + \wh{\ell} ({\rm Id}_{\mathbb{C}^2} \otimes {\bf y})$ on $\mathbb{C}^2 \otimes \mathring{\mathscr{H}}$. Let $f : \mathbb{R}_\Lambda \to \mathbb{C}_\Lambda$ be an admissible function. We define $f({\bf z})$ as the normal operator on $\mathbb{C}^2 \otimes \mathring{\mathscr{H}}$ given in matrix form as follows, for each value of $\Lambda$:
	\begin{align}
		\label{eq:diagonal_operator_-1}
		\Lambda = -1 & ~:~ f({\bf z}) = \mattwo{ \frac{{\rm Id}}{\sqrt{2}}}{\frac{{\rm Id}}{\sqrt{2}}}{\frac{{\rm Id}}{\sqrt{2}}}{-\frac{{\rm Id}}{\sqrt{2}}} \mattwo{f^{(+)}({\bf x} + {\bf y})}{0}{0}{\hspace{-2mm} f^{(-)}({\bf x} - {\bf y})} \mattwo{ \frac{{\rm Id}}{\sqrt{2}}}{\frac{{\rm Id}}{\sqrt{2}}}{\frac{{\rm Id}}{\sqrt{2}}}{-\frac{{\rm Id}}{\sqrt{2}}},                                          \\
		\label{eq:diagonal_operator_1}
		\Lambda = 1  & ~:~ f({\bf z}) = \mattwo{\frac{{\rm Id}}{\sqrt{2}}}{\frac{{\rm Id}}{\sqrt{2}}}{-{\rm i}\frac{{\rm Id}}{\sqrt{2}}}{{\rm i}\frac{{\rm Id}}{\sqrt{2}}} \mattwo{f^{(+)}({\bf x} + {\rm i} {\bf y})}{0}{0}{\hspace{-3mm} f^{(-)}({\bf x} - {\rm i}{\bf y})} \mattwo{\frac{{\rm Id}}{\sqrt{2}}}{{\rm i}\frac{{\rm Id}}{\sqrt{2}}}{\frac{{\rm Id}}{\sqrt{2}}}{-{\rm i}\frac{{\rm Id}}{\sqrt{2}}}, \\
		\label{eq:diagonal_operator_0}
		\Lambda = 0  & ~:~ f({\bf z}) := \mattwo{f({\bf x})}{0}{f'({\bf x}) \, {\bf y}}{f({\bf x})}
	\end{align}
	The operators $f^{(\epsilon)}({\bf x}+\epsilon {\bf y})$, $f^{(\epsilon)}({\bf x}+{\rm i}\epsilon {\bf y})$, $f({\bf x})$, $f'({\bf x}) {\bf y}$ on $\mathring{\mathscr{H}}$, and their domains, are defined by the usual two-variable functional calculus of strongly commuting self-adjoint operators ${\bf x},{\bf y}$ (see e.g. \cite{Sch12} for such a functional calculus).
\end{definition}

We apply the above functional calculus to the exponential function $f(z) = \exp(z) = e^z$, and define the operators
$$
	{\bf Z}^{(\epsilon)}_i := \exp({\bf z}^{(\epsilon)}_i).
$$
Then one can verify the relations
\begin{align*}
	{\bf Z}^{(+)}_i \, {\bf Z}^{(+)}_j & = e^{2\pi {\rm i} \hbar \wh{\ell} \varepsilon_{ij}} \, {\bf Z}^{(+)}_j \, {\bf Z}^{(+)}_i, \cr
	{\bf Z}^{(-)}_i \, {\bf Z}^{(-)}_j & = e^{-2\pi {\rm i} \hbar \wh{\ell} \varepsilon_{ij}} \, {\bf Z}^{(-)}_j \, {\bf Z}^{(-)}_i, \cr
	{\bf Z}^{(+)}_i \, {\bf Z}^{(-)}_j & = {\bf Z}^{(-)}_j \, {\bf Z}^{(+)}_i,
\end{align*}
on suitable domains. One observes that ${\rm Dom}({\bf Z}^{(\epsilon)}_i)$ contains $\mathscr{D}_\Gamma$ and that ${\bf Z}^{(\epsilon)}_i$ preserves $\mathscr{D}_\Gamma$. If one views ${\bf Z}^{(\epsilon)}_i$ as operators $\mathscr{D}_\Gamma \to \mathscr{D}_\Gamma$, then the above relations genuinely hold as operators $\mathscr{D}_\Gamma \to \mathscr{D}_\Gamma$. Therefore there is a unique homomorphism
$$
	\til{\pi}_\Gamma ~:~ \mathcal{A}^\hbar_\Gamma \to {\rm End}_\mathbb{C}(\mathscr{D}_\Gamma )
$$
sending each generator as
$$
	\til{\pi}_\Gamma ( \wh{Z}^{(\epsilon)}_i) = {\bf Z}^{(\epsilon)}_i \restriction  \mathscr{D}_\Gamma ~:~ \mathscr{D}_\Gamma \to \mathscr{D}_\Gamma.
$$
The Schwartz space can then be defined as
$$
	\mathscr{S}_\Gamma := \bigcap_{u\in \mathbb{L}^\hbar_\Gamma} {\rm Dom}(\til{\pi}_\Gamma(u)^*)
$$

This is our solution to (QPL4), i.e. representation for each seed $\Gamma$. One has to make sure that the chosen representations for different seeds $\Gamma$ are compatible with each other, so that the more important intertwiner problem (QPL5) would be solvable. So, just as for (QPu4) and (QPu5) in \S\ref{sec:FG_representations}, (QPL4) and (QPL5) should be thought of as a single problem tied together.

\subsection{A trilogy of quantum dilogarithm functions}
\label{subsec:trilogy}

For the remainder of the present section, we present our solution to the intertwiner problem (QPL5), which constitutes a major new contribution of the present paper. Here we establish a crucial necessary tool for our solution.

\vs

Recall from \S\ref{sec:a_deformation_quantization_problem} that, at the algebraic level, the quantum mutation formula $\mathring{\mu}^\hbar_{\Gamma,\Gamma'}$ for the usual cluster $\mathscr{X}$-varieties carry over to that $\mu^\hbar_{\Gamma,\Gamma'}$ for the $\mathbb{R}_\Lambda$-variety setting, where $q = e^{\pi {\rm i} \hbar}$ should correspond to $q_\Lambda = e^{\pi {\rm i} \ell\hbar}$. At the operator level, while the linear quantum operators $\mathring{\bf x}_i$ defined in  \S\ref{sec:FG_representations} (where each generator $\wh{X}_i$ is represented by $\exp(\mathring{\bf x}_i)$) for the quantum cluster $\mathscr{X}$-variety satisfies $[\mathring{\bf x}_i,\mathring{\bf x}_j] = 2\pi {\rm i} \hbar \varepsilon_{ij}$, the linear quantum operators ${\bf z}^{(\epsilon)}_i$ defined in \S\ref{subsec:representations_on_doubled_Hilbert_spaces} for the quantum $\mathbb{R}_\Lambda$-variety satisfies $[{\bf z}^{(\epsilon)}_i, {\bf z}^{(\epsilon)}_j] = 2\pi {\rm i} \epsilon \hbar \wh{\ell} \varepsilon_{ij}$. In both aspects, one observes that what should play the role of $\hbar$ for the quantum $\mathbb{R}_\Lambda$-variety is $\ell\hbar \in \mathbb{R}_\Lambda$. As the non-compact quantum dilogarithm function $\Phi^\hbar$ was crucially used in \S\ref{sec:FG_representations} for the construction of the mutation intertwiner for the usual quantum cluster $\mathscr{X}$-variety, in our current situation we would like to define a similar tool, heuristically something like $\Phi^{\ell\hbar}$. We will rigorously establish three different functions that will eventually play the role of this sought-for $\Phi^{\ell\hbar}$ for the three values of $\Lambda$.

\vs

To cover the cases $\Lambda=-1,1$ in a uniform manner, we first consider $\Phi^h$ for complex numbers $h$; notice that we use the symbol $h$ instead of $\hbar$. The previous contour integral formula in \S\ref{subsec:Phi_hbar} using the contour $\Omega$ does not always work; when $h$ is purely imaginary, some poles of the integrand lies on the contour. So we introduce a `slanted' version of the Barnes integral.
\begin{definition}[slanted Barnes integral]
	Let $h$ be any nonzero complex number with $\Re(h)\ge 0$. Pick any real numbers $a,\theta$ such that $0<a<\min\{1, \frac{1}{|h|}\}$ and
	$$
		\left\{ \begin{array}{cl}
			-\pi/2<\theta\le 0,\,\,\,\,\,\,\,\,   & \mbox{if ${\rm Im}(h) \ge 0$}, \\
			\,\,\,\,\,\,\,\,\, 0\le \theta<\pi/2, & \mbox{if ${\rm Im}(h) \le 0$},
		\end{array} \right.
	$$
	where we do not allow $\theta=0$ in case ${\rm Re}(h)=0$.

	\vs

	Let $\Omega_a$ be the contour in the complex plane along the real line that avoids the origin along the upper half circle of radius $a$ centered at the origin, with the orientation given by the increasing direction. Let $e^{ {\rm i} \theta} \Omega_a$ be the contour obtained by rotating $\Omega_a$ by the angle $\theta$ about the origin.

	\vs

	For $z\in \mathbb{C}$ living in the slanted strip
	\begin{align}
		\nonumber
		\mathcal{I}_{h,\theta} := & \Big\{ z\in \mathbb{C} ~:~ |\Im(e^{{\rm i} \theta}z)| < \pi (\cos \theta + \Re(h \, e^{{\rm i} \theta})) \, \Big\}            \\
		\nonumber
		=                         & \Big\{ e^{-{\rm i} \theta} w ~:~ w\in \mathbb{C}, ~ |\Im (w)| < \pi((1+\Re(h)) \cos \theta - \Im(h) \, \sin \theta) \, \Big\}
	\end{align}
	define
	\begin{align}
		\label{eq:B_h}
		\Phi^h(z) := \exp \left(
		- \frac{1}{4} \int_{ e^{{\rm i} \theta} \Omega_a } \frac{e^{-{\rm i} pz}}{\sinh(\pi p) \sinh(\pi h p)} \frac{dp}{p} \right).
	\end{align}

\end{definition}

\begin{lemma}
	Fix any $h\in \mathbb{C}^\times$ with ${\rm Re}(h)\ge 0$. Pick corresponding $a,\theta$. The integral in eq.\eqref{eq:B_h} absolutely converges on compact subsets of the slanted strip $\mathcal{I}_{h,\theta}$, so eq.\eqref{eq:B_h} yields a non-vanishing complex analytic function $\Phi^h(z)$ on $\mathcal{I}_{h,\theta}$. The value $\Phi^h(z)$ depends only on $h$ and $z$, but not on $a$ or $\theta$.
\end{lemma}
We omit a proof as it is a straightforward exercise in complex analysis; the corresponding statement is known for the cases $\Re (h)>0$ with $\theta=0$, whose special case ${\rm Im}(h)=0$ was recalled in \S\ref{subsec:Phi_hbar}. In a similar vein with somewhat more care, one can also show the following lemma; for this, one first needs to extend the above construction to $h\in \mathbb{C}^\times$ living in an open neighborhood of the region $\{ h \in \mathbb{C}^\times \, |\, \Re(h)\ge 0\}$ (using the same formulas).
\begin{lemma}
	The dependence of $\Phi^h(z)$ on $h$ is complex analytic.
\end{lemma}
In practice, one would prove this lemma first for the region ${\rm Re}(h)\ge 0$, ${\rm Im}(h)\ge 0$ using some fixed $\theta \in (-\pi/2,0)$, and then prove for the region ${\rm Re}(h)\ge 0$, ${\rm Im}(h)\le 0$ using some fixed $\theta \in (0,\pi/2)$.

\vs

The usual non-slanted contour $\Omega_a$ works for the cases $\Re(h)>0$, which have been already dealt with in the literature; see e.g. \cite{Bar01,FK94,FG09a} and also subsequent works of others, as well as references therein. Among these cases, in an extreme case when $h = \hbar \in \mathbb{R}^+$, with $\theta=0$, the function $\Phi^h(z)$ coincides with the non-compact quantum dilogarithm function $
	\Phi^\hbar(z)$ which is used in \cite{FG09a} as we saw in \S\ref{sec:FG_representations}. For our purpose, define
$$
	\Phi^{-\hbar}(z) := \Phi^\hbar(z)^{-1}, \quad \mbox{for $\hbar \in \mathbb{R}^+$};
$$
this indeed makes sense when one puts $-\hbar$ into the place of $\hbar$ in the contour integral definition of $\Phi^\hbar(z)$. The functions $\Phi^{\pm \hbar}(z)$ for $h=\pm \hbar\in \mathbb{R}^+$ are the versions of the quantum dilogarithm function that we will use in our quantization of the $\mathbb{R}_\Lambda$-variety $\mathscr{X}_{|\Gamma|}(\mathbb{R}_\Lambda^+)$ for the case $\Lambda=-1$.

\vs

We need the slanted contour $e^{{\rm i} \theta} \Omega_a$ to incorporate the case $\Re(h)=0$, which doesn't seem to have been investigated much in the literature but which we definitely need. The function $\Phi^h = \Phi^{\pm {\rm i} \hbar}$ for this case $h = \pm {\rm i} \hbar$ with $\hbar \in \mathbb{R}^+$ are the versions of the quantum dilogarithm function that we use for the case $\Lambda=1$.

\vs

Now we list some characteristic properties of the function $\Phi^h(z)$, for general $h\in \mathbb{C}^\times$ with ${\rm Re}(h)\ge 0$.
\begin{proposition}[properties of the slanted Barnes integral]
	\label{prop:properties_of_the_slanted_Barnes_integral}
	For each $h\in \mathbb{C}^\times$  with ${\rm Re}(h)\ge 0$, the function $\Phi^h(z)$ on $\mathcal{I}_{h,\theta}$ analytically continues to the meromorphic function $\Phi^h(z)$ on the complex plane, which satisfies the following properties.
	\begin{itemize}
		\item[\rm (SB1)] The zeros and poles are at
		      \begin{align*}
			      \mbox{the set of zeros} & = \big\{(2n+1) \pi {\rm i} + (2m+1) \pi {\rm i} h \, | \, n,m\in \mathbb{Z}_{\ge 0} \big\},  \\
			      \mbox{the set of poles} & = \big\{-(2n+1) \pi {\rm i} - (2m+1) \pi {\rm i} h \, | \, n,m\in \mathbb{Z}_{\ge 0} \big\}.
		      \end{align*}
		      If $h \in \mathbb{R}$, then all these zeros and poles are simple if and only if $h\notin \mathbb{Q}$. If $\mathrm{Im}(h)>0$, then all the zeros and poles are simple.

		\item[\rm (SB2)] (difference equations) Each of the functional relations
		      \begin{align*}
			      \left\{
			      \begin{array}{lcr}
				      \Phi^h(z+2\pi {\rm i} h) & = & (1+e^{\pi {\rm i} h} \, e^z) \, \Phi^h(z),     \\
				      \Phi^h(z+2\pi {\rm i})   & = & (1+e^{\pi {\rm i}/h} \, e^{z/h}) \, \Phi^h(z),
			      \end{array}
			      \right.
		      \end{align*}
		      holds, whenever the arguments of $\Phi^h$ are not poles.

		\item[\rm (SB3)] (involutivity) One has
		      $$
			      \Phi^h(z) \, \Phi^h(-z) = c_h \, \exp\left( {z^2}/({4\pi {\rm i} h}) \right),
		      $$
		      whenever $z$ and $-z$ are not poles of $\Phi^h$, where
		      $$
			      c_h := e^{-\frac{\pi {\rm i} }{12} (h+h^{-1})} \in \mathbb{C}^\times.
		      $$
		      In particular, $|c_\hbar|=1$ for $h=\hbar \in \mathbb{R}^+$.

		\item[\rm (SB4)] (ratio of compact quantum dilogarithms) When $\mathrm{Im}(h)>0$, one has
		      \begin{align}
			      \label{eq:B_h_as_ratio}
			      \Phi^h(z) = \frac{\psi^{\exp(\pi {\rm i} h)}(e^z)}{\psi^{\exp(-\pi {\rm i} /h)}(e^{z/h})},
		      \end{align}
		      where $\psi^{\bf q}$ is the compact quantum dilogarithm as in \S\ref{subsec:Phi_hbar}.

		\item[\rm (SB5)] (unitarity) One has
		      $$
			      \overline{\Phi^h(z)} = \Phi^{\overline{h}}(\overline{z})^{-1}
		      $$
		      whenever $z$ and $\overline{z}$ are not poles.

	\end{itemize}

\end{proposition}

The above proposition has been proved and used in the literature in the cases when $\Re(h)>0$, with $\theta=0$. Similar arguments also work for the case $\Re(h)=0$. For example, the item (SB2) implies the analytic continuation and (SB1), while the items (SB3) and (SB4) can be proved by a straightforward residue computation. The item (SB5) can also be easily obtained.

\vs

It remains to introduce a third version of the quantum dilogarithm function for the case $\Lambda=0$. In order to avoid using the ring $\mathbb{R}_0 = \mathbb{R}[\ell]/(\ell^2=0)$, we formulate it as the following two-real-variable function, whose well-definedness is easily seen.

\begin{definition}[the flat quantum dilogarithm]
	Define a function $F_0^\hbar=F_0 : \mathbb{R}^2 \to \mathbb{C}$ by the formula
	$$
		F_0^\hbar(x,y)=F_0(x,y) = (1+e^{x})^{y / (\pi {\rm i} )}
		= \exp\left( \frac{y}{\pi {\rm i} } \, \log(1+e^{x}) \right).
	$$
\end{definition}

As shall be seen in the next subsection in more detail, the uniform expression that appears in all three values of $\Lambda$ is
\begin{align}
	\label{eq:uniform_function_heuristic}
	\Phi^{\ell\hbar}(x+\ell \hbar y) \Phi^{-\ell\hbar}(x-\ell \hbar y),
\end{align}
for $x,y\in \mathbb{R}$. One can view this as a function in the $\mathbb{R}_\Lambda$ variable $z := x + \ell \hbar y$, where $x-\ell \hbar y$ is the $\mathbb{R}_\Lambda$-conjugate of $z$. Although we haven't made sense of the expression $\Phi^{\ell\hbar}$, we can make sense of the above expression as an admissible function $\mathbb{R}_\Lambda \to \mathbb{C}_\Lambda$ in the sense of Def.\ref{def:admissible_function}. In particular, for $\Lambda=-1,1$, in the diagonalized form, in each eigenspace, $\ell$ is represented by the corresponding eigenvalue$\in\{1,-1,{\rm i},-{\rm i}\}$. So, for $\Lambda=-1$, the functions $f^{(\epsilon)}$ corresponding to $\Phi^{\pm \ell\hbar}(z)$ appearing in Def.\ref{def:admissible_function} are $f^{(\epsilon)} = \Phi^{\pm \epsilon \hbar}$, while for $\Lambda =1$, the functions $f^{(\epsilon)}$ for $\Phi^{\pm \ell\hbar}(z)$ are $f^{(\epsilon)} = \Phi^{\pm {\rm i} \epsilon \hbar}$. So the diagonal form of eq.\eqref{eq:uniform_function_heuristic} in the style of Def.\ref{def:admissible_function} is
\begin{align*}
	\begin{pmatrix}
		x+\hbar y & 0 \cr 0 & x- \hbar y
	\end{pmatrix}
	 & \mapsto
	\begin{pmatrix}
		\Phi^\hbar(x+\hbar y) & 0 \cr 0 & \Phi^{-\hbar}(x-\hbar y)
	\end{pmatrix}
	\begin{pmatrix}
		\Phi^{-\hbar}(x-\hbar y) & 0 \cr 0 & \Phi^{\hbar}(x+\hbar y)
	\end{pmatrix}
	\\
	 & =
	\Phi^\hbar(x+\hbar y) \Phi^{-\hbar}(x-\hbar y) \cdot {\rm Id}_{\mathbb{C}^2},
\end{align*}
for $\Lambda = -1$, and
\begin{align*}
	\begin{pmatrix}
		x+{\rm i} \hbar y & 0 \cr 0 & x- {\rm i} \hbar y
	\end{pmatrix}
	 & \mapsto
	\begin{pmatrix}
		\Phi^{{\rm i}\hbar}(x+{\rm i}\hbar y) & 0 \cr 0 & \Phi^{-{\rm i}\hbar}(x-{\rm i} \hbar y)
	\end{pmatrix}
	\begin{pmatrix}
		\Phi^{-{\rm i}\hbar}(x-{\rm i}\hbar y) & 0 \cr 0 & \Phi^{{\rm i}\hbar}(x+{\rm i}\hbar y)
	\end{pmatrix}
	\\
	 & = \Phi^{{\rm i}\hbar}(x+{\rm i}\hbar y) \Phi^{-{\rm i}\hbar}(x-{\rm i}\hbar y) \cdot {\rm Id}_{\mathbb{C}^2},
\end{align*}
for $\Lambda = 1$.

So, in the cases $\Lambda=-1,1$, eq.\eqref{eq:uniform_function_heuristic}  just boils down to $F^\hbar_\Lambda(x,y) \cdot {\rm Id}_{\mathbb{C}^2}$, where
\begin{align*}
	F^\hbar_{-1} (x,y) & := \Phi^\hbar(x+\hbar y) \Phi^{-\hbar}(x-\hbar y),
	\cr
	F^\hbar_1(x,y)     & := \Phi^{{\rm i}\hbar}(x+{\rm i} \hbar y) \Phi^{-{\rm i} \hbar}(x-{\rm i} \hbar y).
\end{align*}
These functions $F^\hbar_{-1}$ and $F^\hbar_1$ are the actual functions that will be used in the construction of the mutation intertwiner ${\bf K}_{\Gamma,\Gamma'}$ in the next subsection, for the cases $\Lambda=-1$ and $\Lambda=1$ respectively. And these are what are in complete alignment with the somewhat isolated-looking function $F_0$. First, if one writes down the expression $\Phi^{\ell\hbar}(x+\ell \hbar y) \Phi^{-\ell\hbar}(x-\ell \hbar y)$ as the exponential of a single contour integral expression, not worrying about being precise but just being heuristic, then for the case $\Lambda=0$ one can observe that it yields the following result
\begin{align}
	\label{eq:F0_contour_integral}
	F_0(x,y) = \exp\left(
	- \frac{y}{2\pi {\rm i} } \int_{\Omega_a} \frac{e^{-{\rm i} p x}}{\sinh (\pi p)} \frac{dp}{p}
	\right),
\end{align}
which in fact makes sense.
One notes that $F_0^\hbar=F_0$ does not depend on $\hbar$; this is because, in the $\Lambda=0$ case, the $\ell$-term captures the derivative in some sense. Moreover, the functions $F^\hbar_\Lambda$ for all three values of $\Lambda$ share common properties, to be used in the next subsection:

\begin{lemma}
	\label{lem:F_Lambda_properties}
	For each $\Lambda = -1,1,0$, for each $x,y\in \mathbb{R}$ one has
	\begin{align*}
		\begin{array}{lrl}
			\mbox{(involutivity)} : & F^\hbar_\Lambda(x,y) F^\hbar_\Lambda(-x,-y) & = e^{xy/(\pi {\rm i})}, \\
			\mbox{(unitarity)} :    & |F^\hbar_\Lambda(x,y)|                      & = 1.
		\end{array}
	\end{align*}
\end{lemma}
The above properties follow from Prop.\ref{prop:properties_of_the_slanted_Barnes_integral} for $\Lambda=-1,1$, and can be easily shown for $\Lambda=0$. To give a preview, we will also establish an operator identity for the functions $F^\hbar_\Lambda$, corresponding to the pentagon equation of $\Phi^\hbar$ as recalled in Prop.\ref{prop:pentagon_Phi_hbar}. Note also that $F_0$ satisfies
\begin{align*}
	\mbox{(difference equation)}:  \quad F_0(x,y+\pi {\rm i}) & = (1+e^{x}) F_0(x,y),
\end{align*}
which is an analog for $\Lambda=0$ of Prop.\ref{prop:properties_of_the_slanted_Barnes_integral}(SB2) for $\Lambda=-1,1$.

\subsection{The mutation intertwiners}\label{subsec:mutation_intertwiner}

We are now ready to describe our solution to the intertwiner problem (QPL5). That is, for each pair of seeds $\Gamma,\Gamma' \in |\Gamma|$ we will construct an intertwining operator
$$
	{\bf K}_{\Gamma,\Gamma'} ~:~ \mathscr{H}_{\Gamma'} \to \mathscr{H}_{\Gamma}
$$
between the quantum Hilbert spaces that represent the quantum coordinate change map $\mu^\hbar_{\Gamma,\Gamma'}$, in the sense of (ITL1)--(ITL2) in \S\ref{subsec:the_case_of_a_cluster_X-variety_and_R_Lambda}. The major nontrivial case is when $\Gamma' = \mu_k(\Gamma)$, in which case we describe the solution as the composition
$$
	{\bf K}_{\Gamma,\Gamma'} = {\bf K}^\sharp_{\Gamma,\Gamma'} \circ {\bf K}'_{\Gamma,\Gamma'} ~:~ \mathscr{H}_{\Gamma'} ~\overset{{\bf K}'_{\Gamma,\Gamma'}}{\longrightarrow}~ \mathscr{H}_{\Gamma} ~\overset{{\bf K}^\sharp_{\Gamma,\Gamma'}}{\longrightarrow}~ \mathscr{H}_{\Gamma},
$$
where the two parts ${\bf K}^\sharp_{\Gamma,\Gamma'}$ and ${\bf K}'_{\Gamma,\Gamma'}$ shall satisfy the intertwining equations for the automorphism part $\mu^\sharp_{\Gamma,\Gamma'}$ and the monomial transformation part $\mu'_{\Gamma,\Gamma'}$ which decompose the quantum coordinate change isomorphism $\mu^\hbar_{\Gamma,\Gamma'}$ as seen in eq.\eqref{eq:quantum_isomorphism}. That is, analogously to eq.\eqref{eq:K_sharp_intertwining_equations}--\eqref{eq:K_prime_intertwining_equations}, we require the following versions of the intertwining equations:
\begin{align}
	\label{eq:K_sharp_Lambda_intertwining_equations}
	( {\bf K}^\sharp_{\Gamma,\Gamma'} \, \pi_{\Gamma'}( u) ) \, \eta & = ( \pi_\Gamma(\mu^\sharp_{\Gamma,\Gamma'}( u)) \, {\bf K}^\sharp_{\Gamma,\Gamma'} ) \, \eta, \qquad \;\;\mbox{$\forall u \in \mu'_{\Gamma,\Gamma'}(\mathbb{L}^\hbar_{\Gamma'})$, $\forall \eta \in {\bf K}'_{\Gamma,\Gamma'}(\mathscr{S}_{\Gamma'})$,}  \hspace{-10mm} \\
	\label{eq:K_prime_Lambda_intertwining_equations}
	( {\bf K}'_{\Gamma,\Gamma'} \, \pi_{\Gamma'}( u') ) \, \eta'     & = ( \pi_\Gamma(\mu'_{\Gamma,\Gamma'}( u')) \, {\bf K}'_{\Gamma,\Gamma'} ) \, \eta', \qquad \mbox{$\forall u' \in \mathbb{L}^\hbar_{\Gamma'}$, $\forall \eta' \in \mathscr{S}_{\Gamma'}$.}
\end{align}

\vs

We first construct the monomial transformation part ${\bf K}'_{\Gamma,\Gamma'} : \mathscr{H}_{\Gamma'} \to \mathscr{H}_{\Gamma}$, by doubling Fock-Goncharov's counterpart $
	\mathring{\bf K}'_{\Gamma,\Gamma'} : \mathring{\mathscr{H}}_{\Gamma'} = L^2(\mathbb{R}^I, \underset{i \in I}{\wedge} dt_i') \to L^2(\mathbb{R}^I, \underset{i \in I}{\wedge} dt_i) = \mathring{\mathscr{H}}_\Gamma
$ constructed in \cite{FG09a} which we reviewed in \S\ref{subsec:FG_mutation_intertwiner}:
$$
	{\bf K}'_{\Gamma,\Gamma'} := {\rm Id}_{\mathbb{C}^2} \otimes \mathring{\bf K}'_{\Gamma,\Gamma'} ~:~ \mathscr{H}_{\Gamma'} = \mathbb{C}^2 \otimes \mathring{\mathscr{H}}_{\Gamma'} \to \mathbb{C}^2 \otimes \mathring{\mathscr{H}}_\Gamma = \mathscr{H}_{\Gamma}.
$$
In particular, one can easily see that ${\bf K}'_{\Gamma,\Gamma'}$ is unitary. From the conjugation action of Fock-Goncharov's $\mathring{\bf K}'_{\Gamma,\Gamma'}$ on the basic operators ${\bf x}_i$ and ${\bf y}_i$ as recalled in Cor.\ref{cor:conjugation_action_of_K_prime_on_x_and_y}, one can easily deduce the conjugation action of ${\bf K}'_{\Gamma,\Gamma'}$ on our linear quantum operators ${\bf z}^{(\epsilon)}_i = ({\rm Id}_{\mathbb{C}^2} \otimes {\bf x}_i) + \epsilon \wh{\ell} \hbar ({\rm Id}_{\mathbb{C}^2} \otimes {\bf y}_i)$, defined in eq.\eqref{eq:our_log_quantum_operator}:
\begin{corollary}
	\label{cor:K_prime_intertwining_equations_on_z}
	When $\Gamma'=\mu_k(\Gamma)$, for each $i\in I$ and $\epsilon\in\{+,-\}$, one has
	$$
		{\bf K}'_{\Gamma,\Gamma'} \, {{\bf z}'}^{(\epsilon)}_{\hspace{-1mm}i} \, ({\bf K}'_{\Gamma,\Gamma'} )^{-1} = \left\{
		\begin{array}{ll}
			- {\bf z}^{(\epsilon)}_k                                                 & \mbox{if $i=k$},     \\
			{\bf z}^{(\epsilon)}_i + [\varepsilon_{ik}]_+ \, {\bf z}^{(\epsilon)}_k, & \mbox{if $i\neq k$,}
		\end{array}
		\right.
	$$
	where $\varepsilon$ denotes the exchange matrix of $\Gamma$. Exponentiating the above conjugation equations, one obtains
	$$
		{\bf K}'_{\Gamma,\Gamma'} \, \pi_{\Gamma'}(\wh{Z}_i^{(\epsilon)}) \, ({\bf K}'_{\Gamma,\Gamma'} )^{-1} = \pi_\Gamma(\mu'_{\Gamma,\Gamma'}(\wh{Z}_i^{(\epsilon)}) ).
	$$
\end{corollary}

Keeping track of the domains, one can deduce:
\begin{proposition}[intertwining equations for the monomial-transformation-part operators]
	\label{prop:K_prime_Lambda_intertwining_equations}
	Eq.\eqref{eq:K_prime_Lambda_intertwining_equations}
	\begin{align*}
		( {\bf K}'_{\Gamma,\Gamma'} \, \pi_{\Gamma'}( u') ) \, \eta' & = ( \pi_\Gamma(\mu'_{\Gamma,\Gamma'}( u')) \, {\bf K}'_{\Gamma,\Gamma'} ) \, \eta', \qquad \mbox{$\forall u' \in \mathbb{L}^\hbar_{\Gamma'}$, $\forall \eta' \in \mathscr{S}_{\Gamma'}$}
	\end{align*}
	holds true.
\end{proposition}

Now we deal with the remaining automorphism part ${\bf K}^\sharp_{\Gamma,\Gamma'} : \mathscr{H}_{\Gamma} \to \mathscr{H}_{\Gamma}$, which constitutes the main problem. The heuristic but intuitive expression for our answer can be written as
$$
	{\bf K}^\sharp_{\Gamma,\Gamma'}
	= \Phi^{\ell\hbar}({\bf z}^{\hbar(+)}_k) \, \Phi^{-\ell\hbar}({\bf z}^{\hbar(-)}_k)
$$
for any value of $\Lambda$, by applying the $\mathbb{R}_\Lambda$-functional calculus on normal operators as introduced in \S\ref{subsec:representations_on_doubled_Hilbert_spaces}, to the (heuristic) functions $\Phi^{\pm \ell\hbar}$ studied in \S\ref{subsec:trilogy}. Based on the arguments of these previous subsections, a more precise way of defining this operator is
$$
	{\bf K}^\sharp_{\Gamma,\Gamma'}
	:= {\rm Id}_{\mathbb{C}^2} \otimes F^\hbar_\Lambda({\bf x}_k,{\bf y}_k).
$$
Here, $F^\hbar_\Lambda({\bf x}_k,{\bf y}_k) : \mathring{\mathscr{H}}_\Gamma \to \mathring{\mathscr{H}}_\Gamma$ is the result of applying the usual two-variable functional calculus of the strongly commuting self-adjoint operators ${\bf x}_k$ and ${\bf y}_k$ (relevant for $\Gamma$) to the function $F^\hbar_\Lambda : \mathbb{R}^2 \to \mathbb{C}$ studied in \S\ref{subsec:trilogy}, see e.g. \cite{Sch12} for a treatment of such a functional calculus. In particular, by the unitarity part of Lem.\ref{lem:F_Lambda_properties}, it follows that $F^\hbar_\Lambda({\bf x}_k,{\bf y}_k)$ is unitary, hence so is ${\bf K}^\sharp_{\Gamma,\Gamma'}$.

\vs

For the case $\Lambda=-1$, our answer ${\bf K}^\sharp_{\Gamma,\Gamma'}$ is the doubling of
\begin{align*}
	F^\hbar_{-1}({\bf x}_k,{\bf y}_k) & = \Phi^\hbar({\bf x}_k + \hbar {\bf y}_k) \Phi^{-\hbar}({\bf x}_k - \hbar {\bf y}_k)        \\
	                                  & = \Phi^\hbar({\bf x}_k + \hbar {\bf y}_k) (\Phi^{\hbar}({\bf x}_k - \hbar {\bf y}_k))^{-1}.
\end{align*}
One can easily recognize that this coincides with Fock-Goncharov's automorphism part operator $\mathring{\bf K}^\sharp_{\Gamma,\Gamma'}=\Phi^\hbar(\mathring{\bf x}_k) (\Phi^\hbar(\mathring{\til{\bf x}}_k))^{-1}$, as recalled in \S\ref{subsec:FG_mutation_intertwiner}. This allows us to use Fock-Goncharov's statements proved in \cite{FG09a} to our case of $\Lambda=-1$. We assert that our way of expressing this operator using the symbols $\Phi^\hbar$ and $\Phi^{-\hbar}$ gives a better understanding of the situation than Fock-Goncharov's way of using $\Phi^\hbar$ and $(\Phi^\hbar)^{-1}$.

\vs

We now have to justify our solution to ${\bf K}^\sharp_{\Gamma,\Gamma'}$, for all three values of $\Lambda$. It suffices to show:
\begin{proposition}[intertwining equations for the automorphism-part operators]
	\label{prop:K_sharp_intertwining_equations}
	Eq.\eqref{eq:K_sharp_Lambda_intertwining_equations}
	\begin{align*}
		( {\bf K}^\sharp_{\Gamma,\Gamma'} \, \pi_{\Gamma'}( u) ) \, \eta & = ( \pi_\Gamma(\mu^\sharp_{\Gamma,\Gamma'}( u)) \, {\bf K}^\sharp_{\Gamma,\Gamma'} ) \, \eta, \qquad \;\;\mbox{$\forall u \in \mu'_{\Gamma,\Gamma'}(\mathbb{L}^\hbar_{\Gamma'})$, $\forall \eta \in {\bf K}'_{\Gamma,\Gamma'}(\mathscr{S}_{\Gamma'})$}
	\end{align*}
	holds true.
\end{proposition}

{\it Proof.}
For the case $\Lambda=-1$, our situation is exactly the doubling of Fock-Goncharov's situation. In particular, the above intertwining equations are proved in Theorem 5.6 of \cite{FG09a}.

\vs

For the case $\Lambda=1$, one would diagonalize the intertwining equations and prove for each eigenspace. For instance, for $i \in I$ with $\varepsilon_{ik} = -1$, one would have to show
\begin{align*}
	 & \Phi^{{\rm i}\hbar}({\bf x}_k + {\rm i} \hbar {\bf y}_k) \Phi^{-{\rm i}\hbar}({\bf x}_k - {\rm i} \hbar {\bf y}_k) e^{{\bf x}_i + {\rm i} \hbar {\bf y}_i} \eta
	\cr
	 & \qquad \qquad = e^{{\bf x}_i + {\rm i} \hbar {\bf y}_i} (1+e^{\pi {\rm i} ({\rm i}\hbar)} e^{{\bf x}_k + {\rm i} \hbar {\bf y}_k})\Phi^{{\rm i}\hbar}({\bf x}_k + {\rm i} \hbar {\bf y}_k) \Phi^{-{\rm i}\hbar}({\bf x}_k - {\rm i} \hbar {\bf y}_k)  \eta
\end{align*}
for vectors $\eta$ living in some subspace of $\mathring{\mathscr{H}}_\Gamma$. Note that ${\bf x}_k-{\rm i} \hbar {\bf y}_k$ strongly commutes with ${\bf x}_k+{\rm i} \hbar {\bf y}_k$ and with ${\bf x}_i + {\rm i} \hbar {\bf y}_i$, one can move around the factor $\Phi^{-{\rm i}\hbar}({\bf x}_k-{\rm i} \hbar {\bf y}_k)$ and cancel from both sides, so that it suffices to show
\begin{align}
	\label{eq:i_hbar_statement}
	\Phi^{{\rm i}\hbar}({\bf x}_k + {\rm i} \hbar {\bf y}_k) e^{{\bf x}_i + {\rm i} \hbar {\bf y}_i} \eta = e^{{\bf x}_i + {\rm i} \hbar {\bf y}_i} (1+e^{\pi {\rm i} ({\rm i} \hbar)} e^{{\bf x}_k + {\rm i} \hbar {\bf y}_k})\Phi^{{\rm i}\hbar}({\bf x}_k + {\rm i} \hbar {\bf y}_k) \eta.
\end{align}
In the meantime, using the known statements about the case $\Lambda=-1$, by the similar arguments we know that
\begin{align}
	\label{eq:h_statement}
	\Phi^{h}({\bf x}_k +  h {\bf y}_k) e^{{\bf x}_i + h {\bf y}_i} \eta = e^{{\bf x}_i + h {\bf y}_i} (1+e^{\pi h} e^{{\bf x}_k + h {\bf y}_k})\Phi^{h}({\bf x}_k +  h {\bf y}_k) \eta
\end{align}
holds for $h = \hbar \in \mathbb{R}^+$. Moreover, for a fixed vector $\eta$, both sides of eq.\eqref{eq:h_statement} depend complex analytically on $h$ (one can take the inner product with another fixed vector $\xi$, to yield functions in $h$). Hence, by analytic continuation for $h$ living in an open neighborhood of the first quadrant ${\rm Re}(h)\ge 0$, ${\rm Im}(h)\ge 0$, $h\neq 0$, eq.\eqref{eq:h_statement} holds for $h = {\rm i} \hbar$, i.e. the sought-for eq.\eqref{eq:i_hbar_statement} holds.  Still for $i$ with $\varepsilon_{ik}=-1$, one would also have to show
\begin{align*}
	 & \Phi^{{\rm i}\hbar}({\bf x}_k + {\rm i} \hbar {\bf y}_k) \Phi^{-{\rm i}\hbar}({\bf x}_k - {\rm i} \hbar {\bf y}_k) e^{{\bf x}_i - {\rm i} \hbar {\bf y}_i} \eta
	\cr
	 & \qquad\qquad = e^{{\bf x}_i - {\rm i} \hbar {\bf y}_i} (1+e^{\pi {\rm i} (-{\rm i}\hbar)} e^{{\bf x}_k - {\rm i} \hbar {\bf y}_k})\Phi^{{\rm i}\hbar}({\bf x}_k + {\rm i} \hbar {\bf y}_k) \Phi^{-{\rm i}\hbar}({\bf x}_k - {\rm i} \hbar {\bf y}_k)  \eta
\end{align*}
This time, one can move around and cancel $\Phi^{{\rm i}\hbar}({\bf x}_k + {\rm i} \hbar {\bf y}_k)$ from both sides, and it suffices to show
\begin{align}
	\label{eq:i_hbar_statement2}
	\Phi^{-{\rm i}\hbar}({\bf x}_k - {\rm i} \hbar {\bf y}_k) e^{{\bf x}_i - {\rm i} \hbar {\bf y}_i} \eta = e^{{\bf x}_i - {\rm i} \hbar {\bf y}_i} (1+e^{\pi {\rm i} (-{\rm i} \hbar)} e^{{\bf x}_k - {\rm i} \hbar {\bf y}_k})\Phi^{-{\rm i}\hbar}({\bf x}_k - {\rm i} \hbar {\bf y}_k) \eta.
\end{align}
By applying the analytic continuation of eq.\eqref{eq:h_statement} for $h$ living in an open neighborhood of the fourth quadrant ${\rm Re}(h)\ge 0$, ${\rm Im}(h)\le 0$, $h\neq 0$, one deduces that eq.\eqref{eq:i_hbar_statement2} holds for $h = -{\rm i} \hbar$ with $\hbar \in \mathbb{R}^+$. Likewise, all intertwining equations for the case $\Lambda=1$ can be proved using this analytic continuation argument from the corresponding statements for $\Lambda=-1$.

\vs

For the case $\Lambda=0$, one can also expect the result by putting $\ell\hbar$ into the place of $h$, if one is willing to deal with the ring $\mathbb{R}_0$ (or $\mathbb{C}_0$). Or, one can prove the intertwining equations directly, which we do now. We begin with:
\begin{lemma}
	\label{lem:F0_conjugation1}
	For $\eta \in \mathring{\mathscr{D}}_\Gamma$ one has
	\begin{align*}
		F_0({\bf x}_k, {\bf y}_k) e^{{\bf x}_k} \, \eta & = e^{{\bf x}_k} F_0({\bf x}_k, {\bf y}_k) \, \eta,                                                                                            \\
		F_0({\bf x}_k, {\bf y}_k) e^{{\bf x}_i} \, \eta & = e^{{\bf x}_i}(1+e^{{\bf x}_k})^{-\varepsilon_{ik}} \, F_0({\bf x}_k,{\bf y}_k) \, \eta \qquad \mbox{if i$\neq k$,}                          \\
		F_0({\bf x}_k, {\bf y}_k) {\bf y}_k \, \eta     & = {\bf y}_k F_0({\bf x}_k, {\bf y}_k) \, \eta,                                                                                                \\
		F_0({\bf x}_k, {\bf y}_k) {\bf y}_i \, \eta     & = ({\bf y}_i - \varepsilon_{ik} {\bf y}_k(1+e^{{\bf x}_k})^{-1} e^{{\bf x}_k})  F_0({\bf x}_k, {\bf y}_k) \, \eta \qquad \mbox{if $i\neq k$}.
	\end{align*}
\end{lemma}

{\it Proof of Lem.\ref{lem:F0_conjugation1}}. The first and the third assertions are obvious, for ${\bf x}_k$ and ${\bf y}_k$ strongly commute. For the second and the fourth assertions, say $n = \varepsilon_{ik}$, so that $[{\bf y}_k,{\bf x}_i] = - n \pi {\rm i} = [{\bf x}_k, {\bf y}_i]$. Then, e.g. by the (generalized) Stone-von Neumann theorem (\cite{vNeu31} \cite[Thm.14.8]{Hal13}), one can assume that
$$
	\textstyle {\bf x}_k = t, \quad {\bf y}_k = s, \quad
	{\bf x}_i = n \pi {\rm i} \frac{\partial}{\partial s}, \quad {\bf y}_i = n \pi {\rm i} \frac{\partial}{\partial t}, \quad \mbox{on the space $L^2(\mathbb{R}^2, dtds)$.}
$$
So the second assertion boils down to showing the equality
$$
	( F_0(t,s) \, e^{ n \pi {\rm i} \frac{\partial}{\partial s} } \eta)(t,s) = ( e^{ n \pi {\rm i} \frac{\partial}{\partial s} } (1+e^{t})^{-n} F_0(t,s) \eta) (t,s)
$$
for $\eta(t,s) \in \mathscr{D} \subset L^2(\mathbb{R}^2, dtds)$, where $\mathscr{D}$ is the nice subspace as defined in eq.\eqref{eq:D_Gamma}. In particular, $F_0(t,s)$ acts as multiplication by $F_0(t,s)$. For $\eta(t,s) \in \mathscr{D}$, one can analytically continue in the $s$ variable to an open neighborhood of the strip $0\le \Im s \le n \pi$ or $n\pi \le \Im s \le 0$, and the resulting function has suitable decaying property in the strip, so that $e^{\pi {\rm i} \frac{\partial}{\partial s}}$ acts as the shift by $n\pi {\rm i}$ in the $s$ argument, i.e.
$$(e^{n\pi {\rm i} \frac{\partial}{\partial s}} \eta)(t,s) = \eta(t,s+n\pi {\rm i}).$$
Likewise, (modulo checking the decaying properties), one has
\begin{align*}
	(e^{n \pi {\rm i} \hbar \frac{\partial}{\partial s}} F_0(t,s) \eta)(t,s)
	 & = F_0(t,s+n\pi {\rm i} ) \eta(t,s+n\pi {\rm i} )                      \\
	 & = (1+e^{t})^{(s+n\pi {\rm i} )/(\pi {\rm i})} \eta(t,s+n\pi {\rm i} ) \\
	 & = (1+e^{t})^n F_0(t,s) \eta(t,s+n\pi {\rm i}).
\end{align*}
Hence
\begin{align*}
	(F_0(t,s) \, e^{ n \pi {\rm i} \frac{\partial}{\partial s} } \eta)(t,s)
	 & = F_0(t,s) \eta(t,s+n\pi {\rm i})                                                         \\
	 & = (1+e^{t})^{-n} (e^{n \pi {\rm i} \hbar \frac{\partial}{\partial s}} F_0(t,s) \eta)(t,s) \\
	 & = (e^{n\pi {\rm i} \hbar \frac{\partial}{\partial s}} (1+e^{t})^{-n} F_0(t,s) \eta)(t,s).
\end{align*}
So the second assertion is proved. Meanwhile, the fourth assertion boils down to
$$
	\textstyle (F_0(t,s) (n\pi {\rm i} \frac{\partial}{\partial t}) \eta)(t,s) = ( (n\pi {\rm i} \frac{\partial}{\partial t} - ns(1+e^{t})^{-1}e^{t}) F_0(t,s) \eta)(t,s)
$$
for $\eta(t,s) \in \mathscr{D} \subset L^2(\mathbb{R}^2,dtds)$. Note
\begin{align*}
	(n\pi {\rm i} \frac{\partial}{\partial t} F_0(t,s) \eta)(t,s)
	 & = n\pi {\rm i} \frac{\partial F_0}{\partial t}(t,s) \eta(t,s) +  F_0(t,s) n\pi {\rm i} \frac{\partial \eta}{\partial t}(t,s)                                \\
	 & = n \pi {\rm i} \frac{s}{\pi {\rm i}} (1+e^{t})^{\frac{s}{\pi {\rm i}} - 1} (e^{t}) \eta(t,s) + F_0(t,s) n\pi {\rm i} \frac{\partial \eta}{\partial t}(t,s) \\
	 & = n s (1+e^{t})^{-1} e^{t}F_0(t,s) \eta(t,s) + F_0(t,s) n\pi {\rm i} \frac{\partial \eta}{\partial t}(t,s),
\end{align*}
proving the fourth assertion.
\renewcommand\qedsymbol{(end of proof of Lem.\ref{lem:F0_conjugation1}) $\square$}
\qed

\vs

Thus, $F_0({\bf x}_k,{\bf y}_k) e^{{\bf x}_k} {\bf y}_k \eta = e^{{\bf x}_k} {\bf y}_k F_0({\bf x}_k,{\bf y}_k) \eta$, and for $i\neq k$ we have
\begin{align*}
	F_0({\bf x}_k, {\bf y}_k) e^{{\bf x}_i} {\bf y}_i \eta
	 & = e^{{\bf x}_i}(1+e^{{\bf x}_k})^{-\varepsilon_{ik}} F_0({\bf x}_k,{\bf y}_k) {\bf y}_i \eta                                                                     \\
	 & = e^{{\bf x}_i}(1+e^{{\bf x}_k})^{-\varepsilon_{ik}}  ({\bf y}_i - \varepsilon_{ik} {\bf y}_k(1+e^{{\bf x}_k})^{-1} e^{{\bf x}_k}) F_0({\bf x}_k,{\bf y}_k) \eta \\
	 & = e^{{\bf x}_i} (1+ e^{{\bf x}_k})^{-\varepsilon_{ik}} {\bf y}_i F_0({\bf x}_k,{\bf y}_k) \eta
	\\
	 & \qquad\qquad\qquad - e^{{\bf x}_i} (1+e^{{\bf x_k}})^{-\varepsilon_{ik}-1} e^{{\bf x}_k} \varepsilon_{ik} {\bf y}_k F_0({\bf x}_k,{\bf y}_k) \eta
\end{align*}
It is a straightforward exercise to deduce the sought-for intertwining equations for ${\bf K}^\sharp_{\Gamma,\Gamma'}$ as stated in Prop.\ref{prop:K_sharp_intertwining_equations}.
\renewcommand\qedsymbol{$\square$}
\qed

\vs

Combining Prop.\ref{prop:K_prime_Lambda_intertwining_equations} and Prop.\ref{prop:K_sharp_intertwining_equations}, one obtains the desired intertwining property for the intertwiner ${\bf K}_{\Gamma,\Gamma'} = {\bf K}^\sharp_{\Gamma,\Gamma'} \circ {\bf K}'_{\Gamma,\Gamma'}$ we constructed for the case $\Gamma' = \mu_k(\Gamma)$. But we need to construct an intertwiner ${\bf K}_{\Gamma,\Gamma'}$ for each pair of seeds $\Gamma,\Gamma' \in |\Gamma|$, not just when the two are related by a single mutation. Another easy elementary case is when $\Gamma' = P_\sigma(\Gamma)$ for a permutation $\sigma$ of the index set $I$. For such a pair, recall from \S\ref{subsec:FG_mutation_intertwiner} that the intertwiner $\mathring{\bf K}_{\Gamma,\Gamma'} : \mathring{\mathscr{H}}_{\Gamma'} = L^2(\mathbb{R}^I) \to L^2(\mathbb{R}^I) = \mathring{\mathscr{H}}_{\Gamma}$ for the quantum cluster $\mathscr{X}$-variety at $\mathbb{R}^+$ is defined as the unitary operator induced by the index-permutation map $\mathbb{R}^I \to \mathbb{R}^I$ corresponding to $\sigma$. For the current situation for quantum $\mathbb{R}_\Lambda$-variety, we assign the doubled version ${\bf K}_{\Gamma,\Gamma'} := {\rm Id}_{\mathbb{C}^2} \otimes \mathring{\bf K}_{\Gamma,\Gamma'} : \mathscr{H}_{\Gamma'} \to \mathscr{H}_\Gamma$. One can then easily show the sought-for intertwining equations hold for this pair of seeds. For a general pair of seeds $\Gamma,\Gamma'$, define ${\bf K}_{\Gamma,\Gamma'}$ as the composition of the intertwiners for these two types of elementary cases, in the style as explained in \S\ref{subsec:FG_mutation_intertwiner}. Then this intertwiner ${\bf K}_{\Gamma,\Gamma'}$ for a general pair of seeds $\Gamma,\Gamma'$ satisfies the intertwining equations as follows.
\begin{theorem}[the intertwining equations for a general pair of seeds]
	For each pair of seeds $\Gamma,\Gamma' \in |\Gamma|$, the operator ${\bf K}_{\Gamma,\Gamma'} : \mathscr{H}_{\Gamma'} \to \mathscr{H}_{\Gamma}$ we constructed in this subsection satisfies the intertwining equations, i.e.
	$$
		( {\bf K}_{\Gamma,\Gamma'} \, \pi_{\Gamma'}( u') ) \, \eta  = ( \pi_\Gamma (\mu^\hbar_{\Gamma,\Gamma'}( u')) \, {\bf K}_{\Gamma,\Gamma'} ) \, \eta
	$$
	holds for all $\eta \in \mathscr{S}_{\Gamma'}$ and all $u' \in \mathbb{L}^\hbar_{\Gamma'}$.
\end{theorem}
Two problems remain, until we can say that we completely solved the quantization problem (QPL5). One is that for a pair of seeds $\Gamma,\Gamma'$, the sequence of mutations and seed automorphisms connecting the two seeds is not unique. So one should make sure that different sequences yield the same intertwining operator ${\bf K}_{\Gamma,\Gamma'}$, at least up to a multiplicative constant. Another problem is to make sure that the consistency equations for the intertwiners hold, i.e.
$$
	{\bf K}_{\Gamma,\Gamma'} \circ  {\bf K}_{\Gamma',\Gamma''} = c_{\Gamma,\Gamma',\Gamma''} \, {\bf K}_{\Gamma,\Gamma''}
$$
for each triple of seeds, for some constant $c_{\Gamma,\Gamma',\Gamma''} \in \mathbb{C}^\times$. These two problems are related to each other, just as in the case of usual cluster $\mathscr{X}$-variety at $\mathbb{R}^+$ as discussed in \S\ref{subsec:FG_mutation_intertwiner}; namely, the former problem implies the latter. We deal with this in the next subsection.

\subsection{Proof of the operator identities for the intertwiners}

In the last subsection we constructed the intertwining operator ${\bf K}_{\Gamma,\Gamma'}$ for elementary pairs of seeds $\Gamma,\Gamma' \in |\Gamma|$, i.e. when $\Gamma' = \mu_k(\Gamma)$ and when $\Gamma' = P_\sigma(\Gamma)$. As in \S\ref{subsec:FG_mutation_intertwiner}, in order to construct the intertwiner for any general pair and to prove their consistency equations, the intertwining operators for the elementary moves $\mu_k$ and $P_\sigma$ must satisfy all the relations satisfied by these moves applied at the seed level; recall Lem.\ref{lem:known_simple_relations} for famous known relations, which come essentially from `rank 1 or 2' cluster algebras and varieties. Recall from Prop.\ref{prop:FST} that, for a seed $\Gamma$ coming from an ideal triangulation of a punctured surface, which yields the space $\mathscr{X}_{|\Gamma|}(\mathbb{R}_\Lambda^+)$ of our main interest to quantize, i.e. related to a certain moduli space of 3d spacetimes, these known relations generate all possible relations. Anyways, our goal of the present subsection is to prove the following:
\begin{theorem}[simple consistency equations for the elementary intertwiners]
	\label{thm:consistency_equations_for_elementary_intertwiners_Lambda}
	For any seed-level relation of mutations and seed automorphisms appearing in Lem.\ref{lem:known_simple_relations}, the operator identity for the corresponding intertwiners ${\bf K}_{\Gamma,\Gamma'}$ constructed in the previous subsection holds up to multiplicative constants.
\end{theorem}

For example, for (R1), when $\Gamma' = \mu_k(\Gamma)$, we have $\Gamma = \mu_k(\Gamma')$, and we must show that
\begin{align}
	\label{eq:R1_operators}
	{\bf K}_{\Gamma,\Gamma'} \circ {\bf K}_{\Gamma',\Gamma} = c_{\Gamma,\Gamma'}{\rm Id}
\end{align}
holds for some constant $c_{\Gamma,\Gamma'}$.

\vs

As mentioned before, for the case $\Lambda=-1$, our intertwiner ${\bf K}_{\Gamma,\Gamma'}$ for a mutation $\mu_k$ and seed automorphism $P_\sigma$ diagonally decomposes into Fock-Goncharov's mutation intertwiner $\mathring{\bf K}_{\Gamma,\Gamma'}$ established in \cite{FG09a}; in particular, Thm.\ref{thm:consistency_equations_for_elementary_intertwiners_Lambda} for $\Lambda=-1$ is proved in Theorem 5.4 of \cite{FG09a}, which is one of the main results of that paper. For the case $\Lambda=1$, our intertwiner ${\bf K}_{\Gamma,\Gamma'}$ has a diagonal decomposition too, which is given in each eigenspace as the product of a factor involving $\Phi^{{\rm i}\hbar}$ and another involving $\Phi^{-{\rm i} \hbar}$. In the style of the proof of Prop.\ref{prop:K_sharp_intertwining_equations} in $\S$\ref{subsec:mutation_intertwiner} , each operator identity for the case $\Lambda=1$, applied to a suitable fixed vector in the Hilbert space, can be proved using analytic continuation arguments from the corresponding identity for $\Lambda=-1$. This settles Thm.\ref{thm:consistency_equations_for_elementary_intertwiners_Lambda} for $\Lambda=1$.

\vs

It only remains to show the consistency equations for the case $\Lambda=0$. We check this directly. Let's begin with (R1), i.e. to show eq.\eqref{eq:R1_operators}; the left hand side is the doubling of
\begin{align*}
	F_0({\bf x}_k, {\bf y}_k) \, \mathring{\bf K}'_{\Gamma,\Gamma'} \, F_0({\bf x}_k', {\bf y}_k') \, \mathring{\bf K}'_{\Gamma',\Gamma}
	= F_0({\bf x}_k, {\bf y}_k)  \, F_0(-{\bf x}_k, -{\bf y}_k) \, \mathring{\bf K}'_{\Gamma,\Gamma'} \, \mathring{\bf K}'_{\Gamma',\Gamma},
\end{align*}
where we used the result of Cor.\ref{cor:conjugation_action_of_K_prime_on_x_and_y} on the conjugation action of $\mathring{\bf K}'_{\Gamma,\Gamma'}$. Recall the involutivity property $F_0(x,y) F_0(-x,-y) = e^{xy/(\pi {\rm i})}$ of Lem.\ref{lem:F_Lambda_properties} which is the counterpart of the involutivity property Prop.\ref{prop:properties_of_the_slanted_Barnes_integral} (SB3) for the quantum dilogarithm functions $\Phi^\hbar$ and $\Phi^{{\rm i}\hbar}$ for the cases $\Lambda=-1,1$. To show that $e^{{\bf x}_k {\bf y}_k/(\pi {\rm i})} \mathring{\bf K}'_{\Gamma,\Gamma'} \, \mathring{\bf K}'_{\Gamma',\Gamma}$ equals a (unitary) scalar operator is a straightforward check which essentially boils down to some simple operator identity on $L^2(\mathbb{R})$; it is proved in \cite{Kim21a}, where a detailed computation is given. In fact, the operator identity for (R1) for the cases $\Lambda=-1,1$ also boils down to showing the same statement, in view of the fact that the involutivity property from Lem.\ref{lem:F_Lambda_properties} is uniform for $F^\hbar_\Lambda$ for all three values of $\Lambda$; and note that the operator identity was already shown in \cite{FG09a} for the case $\Lambda=-1$.

\vs

For (R2), we will show the operator identity corresponding to the identity $\mu_j \mu_i = \mu_i \mu_j$ for $\varepsilon_{ij}=0$; here we used (R1) to transform the original identity into this more symmetric form. Let $\Gamma = \Gamma^{(0)}$ be a seed with $\varepsilon_{ij}=0$. Define $\Gamma^{(1)} = \mu_i (\Gamma^{(0)})$, $\Gamma^{(2)} = \mu_j (\Gamma^{(1)})$, $\Gamma^{(3)} = \mu_j (\Gamma^{(0)})$. Then $\Gamma^{(2)} = \mu_i (\Gamma^{(3)})$, and we shall show
$$
	{\bf K}_{\Gamma^{(0)},\Gamma^{(1)}} \, {\bf K}_{\Gamma^{(1)},\Gamma^{(2)}}
		= {\bf K}_{\Gamma^{(0)},\Gamma^{(3)}} \, {\bf K}_{\Gamma^{(3)},\Gamma^{(2)}}
$$
The operators ${\bf x}_k$, ${\bf y}_k$ acting on the Hilbert space $\mathscr{H}_{\Gamma^{(r)}}$ will be denoted ${\bf x}_k^{(r)}$, ${\bf y}_k^{(r)}$, and the exchange matrix for $\Gamma^{(r)}$ will be denoted $\varepsilon^{(r)}$. Writing each ${\bf K}$ as the composition of two parts, and using the conjugation action of ${\bf K}'$'s, the above sought-for operator identity for $\Lambda=0$ becomes the doubling of
\begin{align*}
	 & F_0({\bf x}^{(0)}_i,{\bf y}^{(0)}_i)
	\, \mathring{\bf K}'_{\Gamma^{(0)},\Gamma^{(1)}}
	F_0({\bf x}^{(1)}_j,{\bf y}^{(1)}_j)
	\, \mathring{\bf K}'_{\Gamma^{(1)},\Gamma^{(2)}}
	\cr
	 & \qquad\qquad = F_0({\bf x}^{(0)}_j,{\bf y}^{(0)}_j)
	\, \mathring{\bf K}'_{\Gamma^{(0)},\Gamma^{(3)}}
	F_0({\bf x}^{(3)}_i,{\bf y}^{(3)}_i)
	\, \mathring{\bf K}'_{\Gamma^{(3)},\Gamma^{(2)}}
\end{align*}
which is equivalent to
\begin{align*}
	 & F_0({\bf x}^{(0)}_i,{\bf y}^{(0)}_i)
	\,
	F_0({\bf x}^{(0)}_j,{\bf y}^{(0)}_j)
	\, \mathring{\bf K}'_{\Gamma^{(0)},\Gamma^{(1)}}
	\, \mathring{\bf K}'_{\Gamma^{(1)},\Gamma^{(2)}}
	\cr
	 & \qquad\qquad = F_0({\bf x}^{(0)}_j,{\bf y}^{(0)}_j)
	\, F_0({\bf x}^{(0)}_i,{\bf y}^{(0)}_i)
	\, \mathring{\bf K}'_{\Gamma^{(0)},\Gamma^{(3)}}
	\, \mathring{\bf K}'_{\Gamma^{(3)},\Gamma^{(2)}}
\end{align*}
Since ${\bf x}_i^{(0)},{\bf y}^{(0)}_i$ strongly commute with ${\bf x}_j^{(0)},{\bf y}^{(0)}_j$, we see that $F_0({\bf x}^{(0)}_i,{\bf y}^{(0)}_i)
	F_0({\bf x}^{(0)}_j,{\bf y}^{(0)}_j)$ equals $F_0({\bf x}^{(0)}_j,{\bf y}^{(0)}_j) F_0({\bf x}^{(0)}_i,{\bf y}^{(0)}_i)$. The equality of the remaining $\mathring{\bf K}'$ operators is a simple linear algebraic check; see \cite{Kim21a}.

\vs

For (R3), we do likewise; we try to show the operator identity corresponding to $P_{(ij)} \mu_j \mu_i = \mu_j \mu_i \mu_j$, applied to $\Gamma$ with $\varepsilon_{ij} = \pm1$. Each mutation intertwiner ${\bf K}$ is decomposed into ${\bf K}^\sharp = F_0(\cdot,\cdot)$ and ${\bf K}'$, and we move all ${\bf K}'$'s to the right, using Cor.\ref{cor:conjugation_action_of_K_prime_on_x_and_y}. In case $\varepsilon_{ij}=1$, the problem then boils down to showing
$$
	F_0({\bf x}_i, {\bf y}_i) F_0({\bf x}_j, {\bf y}_j)
	= F_0({\bf x}_j, {\bf y}_j)  F_0({\bf x}_i+{\bf x}_j, {\bf y}_i+{\bf y}_j)  F_0({\bf x}_i, {\bf y}_i),
$$
and similarly for $\varepsilon_{ij}=-1$. This follows by the following more general form of this operator identity, which we provide a direct proof. In fact, we can formulate the operator identity for all values of $\Lambda$.
\begin{proposition}[the pentagon equation for the functions $F_\Lambda^\hbar$]
	\label{prop:pentagon_identity_F_Lambda}
	If ${\bf x}, {\bf y}, {\bf x}', {\bf y}'$ are self-adjoint operators on a separable Hilbert space satisfying the Weyl-relation-versions of the Heisenberg commutation relations
	$$
		[{\bf x},{\bf x}']=0=[{\bf y},{\bf y}'] = [{\bf x},{\bf y}] = [{\bf x}',{\bf y}'], \qquad
		[{\bf x},{\bf y}'] = \pi {\rm i} \cdot {\rm Id} = [{\bf y},{\bf x}'],
	$$
	then the following operator identity holds as an equality of unitary operators
	$$
		F^\hbar_\Lambda({\bf x}, {\bf y}) F^\hbar_\Lambda({\bf x}', {\bf y}')
		= F^\hbar_\Lambda({\bf x}', {\bf y}')  F^\hbar_\Lambda({\bf x}+{\bf x}', {\bf y}+{\bf y}')  F^\hbar_\Lambda({\bf x}, {\bf y}),
	$$
	where the functions $F^\hbar_\Lambda : \mathbb{R}^2 \to \mathbb{C}$ are as defined in \S\ref{subsec:trilogy}, for $\Lambda=-1,1,0$.
\end{proposition}

{\it Proof.} It remains only to settle the case $\Lambda=0$. By the Stone-von Neumann theorem (\cite{vNeu31} \cite[Thm.14.8]{Hal13}), it suffices to show the statement when
$$
	\textstyle {\bf x} = -t, \quad {\bf y} = \pi {\rm i} \frac{\partial}{\partial s}, \quad
	{\bf x}' = s, \quad {\bf y}' = \pi {\rm i} \frac{\partial}{\partial t}, \quad \mbox{on the space $L^2(\mathbb{R}^2, dtds)$.}
$$
For $\eta(t,s)$ living in the nice subspace $\mathscr{D} \subset L^2(\mathbb{R}^2)$ as in eq.\eqref{eq:D_Gamma}, note that
\begin{align*}
	\textstyle (F_0({\bf x},{\bf y})\eta)(t,s) & = (F_0(-t,\pi{\rm i} \frac{\partial}{\partial s}) \eta)(t,s) = ((1+e^{-t})^{\frac{\partial}{\partial s}} \eta)(t,s) \\
	                                           & = (e^{(\log(1+e^{-t})) \frac{\partial}{\partial s}}\eta)(t,s) = \eta(t, s + \log(1+e^{-t})).
\end{align*}
Likewise, $(F_0({\bf x}',{\bf y}') \eta)(t,s) = \eta(t+ \log(1+e^s),s)$. Note
\begin{align*}
	\textstyle (F_0({\bf x}+{\bf x}', {\bf y} + {\bf y}') \eta)(t,s) & = (F_0(-t+s, \pi {\rm i} (\frac{\partial}{\partial s}+\frac{\partial}{\partial t})) \eta)(t,s) \\
	                                                                 & = ((1+e^{s-t})^{\frac{\partial}{\partial s} + \frac{\partial}{\partial t}}\eta)(t,s)
	= (e^{(\log(1+e^{s-t}))(\frac{\partial}{\partial s} + \frac{\partial}{\partial t})} \eta)(t,s).
\end{align*}
It is not hard to show that the operator $e^{(\log(1+e^{s-t}))(\frac{\partial}{\partial s} + \frac{\partial}{\partial t})}$ acts as
$$
	(e^{(\log(1+e^{s-t}))(\frac{\partial}{\partial s} + \frac{\partial}{\partial t})} \eta)(t,s) = \eta(t+\log(1+e^{s-t}), s + \log(1+e^{s-t})).
$$
One way of seeing it is to use the operator ${\bf S}$ on $L^2(\mathbb{R}^2)$ defined by $({\bf S}\eta)(t,s) = \eta(t,s+t)$. Such operators induced by invertible linear transformations of $\mathbb{R}^2$ are studied in detail in \cite[\S3.3]{Kim21a}; in particular, in this case we have ${\bf S}\, t\, {\bf S}^{-1} = t$, ${\bf S}\,s\, {\bf S}^{-1} = t+s$, ${\bf S}\, {\rm i} \frac{\partial}{\partial t}\,{\bf S}^{-1} = {\rm i}(\frac{\partial}{\partial t} - \frac{\partial}{\partial s})$, ${\bf S} \, {\rm i} \frac{\partial}{\partial s}\,{\bf S}^{-1} = {\rm i} \frac{\partial}{\partial s}$. Hence $e^{(\log(1+e^{s-t}))(\frac{\partial}{\partial s} + \frac{\partial}{\partial t})} = {\bf S}^{-1} e^{(\log(1+e^{s})) \frac{\partial}{\partial t}} {\bf S}$ holds, from which it is easy to obtain the above asserted formula by a straightforward computation. To summarize, for $t,s\in \mathbb{R}$ we shall write $t = \log T$, $s=\log S$ for $T,S>0$. Then
\begin{align*}
	(F_0({\bf x},{\bf y})\eta)(\log T,\log S)                   & = \eta(\log T,\log(S+ST^{-1}))        \\
	(F_0({\bf x}',{\bf y}')\eta)(\log T, \log S)                & = \eta(\log(T+TS), \log S)            \\
	(F_0({\bf x}+{\bf x}',{\bf y}+{\bf y}')\eta)(\log T,\log S) & = \eta(\log(T+S), \log(ST^{-1}(T+S)))
\end{align*}

\vs

Now we prove the sought-for operator equality as follows, when applied to $\eta$.
\begin{align*}
	(F_0({\bf x},{\bf y}) F_0({\bf x}',{\bf y}') \eta)(\log T, \log S) & = (F_0({\bf x}',{\bf y}') \eta)(\log T, \log(S+ST^{-1})) \\
	                                                                   & = \eta(\log(T+TS+S), \log(S+ST^{-1}))                    \\
\end{align*}
\begin{align*}
	 & (F_0({\bf x}',{\bf y}')F_0({\bf x}+{\bf x}',{\bf y}+{\bf y}') F_0({\bf x},{\bf y})\eta)(\log T, \log S) \\
	 & \qquad = (F_0({\bf x}+{\bf x}',{\bf y}+{\bf y}') F_0({\bf x},{\bf y})\eta)(\log(T+TS),\log S)           \\
	 & \qquad = (F_0({\bf x},{\bf y})\eta)(\log(T+TS+S), \log(S(T+TS)^{-1}(T+TS+S)) )                          \\
	 & \qquad = \eta( \log (T+TS+S), \log (S(T+TS)^{-1} (T+TS+S)( 1+(T+TS+S)^{-1}) ) )                         \\
	 & \qquad = \eta( \log(T+TS+S), \log(ST^{-1}(1+S)^{-1}(T+TS+S+1)) )                                        \\
	 & \qquad = \eta(\log(T+TS+S), \log(S T^{-1}(1+T) ) )
	.
\end{align*}
\qed

\vs

The operator identities corresponding to (R4) and (R5) are straightforward to check. They are uniform for all values of $\Lambda$, and they are proven for the case $\Lambda=-1$; see e.g. \cite{Kim21a} for a detailed check. This finishes our proof of Thm.\ref{thm:consistency_equations_for_elementary_intertwiners_Lambda}, hence concludes our solution to the quantization problem (QPL5).

\vs

We remark that, for $\Lambda=-1$, the arguments in \cite{FG09a} actually prove the operator identities of the intertwiners not just for the relations in Lem.\ref{lem:known_simple_relations}, but also for {\em any} seed-level relations of mutations and seed automorphisms. Then analytic continuation yields a corresponding result for $\Lambda=1$, strengthening  Thm.\ref{thm:consistency_equations_for_elementary_intertwiners_Lambda}. In the meantime, it is proved in \cite{Kim21a} that the multiplicative constants appearing in the operator identities for the relations in Lem.\ref{lem:known_simple_relations}, hence also for those in Prop.\ref{prop:FST}, are all equal to $1$, in case $\Lambda=-1$. A corresponding result holds for $\Lambda=1$ by analytic continuation, and for $\Lambda=0$ by direct computations given above.

\subsection{A quantization of moduli spaces of 3d gravity}
\label{subsec:a_deformation_quantization_of_moduli_spaces_of_3d_gravity}

Results of the present section so far settles the quantization problems (QPL3)--(QPL5) to a large extent. As mentioned, for a general seed $\Gamma$, the solutions to the intertwiner problem (QPL5) and the deformation quantization problem (QPL3) are somewhat incomplete, in certain senses. However, for the case when the initial seed $\Gamma$ comes from an ideal triangulation of a punctured surface $S$, which is the main application of the entire constructions of the present paper, these quantization problems are completely solved, as already explained. Namely, our solution to (QPL5) is complete because of Prop.\ref{prop:FST}, and our solution to (QPL3) is concrete and constructive, if we take advantage of Fock-Goncharov's canonical basis \cite{FG06} of classical regular functions on the cluster $\mathscr{X}$-variety $\mathscr{X}_{|\Gamma|}$ and Allegretti-Kim's corresponding quantum canonical basis \cite{AK17}.

\vs

In the meantime, our solution to the quantization problems (QPL1)--(QPL5) for cluster $\mathbb{R}_\Lambda$-varieties, applied to the above special seeds $\Gamma$, does not precisely provide a solution to the problems of quantization of the moduli spaces of 3d gravity. As explained in \S\ref{subsec:geometric_structures_on_the_set_of_R_Lambda_positive_points}, the relevant moduli space $\mathcal{GH}_\Lambda(S\times \mathbb{R})$, which is studied in \cite{MS16} and which is the main space to be quantized in the present paper, is not really the entire $\mathscr{X}_{|\Gamma|}(\mathbb{R}_\Lambda^+)$ but only its symplectic leaf $(\mathscr{X}_{|\Gamma|}(\mathbb{R}_\Lambda^+))_{\rm cusp}$, i.e. the subset of points satisfying constraint equations at punctures. Hence, upon quantization, the monomial elements
$$
	\wh{Z}_p^{(\epsilon)} := \textstyle \prod_{i\in I} (\wh{Z}_i^{(\epsilon)})^{\theta_{p,i}} \in \mathcal{A}^\hbar_\Gamma
$$
corresponding to the constraints at punctures $p$ must be represented by the identity operator; see \S\ref{subsec:generalized_shear_coordinates} for the definition of $\theta_{p,i}$, and \S\ref{subsec:geometric_structures_on_the_set_of_R_Lambda_positive_points} for its generalization in the case of a general seed $\Gamma$. In our representation, the operators
$$
	{\bf Z}_p := \pi_\Gamma(\wh{Z}_p^{(+)}) = \pi_\Gamma(\wh{Z}_p^{(-)})
$$
are self-adjoint operators on $\mathscr{H}_\Gamma$, and one can easily show that they strongly commute with all other operators $\pi_\Gamma(u)$ for $u\in \mathbb{L}^\hbar_\Gamma$; think of the similar statement for the representations of quantum cluster $\mathscr{X}$-variety \cite{FG09a}. Thus one can consider the simultaneous spectral decomposition of all these puncture constraint operators, yielding a direct integral decomposition
$$
	\mathscr{H}_\Gamma = \int_{\lambda \in \mathbb{R}^\mathcal{P}} (\mathscr{H}_\Gamma)_\lambda \, d\lambda
$$
into the slices $(\mathscr{H}_\Gamma)_\lambda$, on which the constraint operator for $p\in \mathcal{P}$ acts as the scalar $\lambda(p)$, where $\mathcal{P}$ is the set of punctures of the punctured surface $S$. However, such an approach using a direct integral does not let us handle each slice, and only provides almost everywhere statements.

\vs

Instead, applying the Stone von-Neumann theorem, one can explicitly build from scratch a new representation space $(\mathscr{H}_{\Gamma})_\lambda$ as the $L^2$ space on a Euclidean space of dimension less than $|I|$, and use suitable linear combinations of the position and momentum (as well as scalar) operators.  A drawback is that such a representation expression is not canonical and needs some extra choice, namely a Lagrangian subspace of the symplectic vector space $\mathbb{R}^I$ whose symplectic form is given by the exchange matrix $(\varepsilon_{ij})_{i,j\in I}$ on the standard basis. Explicit constructions of these slices, for the case of quantum cluster $\mathscr{X}$-varieties, and intertwiners between the different choices are studied in \cite{Kim21b}; a similar construction can be applied here, which could also yield irreducible representations. Another drawback is that one needs to be dealing with more complicated linear combinations of the position and the momentum operators than the representation we constructed. Other than that, most of our solutions to the quantization problem persist. In fact, one thing that could become more complicated is the monomial transformation part operator ${\bf K}'_{\Gamma,\Gamma'}$. Instead of just the kind of operators on $L^2(\mathbb{R}^N)$ induced by linear automorphisms of $\mathbb{R}^N$ as studied in \cite{Kim21a}, one needs some examples of the so-called Shale-Weil intertwiners \cite{Seg63,Sha62,Wei64} which can be thought of as generalizations of the Fourier transform. See \cite{Kim21b} for explanation and review of the necessary methods, and see \cite{Kim_survey} which is a survey on part of the results of the present paper and contains more details on an actual application of these methods to a construction of the slice representation $(\mathscr{H}_\Gamma)_\lambda$ in the case when $\lambda\equiv 0$, which is what we need for the symplectic leaf $(\mathscr{X}_{|\Gamma|}(\mathbb{R}^+_\Lambda))_{\rm cusp}$ of interest.

\vs

We note that, if one chooses a suitable Lagrangian subspace of $\mathbb{R}^I$ for each seed $\Gamma$, a solution to ${\bf K}'_{\Gamma,\Gamma'}$ for the slice representation could be relatively simple. For a seed $\Gamma$ coming from an ideal triangulation $T$ of a surface, there is a geometric choice of a basis of such a Lagrangian subspace. For a fixed triangulation $T$, one notes that the subset $\mathbb{Q}^I \subset \mathbb{R}^I$ can be identified with the set $\mathscr{A}_{|\Gamma|}(\mathbb{Q}^t)$ of tropical rational points of the cluster $\mathscr{A}$-variety $\mathscr{A}_{|\Gamma|}$; each element of $\mathbb{Q}^I$ can be geometrically realized as a rational $\mathscr{A}$-lamination on $S$ \cite{FG06}, i.e. a collection of mutually non-intersecting simple closed curves with rational weights. One can show that the symplectic form on $\mathbb{Q}^I$ given by $\varepsilon$ is an intersection form, i.e. the pairing of two disjoint laminations is zero. So, a collection of simple closed curves forming a pants decomposition of $S$, together with a dual collection of curves, provides a geometric basis of a Lagrangian subspace. When one wants to explicitly write down the slice representations, such a geometric basis will become convenient.

\vs

This concludes our solution to the quantization problem for the leaf $(\mathscr{X}_{|\Gamma|}(\mathbb{R}^+_\Lambda))_{\rm cusp}$, in particular for the moduli space $\mathcal{GH}_\Lambda(S\times \mathbb{R})$. In addition, we note that the above argument also yields a quantization of the other symplectic leaves of $\mathscr{X}_{|\Gamma|}(\mathbb{R}^+_\Lambda)$ corresponding to more general constraints $Z_p = c_p$, where $c_p$ are arbitrary scalars.

\vs
One last remark is on how to interpret our solution to the quantization problem of the moduli spaces of 3d gravity, or in fact on the very formulation of the problem. What did we quantize? We chose to quantize a special class of functions. For $\Lambda=-1$, they are real-valued universally Laurent functions in terms of some real coordinate system, and for $\Lambda=1$ they are complex-valued universally Laurent functions. For $\Lambda=0$, we only worked with $\mathbb{R}_0$-valued universally Laurent functions; to describe the final results in terms of real-valued observables, one may take for example the $\Lambda$-real and the $\Lambda$-imaginary parts, in both classical and quantum settings. One can then describe our quantization for these real-valued functions. Indeed, all our unitary intertwining operators ${\bf K}_{\Gamma,\Gamma'} : \mathscr{H}_{\Gamma'} = \mathbb{C}^2 \otimes \mathring{\mathscr{H}}_{\Gamma'} \to \mathbb{C}^2 \otimes \mathring{\mathscr{H}}_\Gamma = \mathscr{H}_\Gamma$ are of the doubled form ${\rm Id}_{\mathbb{C}^2} \otimes \check{\bf K}_{\Gamma,\Gamma'}$ for a unitary operator $\check{\bf K}_{\Gamma,\Gamma'} : \mathring{\mathscr{H}}_{\Gamma'} \to \mathring{\mathscr{H}}_\Gamma$. So the operators on the doubled Hilbert space $\mathscr{H}_\Gamma = \mathring{\mathscr{H}}_\Gamma \oplus \mathring{\mathscr{H}}_\Gamma$ can be studied completely separately for each direct summand $\mathring{\mathscr{H}}_\Gamma$, i.e. splits into the $\Lambda$-real and the $\Lambda$-imaginary parts. However, in the algebraic aspect, unlike the cases for $\Lambda=-1,1$, when $\Lambda=0$ these real-valued observables do not form a ring; they are not closed under product. One way of interpreting this natural class of classical observables in the case $\Lambda=0$ is perhaps as some semi-direct product of algebras. We leave a further investigation of this interesting phenomenon to the future.

\section*{Acknowledgments}

\noindent{For H.K.: This research was supported by Basic Science Research Program through the National Research Foundation of Korea (NRF) funded by the Ministry of Education (grant number 2017R1D1A1B03030230). This work was supported by the National Research Foundation of Korea (NRF) grant funded by the Korea government(MSIT) (No. 2020R1C1C1A01011151). H.K. acknowledges the support he received from the Associate Member Program of Korea Institute for Advanced Study (Open KIAS program) during 2017--2020. H.K. has been supported by a KIAS Individual Grant (MG047203) at Korea Institute for Advanced Study. H.K. thanks Sung-Jin Oh for helpful discussions.}

\noindent{For C.S.: This research was supported by the National Research Foundation of Korea (NRF) funded by the Korea government (MSIT) (grants no. 2019R1F1A1060827 and no. 2019R1A6A1A11051177). C.S. was also supported by a KIAS Individual Grant (SP036102) via the Center for Mathematical Challenges of Korea Institute for Advanced Study.}

\section*{Declarations}

\subsection*{Conflict of interest}
The authors have no relevant financial or non-financial interests to disclose.

\subsection*{Data Availability}

Data sharing is not applicable to this article as no datasets were generated or analysed during the current study.

\end{document}